\documentclass[aps,prd,amsmath,amssymb,preprintnumbers,nofootinbib,preprint,superscriptaddress]{revtex4-1}
\pdfoutput=1
\usepackage[utf8]{inputenc} 
\usepackage{graphicx}
\usepackage{url}
\usepackage[bookmarks, pagebackref=false]{hyperref}
\usepackage[usenames,dvipsnames]{xcolor}
\definecolor{orange}{cmyk}{0,0.5,1,0}
\definecolor{rossoCP3}{cmyk}{0,.88,.77,.40}
\definecolor{graa}{rgb}{0.8,0.8,0.8}
\definecolor{blaa}{rgb}{0.2,0.2,0.6}
\hypersetup{
	colorlinks, 
	bookmarksopen, 
	bookmarksnumbered,
	citecolor=blaa, 		
	linkcolor=rossoCP3,	
	urlcolor=rossoCP3,			
	    }

\usepackage{amsthm}
\usepackage{bm}
\usepackage{bbm}
\usepackage{pxfonts}

\usepackage{amsmath,amssymb,amsfonts}
\usepackage{color}
\usepackage{float}
\usepackage{hyperref}
\usepackage[Symbolsmallscale]{upgreek}
\usepackage{amsmath}
\usepackage{amsfonts}
\usepackage{amssymb,dsfont}
\usepackage{graphicx}
\usepackage{amssymb}
\usepackage[vcentermath]{youngtab}
\usepackage[all]{xy}
\usepackage{pstricks}
\usepackage{dsfont}%
\usepackage{multirow}

\usepackage{cases}
\usepackage{comment}
\usepackage{enumerate}

\usepackage{placeins}
\usepackage{xspace}
\usepackage{cancel} 

\usepackage{slashed}
\usepackage[caption=false, labelformat=simple, listofformat=subsimple, labelfont=default, margin=5pt,
    justification=raggedright]{subfig}

\makeatletter
	\renewcommand{\p@subfigure}{}
\makeatother

\usepackage{natbib}

\usepackage{feynmf}

\newcommand{\beq}{\begin{eqnarray}}
\newcommand{\eeq}{\end{eqnarray}}

\newcommand{\bmp}{\noindent\begin{minipage}{16cm}}
\newcommand{\emp}{\end{minipage}\vskip 7mm} 

    \newcommand{\ii}{\mathrm{i}}

    \newcommand{\Sp}{\mathrm{Sp}}
    \newcommand{\SO}{\mathrm{SO}}
    \newcommand{\Tr}{\mathrm{Tr}}
    \newcommand{\SU}{\mathrm{SU}}
    \newcommand{\SUF}{\mathrm{SU}(N_{\mathrm{F}})}
    \newcommand{\SpF}{\mathrm{Sp}(N_{\mathrm{F}})}
    \newcommand{\SOF}{\mathrm{SO}(N_{\mathrm{F}})}
    \newcommand{\NF}{N_{\mathrm{F}}}
    \newcommand{\GF}{G_{\mathrm{F}}}
    \newcommand{\HF}{H_{\mathrm{F}}}
    
    \newcommand{\EW}{\mathrm{EW}}
    
    \newcommand{\XiF}{\Xi_{\mathrm{F}}}
    \newcommand{\XiS}{\Xi_{\mathrm{S}}}
    \newcommand{\XiA}{\Xi_{\mathrm{A}}}
    \newcommand{\XiSA}{\Xi_{\mathrm{S}/\mathrm{A}}}
    \newcommand{\XiAS}{\Xi_{\mathrm{A}/\mathrm{S}}}
    \newcommand{\XiAd}{\Xi_{\mathrm{Adj}}}

    \newcommand{\SUR}{\SU(2)_{\mathrm{R}}}

\def\lsim{\mathrel{\rlap{\lower4pt\hbox{\hskip1pt$\sim$}}
    \raise1pt\hbox{$<$}}}                
\def\gsim{\mathrel{\rlap{\lower4pt\hbox{\hskip1pt$\sim$}}
    \raise1pt\hbox{$>$}}}                

\begin{document}

\title{\texorpdfstring{\Large\color{rossoCP3}  Classification of 
NLO operators for composite Higgs models}{Classification of the different breaking schemes}}
\author{Tommi~{\sc Alanne}}
\email{tommi.alanne@mpi-hd.mpg.de}
\affiliation{${\text{Max-Planck-Institut f\"{u}r Kernphysik, Saupfercheckweg 1, 69117 Heidelberg, Germany}}$}
\author{Nicolas~{\sc Bizot}}
\email{n.bizot@ipnl.in2p3.fr}
\author{Giacomo~{\sc Cacciapaglia}}
\email{g.cacciapaglia@ipnl.in2p3.fr}
\affiliation{Universit\'e de Lyon, Universit\'e Lyon 1, CNRS/IN2P3, UMR5822 IPNL, F-69622 Villeurbanne, France}
\author{Francesco~{\sc Sannino}}
\email{sannino@cp3.dias.sdu.dk}
\affiliation{CP3-Origins \& Danish IAS, University of Southern Denmark, Campusvej 55, 5230 Odense, Denmark}

\begin{abstract}
We provide a general  classification of template operators, up to next-to-leading order, that appear in chiral perturbation theories based on the two flavour patterns of spontaneous symmetry breaking  $\SUF/\SpF$ and $\SUF/\SOF$.
All possible explicit-breaking sources parametrised by spurions transforming in the fundamental and in the 
two-index representations of the flavour symmetry are included. 
While our general framework can be applied to any model of strong dynamics, we specialise to composite-Higgs models, where the main explicit breaking sources are a current mass, the gauging of flavour symmetries, and the Yukawa couplings (for the top). For the top, we consider both bilinear couplings and linear ones {\it \`a la} partial compositeness. Our templates provide a basis for lattice calculations in specific models. 
As a special example, we consider the $\SU(4)/\Sp(4)\cong \SO(6)/\SO(5)$ pattern which corresponds to  the minimal fundamental composite-Higgs model.
We further revisit issues related to the misalignment of the vacuum.
In particular, we shed light on the physical properties of the singlet $\eta$, showing that it cannot develop a vacuum expectation value without explicit CP violation in the underlying theory.
\\
[.3cm]
{\footnotesize  \it Preprint: CP$^3$-Origins-2018-002 DNRF90 \& LYCEN-2018-02}
\end{abstract}

\maketitle
\tableofcontents
\clearpage



\section{Introduction} 

The discovery of a Higgs-like boson~\cite{Englert:1964et,Higgs:1964pj} at the LHC experiments is one of the most remarkable scientific successes of the beginning of the century, as it concludes a 50-year-long difficult quest~\cite{Ellis:1975ap}.
While our knowledge of the properties of the new particle is increasing thanks to the extraordinary effort of the experimental collaborations~\cite{Aad:2012tfa,Chatrchyan:2012xdj,Aad:2015zhl}, its true nature is still as elusive as ever.
The lack of signals of new physics in other searches at the LHC (and other experiments) may be telling us that the Standard Model (SM) is the correct model after all, or it may be telling us that new physics may be either light and lurking in signatures that are difficult to access, or heavy and difficult to produce at the LHC. The latter possibility can be seen as an indirect support for theories where electroweak (EW) symmetry breaking is induced by a confining force at a few TeV scale. The time-honoured idea of technicolor~\cite{Weinberg:1975gm,Susskind:1978ms}, in fact, predicts that new resonances besides the Nambu--Goldstone bosons (NGB) eaten by the massive EW gauge bosons should appear above a few TeV and be weakly coupled to the SM (thus difficult to produce).
While early proposals did not have a light scalar that could play the role of the 125 GeV Higgs, such a light scalar can be obtained either as an additional pseudo-NGB (pNGB)~\cite{Kaplan:1983sm,Kaplan:1983fs,Dugan:1984hq} or as a light resonance (whose lightness may derive from an approximate infrared conformal behaviour of the theory~\cite{Yamawaki:1985zg,Holdom:1986ub,Holdom:1987yu,Dietrich:2005jn,Sannino:2005dy,Dietrich:2006cm}).
The idea of a pNGB Higgs has recently been revived via holographic realisations in extra dimensions~\cite{Contino:2003ve}, which share common traits to gauge-Higgs unification models~\cite{Hosotani:1983xw,Hatanaka:1998yp,Dvali:2001qr}.

While most of the recent progress has been based either on holography or on effective theories (see, for instance, Refs~\cite{Contino:2010rs,Bellazzini:2014yua,Panico:2015jxa}), models that can be based on an underlying theory have a special role to play. On the one hand, they may truly be addressing the hierarchy problem as no scalars are present in the theory\footnote{This statement is, of course, incomplete unless a theory that generates the coupling of fermions is also specified.}. On the other hand, they can be studied on the lattice, thus providing quantitative predictions for the phenomenology of the Higgs boson.
In addition, the symmetry-breaking pattern is linked to the properties of the representation of the underlying fermions~\cite{Vafa:1983tf,Kosower:1984aw}: only three cases exist, $\SU(\NF)/\Sp(\NF)$, $\SU(\NF)/\SO(\NF)$ and $\SU(\NF)\times \SU(\NF)/\SU(\NF)$ for pseudo-real, real and complex representations respectively. The minimal composite-Higgs model can be achieved for the first class with $\NF=4$~\cite{Cacciapaglia:2014uja}.
A simple underlying theory based on a gauged $\SU(2)$ has been proposed in Refs~\cite{Ryttov:2008xe,Galloway:2010bp}, and studied on the lattice~\cite{Lewis:2011zb,Hietanen:2014xca,Arthur:2014zda,Arthur:2014lma,Arthur:2016dir,Arthur:2016ozw,Drach:2017jsh} (preliminary results for an underlying $\Sp(4)$ theory can be found in Ref.~\cite{Bennett:2017kga}).  
Other theories widely studied on the lattice are the ones that feature a light CP-even scalar resonance~\cite{Appelquist:2016viq,Aoki:2016wnc} (where the lightness is defined by comparison to the other resonances, such as spin-1 ones). This state has been proposed as a candidate for the discovered Higgs-like boson~\cite{Yamawaki:1985zg,Dietrich:2005jn,Appelquist:2010gy}, even though it is not clear if its couplings can really mimic the ones of the SM Higgs~\cite{Belyaev:2013ida}. A next-to-leading order (NLO) chiral Lagrangian including the singlet has been presented in Ref.~\cite{Hansen:2016fri}.

Motivated by the great progress on the lattice, in this work we focus on the construction of effective theories up to NLO, which include the effect of spurions that explicitly break the global symmetry of the theory. We limit our study to spurions in up to two-index representations of the global symmetry, and provide a complete list of template operators that can be used to construct the NLO counterterm operators once the nature of the spurions is specified.
We then focus on spurions relevant for composite-Higgs models, namely a current mass for the underlying fermions, the gauging of the EW symmetry (embedded in the global symmetry), and the sources generating the Yukawa coupling for the top quark.
The latter play an important role, as they usually are the most relevant spurions in the theory.
There are two distinct ways to introduce such coupling: either via bilinear couplings to a scalar operator, or by linear couplings to fermionic operators. The former follows the old proposal of extended technicolor interactions~\cite{Eichten:1979ah}, while the latter is based on the idea of partial compositeness~\cite{Kaplan:1991dc} which was also realised in holographic models. 
In this work we will consider both: note that, in terms of an underlying theory, both appear as four-fermion interactions involving underlying fermions and elementary ones. Realising partial compositeness in an underlying theory often requires the presence of two distinct representations of the underlying gauge group, with chromodynamics (QCD) interactions sequestered by one and the job of EW symmetry breaking assigned to the other~\cite{Barnard:2013zea,Ferretti:2013kya}.
An NLO chiral Lagrangian for this situation has been constructed in Ref.~\cite{DeGrand:2016pgq}, while preliminary lattice results for the specific model of Ref.~\cite{Ferretti:2014qta} can be found in Refs~\cite{DeGrand:2016mxr,Ayyar:2017uqh,Ayyar:2017qdf}.
The main role of the spurions for the phenomenology of the composite Higgs is to misalign the vacuum toward EW symmetry breaking.

Up to now, the global symmetry $\GF$ has been assumed to be only spontaneously 
broken by the condensation of the strong sector to a subgroup $\HF$. All alignments of $\HF$ within the global symmetry are equivalent from the point of view of the confining force.
However, when explicit breaking sources external to the strong dynamics are present, one direction may be preferred. Furthermore, the sources may also break $\HF$ explicitly:
the prime example is QCD where the current masses and the gauging of electromagnetism explicitly break $\SUF_{\mathrm{V}}$  down to $\mathrm{U}(1)_{\mathrm{EM}}$, generating a mass for the pNGBs, i.e. the pions.
In composite-Higgs models, the explicit breaking sources are crucial to misalign the vacuum with respect to the EW gauge sector and, therefore, to drive EW symmetry breaking and give mass to the Higgs (and additional pNGBs).

The misalignment between the EW preserving and physical vacua is conveniently parametrised by an angle,
$\theta$~\cite{Dugan:1984hq},
and the physical vacuum, $E_\theta$, can be written as
\begin{equation}
E_\theta=U_\theta E U_\theta^T,
\end{equation}
where $E$ is an EW preserving vacuum, and $U_\theta$ is a rotation matrix of $\GF$ connecting the two vacua. 
In the above equation, we have assumed that the underlying fermions are pseudo-real or real, in which case the vacuum is an antisymmetric or symmetric matrix.
The interpretation of the angle $\theta$ is simple, as it can be directly linked to the electroweak scale as $\sin \theta = v/f$, $f$ being the decay constant of the pNGBs.
Thus, the limit $\theta \ll 1$ corresponds to a pNGB Higgs, while for $\theta=\pi/2$ we have a technicolor model where $v=f$.
The value of the angle $\theta$ (as well as the form of the EW preserving vacuum $E$) will be determined by the interplay between the spurions of the theory.

In general, the vacuum may be misaligned along more than one direction, and not just along the Higgs one. This can easily be implemented by rotating the vacuum $E$ (or $E_\theta$) with other rotations in $\GF$ parametrised by the appropriate (broken) generators. Loosely, the misalignment can be thought of as a vacuum expectation value for some of the pNGBs, even though this formalism does not respect the shift symmetry of the theory along the rotated vacuum and is thus dangerous.

The paper is organised as follow.
In Sec.~\ref{Chiral perturbation theory}, we present the chiral perturbation theory based on the two patterns of symmetry breaking: $\SU(\NF)/\SO(\NF)$ and $\SU(\NF)/\Sp(\NF)$.
We introduce  generic spurions belonging to  the fundamental and to the two-index representations of the flavour symmetry and  classify, up to NLO, the  non-derivative operators.
We then specialise to the three main explicit breaking sources in composite-Higgs models. 
In Sec.~\ref{Minimal Fundamental Composite (Goldstone) Higgs }, we give a concrete example with the minimal fundamental composite-Higgs model based on  $\SU(4)/\Sp(4)$.
We discuss the vacuum alignment when NLO contributions are included as well as the properties of the additional pNGB singlet, $\eta$. 
We finally present our conclusions in Sec.~\ref{Conclusion}.
More details about the classification of the relevant operators  and a complete list of templates are given in  the Appendices.


\section{Chiral perturbation theory for pseudo-real and real representations}
\label{Chiral perturbation theory}

The chiral perturbation theory that we introduce in this section is intended to parametrise the low-energy physics of some strongly coupled 
hypercolour (HC) gauge theories.
We focus on the sector of the theory that is responsible for the breaking of the EW sector of the SM with the aim of providing a dynamical symmetry breaking  and solving the hierarchy problem of the Higgs mass.
Thus, the matter content consists of fermions, and their representation under the HC interactions completely determines the pattern of global symmetry breaking \cite{Vafa:1983tf} of the model.
In particular, $\NF$ Weyl fermions in a real or pseudo-real representation lead to a global symmetry $\GF=\SU(\NF)$ which can only be spontaneously broken to $\HF=\SO(\NF)$ or $\HF=\Sp(\NF)$, respectively.
Note that $\NF$ is necessarily even for pseudo-real representations to avoid Witten anomalies~\cite{Witten:1982fp}, while no constraint applies to the real case.
If the representation is complex, $\NF$ Weyl fermions need to be accompanied by equally many anti-fermions to cancel gauge anomalies\footnote{We assume that the theory is vector-like with respect to the SM gauge quantum numbers, so that an EW preserving vacuum is allowed.}, and the global $\GF = \SU(\NF)\times \SU(\NF)$ can only be spontaneously broken to the diagonal subgroup $\SU(\NF)_{\mathrm{D}}$, exactly as in QCD.
After spontaneous breaking, the Weyl fermions pair into massive Dirac fermions, with the dynamical mass leaving the HC symmetry, $G_{\mathrm{HC}}$, unbroken.
In the case of an odd number of Weyl fermions\footnote{However, this class of models can no longer be considered as vector-like gauge theories \cite{Vafa:1983tf,Kosower:1984aw}.} in a real representation of $G_{\mathrm{HC}}$, instead, the dynamics generates a gauge-invariant Majorana mass. 

Note that  underlying models with a different gauge group and fermionic representations may lead, at low energy, to the same chiral perturbation theory, i.e. to the same global symmetry-breaking pattern. 
Furthermore, the number of fermions $\NF$ is constrained by the fact that the unbroken global symmetry needs to contain the EW gauge symmetry of the SM extended to the full custodial symmetry, $G_{\mathrm{EW}}=\SU(2)_{\mathrm{L}}\times \mathrm{U}(1)_\mathrm{Y} \subset \SU(2)_\mathrm{L} \times \SU(2)_\mathrm{R} \subset \HF$, and a Higgs doublet candidate in the coset.
Under these conditions, the minimal coset with an underlying fermionic origin is $\SU(4)/\Sp(4)$~\cite{Cacciapaglia:2014uja}, which can be  generated by a $G_{\mathrm{HC}} = \SU(2)$ gauge group with four Weyl fermions transforming as doublets~\cite{Ryttov:2008xe,Galloway:2010bp}.
The next-to-minimal cosets are  $\GF/\HF = \SU(6)/\Sp(6)$~\cite{Low:2002ws}, $\SU(5)/\SO(5)$~\cite{Dugan:1984hq,Ferretti:2014qta} and $\SU(4)\times \SU(4)/\SU(4)_{\mathrm{D}}$~\cite{Ma:2015gra} for the pseudo-real, real and complex cases respectively.
We focus here on the real and pseudo-real cases as they can be described by a chiral Lagrangian of the same form, and they are 
associated with the smallest viable cosets.
Moreover, the complex case is the one associated with QCD, and it has already been explored in great detail in the literature~\cite{Dashen:1969ez,Dashen:1969eg,Dashen:1970et,Gasser:1983yg}.
We leave the number of flavours, $\NF$, free in order to remain as general as possible.

Besides the spontaneous breaking of $\GF$ due to the strong dynamics, the global symmetries are also explicitly broken by the interactions with the elementary states of the SM: the EW gauge interactions and the interactions giving rise to the top mass are the prime examples. Note also that, as the theories we study are vector-like with respect to the SM gauge interactions (and also non-chiral with respect to the HC interactions), a bare mass term for the fermions can (and should) always be added.
The explicit breaking terms can
be thought of as spurions that transform under both the global symmetry, 
$\GF$, and the
SM symmetries (both gauged and global). The fact that they are not 
dynamical fields
explicitly breaks $\GF$ . They will play a crucial role for the alignment of the condensate with respect to the EW symmetries.

In the following, we first present the chiral Lagrangian associated with the real and pseudo-real cases~\cite{Appelquist:1999dq,Duan:2000dy} up to NLO in the chiral expansion. 
Then, we parametrise the effect of the explicit breaking interactions in the chiral perturbation theory through  generic spurionic fields.
Finally, we specialise to the explicit-breaking sources appearing in composite-Higgs models, namely: 
a current mass for the fundamental fermions, 
the gauging of $G_{\mathrm{EW}}$, and the linear or bilinear couplings between the elementary top quark and the strong sector.

\subsection{Chiral Lagrangian up to NLO}
\label{Usual chiral Lagrangian at NLO}

In this section, we present the NLO chiral perturbation theory for the real and pseudo-real cases. 
Both give rise to an $\SU(\NF)$ global symmetry, $\NF\geq 4$,
and they can, therefore, be described within a unified framework.
We will parametrise the NGBs in terms of a linearly transforming matrix, $\Sigma$, which is symmetric under flavour indices of $\GF$ for the real case and antisymmetric for the pseudo-real one. We finally remind the reader that the chiral expansion is in terms of powers of the momentum $p_\mu$ of the NGBs. 
At LO, i.e. order $p^2$, the chiral Lagrangian reads:
\begin{equation}
    {\cal L}_2 =\dfrac{f^2}{8 c_r^2} \mathrm{Tr} [(D_\mu \Sigma)^\dagger D^\mu \Sigma] +\dfrac{f^2}{8 c_r^2} \mathrm{Tr}[ \chi \Sigma^\dagger + \Sigma \chi^\dagger]~,
    \label{L2-general}
\end{equation}
where $f$ is related to the decay constant of the NGBs\footnote{By expanding the kinetic term in Eq.~(\ref{L2-general}), one obtains the relation to the decay constant defined by:
\begin{equation}
    \langle \mathrm{vac} |{\cal J}_\mu^{\hat{A}}(0) | G^{\hat{B}}(p) \rangle =i p_\mu  \dfrac{f} {\sqrt{2} c_r} \ \delta^{\hat{A}\hat{B}},
    \quad p^2=0.
    \label{decay-constant-FG}
\end{equation}
}, and the complex matrix $\chi$ (and the covariant derivative $D_\mu$) contain scalar (and axial/vector) sources.
Following Ref.~\cite{Belyaev:2016ftv}, we introduce a normalisation factor $c_r$ (equal to $\sqrt{2}$ for real representations, and $1$ for pseudo-real) so that the relation between $f$ and the EW scale, $v$, is the same for all models.\footnote{In the technicolor limit, $f=v$. This is valid if the EW symmetry is embedded in $\Sp(2) \sim \SU(2)$ subgroups of $\Sp(\NF)$ and $\SO(4)$ subgroups of $\SO(\NF)$.}

The NGBs, $G^{\hat{A}}$, are parametrised by the  matrix $\Sigma$ as follows:
\begin{equation}
    \Sigma\equiv \exp(2\sqrt{2} c_r i~ G^{\hat{A}} X^{\hat{A}}/f) ~E,
    \quad
    \Sigma \rightarrow g \Sigma g^T,
    \label{sigma-def1}
\end{equation}
where $E$ is a matrix giving the orientation of the vacuum within $\GF$, and $X^{\hat{A}}$ are the corresponding broken generators. 
In the absence of explicit breaking of the global symmetry, all the vacua are equivalent. 
The broken $X^{\hat{A}}$ and unbroken $S^A$ generators  are defined by the following relations:
\begin{equation}
X^{\hat{A}} E -E (X^{\hat{A}})^T=0,
\quad
S^A E + E (S^A)^T=0,
\label{property-generators}
\end{equation}
and are normalised according to $\mathrm{Tr}[S^A S^B]=1/2~ \delta^{AB}$ and $\mathrm{Tr}[X^{\hat{A}} X^{\hat{B}}]=1/2~ \delta^{\hat{A}\hat{B}}$.

The covariant derivative
 is defined as follows:
\begin{equation}
D_\mu \Sigma \equiv \partial_\mu \Sigma-i j_\mu \Sigma-i \Sigma j_\mu^T,
\quad
j_\mu\equiv v_\mu ^A S^A+ a_\mu^{\hat{A}} X^{\hat{A}}, 
\label{covariant-derivative-sigma}
\end{equation}
where $v_\mu$ and $a_\mu$  are the vector and axial sources, respectively.
It is  convenient to define the field strength tensor $j_{\mu\nu}=\partial_\mu j_\nu- \partial_\nu j_\mu \rightarrow g j_{\mu\nu} g^\dagger$.

Note that, apart from the NGB matrix, $\Sigma$, the other fields appearing in the Lagrangian are external sources that transform in complete representations of $\GF$. They should not be confused with the spurions, that we will introduce in the next section, because they do not break the global symmetries of the strong dynamics.
The transformation properties under $\GF$ of the NGB matrix and of the external sources, as well as their chiral counting, are summarised in Table~\ref{tab0}.

\begin{table}[tb]
\renewcommand{\arraystretch}{1.}
\begin{center}
\begin{tabular}{ c c  c c c }
\hline\hline
\multirow{2}{*}{Fields} &  Transformation & \quad $\HF$ reps (PR)\quad & \quad $\HF$ reps (R)\quad & \quad \multirow{2}{*}{Counting}
\\
 & under $\GF$  & $\SU(4)/\Sp(4)$ & $\SU(5)/\SO(5)$ &  \\
\hline 
$\Sigma$ & $ \Sigma\rightarrow g \Sigma g^T$ & \multirow{2}{*}{${\Yvcentermath1 \tiny \yng(1,1)} = 5$} & \multirow{2}{*}{${\Yvcentermath1 \tiny \yng(2)} = 14$} & ${\cal O} (p^0)$ \\
$D_\mu \Sigma$ & $D_\mu \Sigma\rightarrow g (D_\mu \Sigma) g^T$  & & &  ${\cal O} (p)$ 
\\
\hline
$\chi$ & $\chi\rightarrow g \chi g^T$  & ${\Yvcentermath1 \tiny \yng(1,1)} = 5$ & ${\Yvcentermath1 \tiny \yng(2)} = 14$ &${\cal O} (p^2)$ 
\\
$j_\mu = v_\mu$ & \multirow{2}{*}{$j_\mu\rightarrow g j_\mu g^\dagger +i g(\partial_\mu g)^\dagger$} & ${\Yvcentermath1 \tiny \yng(2)} = 10$ & ${\Yvcentermath1 \tiny \yng(1,1)} = 10$ & \multirow{2}{*}{${\cal O} (p)$} \\
$j_\mu = a_\mu$ & & ${\Yvcentermath1 \tiny \yng(1,1)} = 5$ & ${\Yvcentermath1 \tiny \yng(2)} = 14$ & \\
$j_{\mu \nu}$ & $j_{\mu \nu}\rightarrow g j_{\mu \nu} g^\dagger$ & \multicolumn{2}{c}{Same as $j_\mu$} & ${\cal O} (p^2)$
\\
\hline\hline
\end{tabular}
\end{center}
\caption{Properties of the NGB matrix, $\Sigma$, and of the external sources, $a_\mu$, $v_\mu$ and $\chi$.
The representations under the unbroken global symmetry, $\HF$, are shown in general for the real (R) and pseudo-real (PR) cases as well as for the minimal composite-Higgs models based on $\SU(4)/\Sp(4)$ and $\SU(5)/\SO(5)$.
The chiral counting is also given in the last column.}
\label{tab0}
\end{table}

The NLO chiral Lagrangian at order ${\cal O} (p^4)$ is given by \cite{Bijnens:2009qm}:
\begin{eqnarray}
    {\cal L}_4 &=&L_0 \mathrm{Tr} [D_\mu \Sigma (D_\nu \Sigma)^\dagger D^\mu \Sigma (D^\nu \Sigma)^\dagger] 
    +L_1 \mathrm{Tr} [D_\mu \Sigma(D^\mu \Sigma)^\dagger]^2 
    + L_2 \mathrm{Tr} [D_\mu \Sigma (D_\nu \Sigma)^\dagger] \mathrm{Tr}[D^\mu \Sigma (D^\nu \Sigma)^\dagger]
    \nonumber
    \\
    && + L_3 \mathrm{Tr} [D_\mu \Sigma (D^\mu \Sigma)^\dagger D_\nu \Sigma (D^\nu \Sigma)^\dagger]
    +L_4 \mathrm{Tr}[(D_\mu \Sigma)(D^\mu \Sigma)^\dagger] \mathrm{Tr}[\chi \Sigma^\dagger+\Sigma \chi^\dagger]
    \nonumber
    \\
    &&
    +L_5 \mathrm{Tr}[(D_\mu \Sigma)(D^\mu \Sigma)^\dagger (\chi \Sigma^\dagger+\Sigma \chi^\dagger)]
    +L_6 \mathrm{Tr}[\chi \Sigma^\dagger + \Sigma \chi^\dagger]^2
    +L_7 \mathrm{Tr}[\chi \Sigma^\dagger - \Sigma \chi^\dagger]^2
    \\
    &&
    + L_8 \mathrm{Tr}[\chi \Sigma^\dagger \chi \Sigma^\dagger + \Sigma \chi^\dagger \Sigma \chi^\dagger]
    -i L_9 \mathrm{Tr} [j_{\mu \nu} D^\mu \Sigma (D^\nu \Sigma)^\dagger-j_{\mu\nu}^T (D^\mu \Sigma)^\dagger D^\nu \Sigma]
    \nonumber
    \\
    &&
     + L_{10} \mathrm{Tr} [\Sigma j_{\mu\nu}^T \Sigma^\dagger j^{\mu\nu}] +2 H_1 \mathrm{Tr} [j_{\mu\nu} j^{\mu\nu}]
     + H_2 \mathrm{Tr} [\chi \chi^\dagger],
    \nonumber
    \label{Gasser-Leutwyler-Lagrangian}
\end{eqnarray}
where the coefficients $L_i$ and $H_i$ are low-energy constants (LEC) that only depend on the strong dynamics and can be computed on the lattice once the details of the underlying theory are specified.
The above Lagrangian is expressed in a particular basis where we remove the redundant operators\footnote{When the number of flavours, $\NF$, is small, the Caley--Hamilton relations 
may be used to remove additional redundant operators.
The equations of motion have also been used to remove two other operators.} in complete analogy with the Gasser and Leutwyler~\cite{Gasser:1983yg} Lagrangian for the complex case.

\subsection{Generic spurionic operators}
\label{Introduction of spurions}

The chiral Lagrangian can be completed by introducing explicit breaking terms of the flavour symmetry, $\GF$: in the following, we will employ the spurion technique by defining non-dynamical spurions, $\Xi$, that transform as complete representations of $\GF$.
We will limit ourselves to the lowest-dimensional representations with up to two indices, so that the subscripts F, A, S and Adj indicate, in the following, the fundamental, antisymmetric, symmetric and adjoint representation, respectively.
The spurions also carry quantum numbers related to the SM gauge and global symmetries.
Being agnostic of their origin, we will overlook this in this section, together with their proper counting in the chiral expansion: we will, thus, classify the operators based on the number of spurions. We will then specialise to the quantum numbers and chiral counting for various models of composite Higgs in the next section.
Note that, sometimes, it will be convenient to embed one, or more, elementary SM fields in the definition of the spurion, as we will see in concrete examples in the next section.

\begin{table}[tb]
    \renewcommand{\arraystretch}{1.}
    \begin{center}
	\begin{tabular}{ c c  c c }
\hline\hline
Spurions~~~~ & Transformation ~~~~ & Convenient form
\\
\hline 
$\XiF$ & $\XiF \rightarrow g \XiF$   &  $\XiF \ast \XiF^\dagger$~, ~~ $\Sigma \XiF^* \ast \XiF^T \Sigma^\dagger$
\\
 & & $\XiF \ast \XiF^T \Sigma^\dagger$, ~~$\Sigma \XiF^* \ast \XiF^\dagger$
\\
\hline
$\XiSA$ & $\XiSA \rightarrow g \XiSA g^T$ & $ \XiSA \Sigma^\dagger$, ~~$\Sigma \XiSA^\dagger$
\\
$\XiAd$ & $\XiAd\rightarrow g\XiAd g^\dagger$ & $\XiAd$, ~~$\Sigma ~\XiAd^T\Sigma^\dagger$
\\
\hline\hline
	\end{tabular}
    \end{center}
    \caption{Transformations under $\GF$ of the generic spurions in the fundamental and two-index representations. 
Convenient combinations of these spurions with the NGB matrix, $\Sigma$, are also shown, transforming as $X\rightarrow g X g^\dagger$, as they allow one to easily construct the explicit operators. The tensor product $(\; \ast \; )$ allows one to define a two-index matrix out of a fundamental and an anti-fundamental, $\mathrm{F} \ast \mathrm{F}^\dagger$.}
\label{tabspurions}
\end{table}

Spurions in representations up to two $\GF$ indices are sufficient to describe all the composite-Higgs models we are interested in.
In Tab.~\ref{tabspurions} we list the spurions with their transformation properties. We found it useful to construct, out of these spurions, objects that transform like the adjoint, as they are convenient building blocks for $\GF$-invariant operators.
We also truncate the classification to operators with up to four spurions.
Without a proper chiral counting this restriction may seem arbitrary. However, as we will see in the following, it is enough to derive all the NLO operators we are interested in.
We are now armed to build a general basis of operators: the results we present here and in the Appendices are for the case of pseudo-real representations for which $\Sigma$ is an antisymmetric matrix. The case of real representations, for which $\Sigma$ is symmetric, can be easily derived by exchanging $\XiA \leftrightarrow \XiS$ in the operators.
Using the properties of the spurions, one finds that the only operator involving one spurion is 
$\mathrm{Tr}[\XiA\Sigma^\dagger] +\mathrm{h.c.}$.
Using the convenient forms, operators with two spurions can be straightforwardly constructed, and they are listed in Tab.~\ref{tab1}.
The classification of operators with three and four spurions is more involved, thus we reported details and results in App.~\ref{General classification of the spurionic operators}.

\begin{table}[tb]
    \renewcommand{\arraystretch}{1.}
    \begin{center}
	\begin{tabular}{ c c  c c }
	    \hline\hline
	    & No $\Sigma$ &\qquad\qquad Linear in $\Sigma$\qquad\qquad & Quadratic in $\Sigma$  
	    \\
	    \hline 
	    One spurion &  & $\mathrm{Tr}[\XiA\Sigma^\dagger] + \mathrm{h.c.}$ & 
	    \\
	    \hline
	    Two spurions & $\mathrm{Tr}[ \XiSA \XiSA^\dagger]$ & $\mathrm{Tr}[ \XiSA \Sigma^\dagger \XiAd]+ \mathrm{h.c.}$  
	    & $\mathrm{Tr}[ \XiA\Sigma^\dagger]  \mathrm{Tr}[\Sigma \XiA^\dagger]$    
	    \\
	    & $\mathrm{Tr}[  \XiAd^2]$ & $\XiF^T \Sigma^\dagger \XiF+ \mathrm{h.c.}$ & $\mathrm{Tr}[\XiA \Sigma^\dagger]^2 + \mathrm{h.c.}$
	    \\
	    & $\XiF^\dagger \XiF$ && $\mathrm{Tr}[ \XiSA\Sigma^\dagger\XiSA\Sigma^\dagger]  + \mathrm{h.c.}$
	    \\
	    &  && $\mathrm{Tr}[ \XiAd\Sigma \XiAd^T \Sigma^\dagger]$
	    \\
	    \hline\hline
	\end{tabular}
    \end{center}
    \caption{Operators with one or two spurions $\XiSA$, $\XiF$ and $\XiAd$ and no derivatives.}
    \label{tab1}
\end{table}

Explicit models can contain more than one spurion transforming under the same $\GF$ representation that are distinguished by their SM quantum numbers, thus the list of operators we present here are to be considered a template to build explicit operators in specific models.
Operators that are singlet under the SM symmetries correspond to a potential for the NGBs that will fix the alignment of the vacuum in the $\GF$ space.
Operators that are not singlets, however, need to be coupled to SM fields; alternatively, one can embed the SM fields in dynamical spurions that, therefore, may carry Lorentz indices and spin.
Operators containing derivatives can, in principle, be inferred systematically from the non-derivative ones as explained in App.~\ref{General classification of the spurionic operators}.
We finally remark that the list of operators derived from the above templates may contain redundant operators, which need to be eliminated case by case if one wants to identify the minimal number of independent LECs in the model.
As already mentioned, the chiral counting of each operator crucially depends on the physical origin of the spurions, and this will be discussed in the following section.

\subsection{Explicit breaking sources in composite-Higgs models}
\label{Explicit breaking sources in composite Higgs models}

Having at our disposal a complete basis of 
non-derivative operators involving up to four spurions (see Tab.~\ref{tab1} and App.~\ref{General classification of the spurionic operators}), we now specify the sources of explicit breaking that are relevant in the context of composite-Higgs models.
We focus on the  following possibilities:
\begin{itemize}
    \item[(i)] A  current mass for the underlying fermions $\psi$. In general, this spurion transforms in the same representation of the NGB matrix, $\Sigma$. The maximally symmetric case corresponds to a common mass with the flavour structure aligned to the EW preserving vacuum, $E$.
    \item[(ii)] The gauging of the EW symmetry, $G_{\EW}=\SU(2)_{\mathrm{L}} \times \mathrm{U}(1)_Y \subset \GF$. Note that, in general, additional gauging is allowed if the flavour symmetry, $\GF$, is large enough: for instance, the $\SU(3)_{\mathrm{c}}$ of QCD may be included~\cite{Vecchi:2015fma}, or additional non-SM gauge symmetries. Examples of the latter are a $\mathrm{U}(1)$ symmetry broken on the EW-preserving vacuum, $E$, in the $\SU(4)/\Sp(4)$ case~\cite{Cacciapaglia:2014uja}, or duplicates of the SM gauge symmetries in little-Higgs models~\cite{ArkaniHamed:2002qy,Low:2002ws}.
    \item[(iii)] A SM-like bilinear coupling between the elementary top quark multiplets and the strong dynamics: 
	$Q t^c$ couples to a scalar operator of the strong sector ${\cal O}_{Qt}$ that has the same quantum numbers as the Higgs doublet in the SM. Note that  the coset may allow for more than one doublet, so that multiple choices for ${\cal O}_{Qt}$ within the NGB matrix are allowed.
    \item[(iv)]  Linear couplings {\it \`{a} la} partial compositeness \cite{Kaplan:1991dc} between the elementary top quark multiplets 
	and the strong dynamics: $Q$ and $t^c$ couple separately to the fermionic operators ${\cal O}_Q$ 
	and ${\cal O}_t$, respectively.
\end{itemize}

A detailed list of all the relevant spurions can be found in Table~\ref{tab-spurions}.
These sources of explicit breaking  generate masses for the gauge bosons and SM fermions, as well as a potential for the NGBs and in particular for the Higgs boson. 
Four-fermion interactions among the SM fermions are also generated in the same formalism.
The potential determines the alignment of the vacuum within the flavour symmetry, $\GF$, thus allowing for a spontaneous breaking of the EW symmetry and for mass generation for some of the NGBs (that thus become pseudo-NGB, or pNGB in the following).
The time-honoured result is that the top loops, associated with $\mathrm{(iii)}$ and $\mathrm{(iv)}$, have the correct sign to destabilise 
the Higgs potential, while the current mass and the EW gauging cannot break $G_{\EW}$ alone.
The vacuum alignment in the presence of the above spurions and in the context of the minimal $\SU(4)/\Sp(4)$ model will be 
discussed in detail in  Sec.~\ref{Vacuum alignment in the minimal model}.

The underlying fundamental theory involving the hyper-fermions, $\psi$, dictates the form and the properties of the spurions.
We start, therefore, from the fundamental interactions in order to derive the chiral counting of the spurions as well as their general properties. 
These underlying properties imply that a large number of operators present in the general classification are not anymore allowed in these specific cases. 
In the following, we describe in detail the underlying properties of the composite-Higgs spurions.
The complete NLO basis  of non-derivative operators is reported in App.~\ref{Operators involving mass, gauge and top bilinear spurions} and ~\ref{Spurionic operators generating the top quark mass} 
 where more details on its derivation are given.
%

\begin{table}[tb]
    \renewcommand{\arraystretch}{1.}
    \begin{center}
	\begin{tabular}{ c c c c  c}
	    \hline\hline
	    Explicit breaking
	      & ~~General form   & \quad Explicit form \quad& $G_{\mathrm{SM}}$ & ~~Counting  \\
	    \hline
	    Current mass &  $\XiA$ & $\chi$ & $(1,1)_0$ & ${\cal O}(p^2)$
	    \\
	    \hline
	    Gauging $\SU(2)_{\mathrm{L}}$ &  $\XiAd^\mu$ & $g T^A_{\mathrm{L}} W^\mu_A$ &$(1,1)_0$  & ${\cal O}(p)$
	    \\
	     &  $\XiAd^A$  & $g T^A_{\mathrm{L}}$ & $(1,3)_0$ & ${\cal O}(p)$
	      \\
	    Gauging  $\mathrm{U}(1)_Y$ &  $\XiAd^\mu$ & $g^\prime T_Y B^\mu$ & $(1,1)_0$ &${\cal O}(p)$
	     \\ 
	     &  $\XiAd$  & $g^\prime T_Y$ & $(1,1)_0$  &${\cal O}(p)$
	    \\
	    \hline
	    Top bilinear &  $\XiA^\dagger$ & $y_t P^\alpha (Q_\alpha t^c)^\dagger$ & $(1,1)_0$ & ${\cal O}(p^2)$
	    \\
	    &    $\XiA^{\alpha, \dagger}$ & $y_t P^\alpha$ & $(1,2)_{-1/2}$ &${\cal O}(p)$
	    \\
	    \hline
	    Partial compositeness &  $\Xi_{i}$ & $y_{tL} P_q^\alpha Q_\alpha^\dagger$ & $(1,1)_0$ &${\cal O}(p)$
	    \\
	     &  $\Xi_{i}^\alpha$ & $y_{tL} P_q^\alpha$ & $(3,2)_{1/6}$ & ${\cal O}(\sqrt{p})$
	     \\
	     &  $\Xi_{i}$ & $y_{tR} P_t t^{c \dagger}$ & $(1,1)_0$ &${\cal O}(p)$
	    \\
	     &  $\Xi_{i}$ & $y_{tR} P_t$ & $(\overline{3},1)_{-2/3}$ &${\cal O}(\sqrt{p})$
	     \\
	    \hline\hline
	\end{tabular}\end{center}
    \caption{Spurions parametrising the explicit breaking sources appearing in composite-Higgs models. The representation of the partial-compositeness spurions depends on the trilinear baryon involved in the linear couplings, i.e. on the flavour representation of ${\cal O}_{q,t}$. Then ${i=\{\mathrm{F,A,S,Adj}\}}$. }
    \label{tab-spurions}
\end{table}

\subsubsection{Current mass}

Let us start with the simplest source of explicit breaking, namely a current mass for  the hyper-fermions. 
At the fundamental level, the relevant Lagrangian is given by
\begin{equation}
    {\cal L}_m=-\dfrac{1}{2} ( \psi{\cal M}^\dagger \psi+ \psi^\dagger {\cal M} \psi^\dagger)=-\dfrac{1}{2} ( m^\ast \ \psi E  \psi
	+ m\ \psi^\dagger E^\dagger \psi^\dagger)\,.
    \label{fundamental-mass-Lagrangian}
    \end{equation}
In the second equality, we consider explicitly the maximally symmetric case where the mass matrix is aligned to the EW preserving vacuum $E$: this is not an arbitrary choice, as the mass term itself generates a potential that aligns the vacuum inside $\GF$. In other words, it is the mass term that fixes the matrix $E$. With the maximally symmetric choice, the mass term explicitly breaks $\GF$ to $\HF$, thus giving mass to all the NGBs.
Additional terms further breaking $\HF$ are also possible, and they can be parametrised as additional mass parameters proportional to other EW preserving directions in the vacuum $E'_i$, so that in general the mass term can be written as ${\cal M} = m E + \sum_i \delta m_i E'_i$.
In chiral perturbation theory, the spurion associated with the mass transforms as the NGB matrix $\Sigma$ (as it can be inferred from Eq.~(\ref{fundamental-mass-Lagrangian})), and it can be introduced as a vacuum expectation value for the scalar source, $\chi$, that we introduced in Eqs~(\ref{L2-general}) and (\ref{Gasser-Leutwyler-Lagrangian}).
It is defined as follows
\begin{equation}
    \chi\equiv 2 B_0 {\cal M}=2 m B_0 E~,
    \label{def-chi1}
\end{equation}
with $B_0$ being a positive LEC.

From Tab.~\ref{tab1}, we derive the LO operator (with one spurion) involving $\chi$.
Note that this operator has the same form as the second term in Eq.~(\ref{L2-general}) once we  replace the scalar source with the spurion defined in Eq.~(\ref{def-chi1}).
Expanding  to second order in the Goldstone fields, we get for the  pNGB mass $M_G^2=2 B_0 m$, and thus $\chi$ counts as ${\cal O}(p^2)$.
The  NLO operators involving the mass spurion, $\chi$, can be derived in the same way starting from our general basis of operators.
Due to the counting of $\chi$, a great simplification appears at NLO: only the operators with two spurions need to be considered.
The result is reported in App.~\ref{Operators involving mass, gauge and top bilinear spurions}  and is in agreement with Eq.~(\ref{Gasser-Leutwyler-Lagrangian}) providing a first check of our procedure to derive all the non-derivative NLO operators starting from our template list.

\subsubsection{Gauging of flavour symmetries}
\label{Gauging of the EW symmetry}

We now turn to the second obvious source of explicit breaking, i.e. the gauging of the EW
symmetry, $G_{\EW} \subset \GF$. 
At the fundamental level, the fermions are minimally coupled to the SM gauge bosons via a covariant derivative 
\begin{equation}
    {\cal L}_{\mathrm{gauge}}=i \psi^\dagger \overline{\sigma}^\mu D_\mu \psi,
    \quad
    D_\mu=\partial_\mu -ig T^A_{\mathrm{L}} W_\mu^A -i g^\prime T_Y B_\mu.
\end{equation}
Note that the generators $T_L^A$ and $T_Y$ are written as matrices in the $\GF$ space. However, they are not normalised as the $\GF$ generators in Eq.~(\ref{property-generators}) but in order to reproduce the correct transformation properties of each of the components of $\psi$.
For each gauged generator, thus, one can define a spurion transforming as the adjoint of $\GF$, $\XiAd^A = g T^A_{\mathrm{L}}$ and $\XiAd^Y = g' T_Y$, that also transforms as the adjoint representations of the gauge groups. In the chiral expansion, they inherit the same counting as derivatives, i.e. ${\cal O}(p)$.
It is also convenient to define a spurion that contains the gauge fields, i.e. $\XiAd^\mu = g T^A_{\mathrm{L}} W_\mu^A + g^\prime T_Y B_\mu$, that can be introduced by replacing the vector and axial sources as $j^\mu \to \XiAd^\mu$; see Eq.~(\ref{covariant-derivative-sigma}).
The LO and NLO operators containing EW gauge fields can easily be read off from Eqs~(\ref{L2-general}) and (\ref{Gasser-Leutwyler-Lagrangian}).

The effect of the gauging also appears in non-derivative operators that can be built in terms of the spurions $\XiAd^A$ and $\XiAd^Y$: technically, they should be thought of as counterterms necessary to regulate loops of gauge bosons. Thus, besides the counting of the chiral expansion, one needs to add loop suppression factors in order to correctly estimate the impact of such operators.
The LO operators, containing two spurions and thus appearing at ${\cal O}(p^2)$ read
\begin{equation}
V^{(2)}_g=  C_g  \left( g^2\ \mathcal{F}_{\rm loop}^{\rm SU(2)} f^2 ~\mathrm{Tr}[T^A_{\mathrm{L}} \Sigma (T_{\mathrm{L}}^A)^T \Sigma^\dagger]+
  g^{\prime\,2}\ \mathcal{F}_{\rm loop}^{\rm U(1)}  f^2~\mathrm{Tr}[T_Y\Sigma (T_Y)^T \Sigma^\dagger] \right)~,
  \label{gauge-operators}
\end{equation}
where there appears a single LEC, $C_g$, that depends on the HC dynamics.
The two factors, $\mathcal{F}_{\rm loop}$, contain the details of the loop of elementary gauge bosons: as for both groups we need to consider massless gauge bosons, the loop factors are approximately the same, and they can be estimated to be~\cite{Das:1967it}
\begin{equation}
\mathcal{F}_{\rm loop} \sim \frac{1}{16 \pi^2} \Lambda_{\rm HC}^2 \sim f^2\,,
\label{F-loop}
\end{equation}
where the $\Lambda_{\rm HC}^2 \sim (4 \pi f)^2$ factor comes from the quadratic divergence of the loop. Thus, the loop suppression is compensated by the quadratic sensitivity to the cut-off of the effective theory, and the operators can be estimated to
\begin{equation}
V^{(2)}_g=  C_g  \left( g^2  f^4 ~\mathrm{Tr}[T^A_{\mathrm{L}} \Sigma (T_{\mathrm{L}}^A)^T \Sigma^\dagger]+
  g^{\prime\,2}   f^4~\mathrm{Tr}[T_Y\Sigma (T_Y)^T \Sigma^\dagger] \right)~.
  \label{gauge-operators2}
\end{equation}
It is natural to expect that the contribution of the gauge bosons cannot break the gauge symmetry by misaligning the vacuum~\cite{Peskin:1980gc,Preskill:1980mz}; thus we can assume $C_g > 0$.
Following the chiral counting, NLO terms are generated with four spurions and are proportional to the gauge couplings to the fourth power. However, due to the smallness of the gauge couplings, one can realistically restrict to order $g^2$ and $g^{\prime\,2}$, as it is done in QCD~\cite{Urech:1994hd,Knecht:1999ag}.
Furthermore, we would like to remind the reader that the gauging of additional gauge interactions can be introduced in a similar way as done for the EW ones.
A complete list of NLO non-derivative operators containing gauge spurions can be found in Tab.~\ref{tab-mass-gauge-spurions} in App.~\ref{Operators involving mass, gauge and top bilinear spurions}.

Finally, we would like to point out that the effect of the gauging of additional symmetries within $\GF$, such as QCD or beyond-the-SM symmetries, can be included by adding appropriate terms to Eqs~(\ref{gauge-operators}) and (\ref{gauge-operators2}). No additional LECs are needed, as long as the masses of the additional gauge bosons are generated by the condensation itself.

\subsubsection{Top couplings}
\label{Top couplings}

The third source of explicit breaking relevant  for composite-Higgs models that we consider is due to couplings between the 
elementary top quarks and the strong sector.
Two main possibilities are available: couplings that are either bilinear or linear in the SM fields, with the latter realising the partial compositeness paradigm.
Linear couplings, however, always need an extension of the underlying theory as, minimally, hyper-fermions charged under QCD are needed in order to generate QCD-coloured bound states. This can be done either by sequestering the QCD interactions to a sector containing a different HC representation~\cite{Barnard:2013zea,Ferretti:2013kya}, or by adding heavy-flavours in QCD-like theories~\cite{Vecchi:2015fma}.
In either case, the fermionic operators that couple linearly to tops are made of three hyper-fermions.
Another possibility to achieve partial compositeness is to add hypercoloured scalars, so that the linear couplings arise as renormalisable Yukawa couplings in the underlying theory; see Refs~\cite{Sannino:2016sfx,Cacciapaglia:2017cdi,Sannino:2017utc}.
In all cases, the top partners always appear in a representation of $\GF$ with one or two indices, and thus we will restrict ourselves to these.
\\

$\bullet$ \textbf{Bilinear couplings}

At the fundamental level, we assume that the top mass is generated by the following operators:
\begin{equation}
    {\cal L}_t^{\mathrm{bilinear}}=\sum_i \ \dfrac{y_{t,i}}{\Lambda_{t,i}^{n}} \ (Q_{\alpha} t^c)^\dagger \ {\cal O}_{Qt,i}^\alpha + \mathrm{h.c.} 
	= \sum_i\  \dfrac{y_{t,i}}{\Lambda_{t,i}^2} \ (Q_{\alpha} t^c)^\dagger\  \psi^T P_i^\alpha \psi +\mathrm{h.c.},
    \end{equation}
where $\Lambda_{t,i} \geq \Lambda_\mathrm{HC}$ are scales independent from the strong sector\footnote{The scales $\Lambda_{t,i}$ need to be, at least, larger than the cut-off of the effective theory, because they correspond to additional interactions that may affect the low-energy properties of the strong dynamics. For an explicit example, see Ref.~\cite{Cacciapaglia:2015yra}.}, and $\alpha=1,2$ stands for the index of an $\SU(2)_{\mathrm{L}}$ doublet.
In the second equality, we assume that the scalar operators, ${\cal O}_{Qt,i}$, originating from the strong sector, are fermionic 
bilinears, thus leading to four-fermion operators.
The projectors, $P_i^\alpha$, select the $\SU(2)_{\mathrm{L}}$-doublet components of $\psi^T \psi$ with hypercharge $-1/2$: in general there may be several possibilities, and for an explicit example with four independent couplings, we refer the reader to Ref.~\cite{Ma:2015gra}.
Note that one can write different types of operators where the spurion transforms as $\Sigma$, with ${\cal O}_{Qt,i} =  \psi^\dagger \bar{P}_i^\alpha \psi^\ast$; however the physical results are the same as the matrix $\Sigma$ is always symmetric or antisymmetric.

The spurion encoding the explicit breaking is $\XiA^{\alpha,\dagger} = \sum_i \ y_{t,i} P_i^\alpha$, transforming as a doublet of $\SU(2)_{\mathrm{L}}$ with hypercharge $-1/2$, so that it always needs to appear in pairs in order to build gauge-invariant operators.
Similar to what we did for the gauging, we define a single spurion including elementary fields that reads $\XiA^{Qt,\dagger} = \sum_i y_{t_i} P_i^\alpha (Q_{\alpha} t^c)^\dagger$.
Then, the LO operators associated with the bilinear spurion are given by~\cite{Galloway:2010bp,Cacciapaglia:2014uja}:
\begin{equation}
\begin{split}
{\cal L}_t^{(2)}=&   C_{y} f \  \left( \sum_i y_{t,i} \mathrm{Tr} [\Sigma P_i^\alpha ] (Q_{\alpha} t^c)^\dagger +\mathrm{h.c.} \right)\\
&-C_t f^2 \mathcal{F}_{\rm loop}^{\rm top}\ \left( \sum_i y_{t,i} \mathrm{Tr}[ \Sigma P_i^\alpha] \right) \left( \sum_i y_{t,i}^* \mathrm{Tr}[ P_{i, \alpha}^\dagger \Sigma^\dagger] \right)\,,
\end{split}
    \label{top bilinear operators}
\end{equation}
where the form of the second operator derives from the fact that it is generated by loops of elementary tops. Only two LECs are needed: one relative to the operators generating a mass for the top (the former), and one for the NGB potential (the latter). Note that the dependence on the scales $\Lambda_{t,i}$, which may contain a large anomalous exponent if the theory is conformal above $\Lambda_\mathrm{HC}$, can be embedded in a redefinition of the couplings $y_{t,i}$ without loss of information.
In some cases the presence of many possible alignments of the doublet within the NGB matrix is superfluous as a transformation of $\GF$ may be used to change basis (i.e. reshuffling the hyper-fermions $\psi$) and write a smaller number of couplings without affecting other spurions.
Assuming that the top, $m_t$, and Higgs (pNGB) masses are naturally of the same origin, we can impose the following counting for the spurions: $y_t P^\alpha (Q_{\alpha} t^c)^\dagger\sim {\cal O}(p^2)$ and $y_t P^\alpha \sim {\cal O}(p)$.
Note that the two spurions do not have the same counting contrary to the gauge ones.
Similar to the gauge boson loops, the loop factor for massless tops can be approximated by Eq.~(\ref{F-loop}), where colour and other factors are embedded in the LEC, $C_t$.

At NLO, the $\mathcal{O} (p^4)$ Lagrangian contains five new operators contributing to the potential (for simplicity we omit the sums, so that $\sum_i y_{t,i} P_i^\alpha \to y_t P^\alpha$):
\begin{equation}
\begin{split}
{\cal L}_t^{(4)} \supset - \frac{y_t^4 f^6}{\Lambda_\mathrm{HC}^2} 
&\left\{ C_{t1}\ \left(  \mathrm{Tr}[\Sigma P^\alpha]  \mathrm{Tr}[P^\dagger_\alpha \Sigma^\dagger] \right)^2 
+ C_{t2}\  \mathrm{Tr}[\Sigma P^\alpha \Sigma P^\beta] \mathrm{Tr}[P^\dagger_\alpha \Sigma^\dagger P^\dagger_\alpha \Sigma^\dagger] \right. \\
&~\  + C_{t3}\  \mathrm{Tr}[P_\alpha^\dagger \Sigma^\dagger P_\beta^\dagger P^\gamma \Sigma P^\delta] (\delta^\gamma_\alpha \delta^\delta_\beta 
    + \delta^\gamma_\beta \delta^\delta_\alpha)\\
&~\  +  \left(  C_{t4}\  \mathrm{Tr}[\Sigma P^\alpha]   
    \mathrm{Tr}[\Sigma P^\beta]   \mathrm{Tr}[P^\dagger_\alpha \Sigma^\dagger P^\dagger_\beta \Sigma^\dagger] + \mathrm{h.c.}  \right)   \\
&~\  +\left.  \left( C_{t5} \ \mathrm{Tr}[\Sigma P^\alpha \Sigma P^\beta]   
    \mathrm{Tr}[P^\dagger_\alpha \Sigma^\dagger P^\dagger_\beta \Sigma^\dagger] + \mathrm{h.c.}  \right)  \right\} \,,
    \end{split}
\end{equation}
where the first three operators are self-hermitian. Three additional operators contain one insertion of the hyper-fermion mass spurion:
\begin{equation}
\begin{split}
{\cal L}_t^{(4)} \supset - \frac{y_t^2 f^4}{\Lambda_\mathrm{HC}^2} &\left\{ C_{t6}\ \left(  \mathrm{Tr}[\chi \Sigma^\dagger P^\dagger_\alpha P^\alpha]    \mathrm{Tr}[P^\dagger_\alpha \Sigma^\dagger \chi P^\alpha] \right) + \right. \\
&~\ +\left. C_{t7}\  \mathrm{Tr}[\chi \Sigma^\dagger]  \mathrm{Tr}[P^\dagger_\alpha \Sigma^\dagger] \mathrm{Tr}[ \Sigma P^\alpha] + C_{t8}\ \mathrm{Tr}[P^\dagger_\alpha \Sigma^\dagger] \mathrm{Tr}[\Sigma \chi^\dagger \Sigma P^\alpha]  + \mathrm{h.c.} \right\}\,.
\end{split}
\end{equation}
Other operators involving gauge couplings are also present, and listed in Table~\ref{tab-bilinear-potential} in App.~\ref{Operators involving mass, gauge and top bilinear spurions}.\\

$\bullet$ \textbf{Linear couplings  {\it \`{a} la} partial compositeness}

%
Let us now consider the second  way of giving mass to the top quark by means of linear couplings of the elementary top fields to fermionic operators of the strong dynamics (partial compositeness),
\begin{equation}
    {\cal L}_t^{\mathrm{PC}}= \sum_i\  \dfrac{y_{t_{\mathrm{L},i}}}{\Lambda_{t,i}^{n}}\ Q_{\alpha} {\cal O}_{Q,i}^\alpha 
    + \sum_j\ \dfrac{y_{t_{\mathrm{R},j}}}{\Lambda_{t,j}^{n}} \ t^c {\cal O}_{t,j}+ \mathrm{h.c.},
\label{LtPC}
\end{equation}
where the sums span over all the possible operators and, as for the bilinear case, the interactions are generated at scales $\Lambda_{t,i} \geq \Lambda_\mathrm{HC}$.
We will assume that the operators are made of three underlying fermions, as it happens in all explicit examples~\cite{Barnard:2013zea,Ferretti:2013kya,Ferretti:2014qta,Vecchi:2015fma}; the linear couplings will thus correspond to four-fermion operators.\footnote{There is also the possibility of a hyper-fermion/hyper-gluon bound state. However this is unlikely because it would require the hyper-fermion to be in the adjoint representation of HC, thus making the theory lose asymptotic freedom.} 
As previously mentioned, the operators need to contain at least one hyper-fermion that carries QCD colour, which we denote as $X$, and which corresponds to a different HC representation or to heavy flavours. As a consequence, either one or two $\psi$'s are allowed: the former case corresponds to the fundamental of $\GF$, while the latter corresponds to two-index representations.
The fundamental can also be obtained in models with scalars~\cite{Sannino:2016sfx,Cacciapaglia:2017cdi}.

Spelling out the various cases, the linear couplings can thus be rewritten as follows:
\begin{equation}
    {\cal L}_t^{\mathrm{PC}}= 
    \left\{
\begin{array}{l}
    \dfrac{y_{t_{\mathrm{L}}}}{\Lambda_t^{2}} Q_{\alpha}^\dagger (\psi^\dagger P_Q^\alpha \psi^\ast X^\dagger) 
    + \dfrac{y_{t_{\mathrm{R}}}}{\Lambda_t^{2}} t^{c,\dagger} (\psi^\dagger P_t \psi^\ast X^\dagger) + \mathrm{h.c.},
    \quad  
    \XiAS^{Q,\alpha} = y_{t_{\mathrm{L}}} P_Q^\alpha, \ \XiAS^t = y_{t_{\mathrm{R}}} P_t;
    \\
    \dfrac{y_{t_{\mathrm{L}}}}{\Lambda_t^{2}} Q_{\alpha}^\dagger (\psi^T P_Q^\alpha \psi X^\dagger) 
    + \dfrac{y_{t_{\mathrm{R}}}}{\Lambda_t^{2}} t^{c, \dagger} (\psi^T P_t \psi X^\dagger) + \mathrm{h.c.},
    \quad  
    \XiAS^{Q,\alpha,\dagger} = y_{t_{\mathrm{L}}} P_Q^\alpha, \ \XiAS^{t,\dagger} = y_{t_{\mathrm{R}}} P_t;
    \\
\dfrac{y_{t_{\mathrm{L}}}}{\Lambda_t^{2}} Q_{\alpha}^\dagger (\psi^\dagger P_Q^\alpha \psi X^\dagger) 
    + \dfrac{y_{t_{\mathrm{R}}}}{\Lambda_t^{2}} t^{c, \dagger} (\psi^\dagger P_t \psi X^\dagger) + \mathrm{h.c.},
    \quad  
    \XiAd^{Q,\alpha} = y_{t_{\mathrm{L}}} P_Q^\alpha, \ \XiAd^t = y_{t_{\mathrm{R}}} P_t;
    \\
 \dfrac{y_{t_{\mathrm{L}}}}{\Lambda_t^{2}} Q_{\alpha}^\dagger (P_Q^\alpha  \psi^\dagger X^\dagger X^\dagger) 
    + \dfrac{y_{t_{\mathrm{R}}}}{\Lambda_t^{2}}  t^{c, \dagger} ( P_t \psi^\dagger X^\dagger X^\dagger) + \mathrm{h.c.},
    \quad
    \XiF^{Q,\alpha} = y_{t_{\mathrm{L}}} P_Q^\alpha, \ \XiF^t = y_{t_{\mathrm{R}}} P_t.
    \end{array}
\right.
\label{LtPC2}
\end{equation}
Note that with the preceding definitions the spurions have the same transformation properties as the left-handed composite operators and of the left-handed SM quark fields.
We recall that for each operator representation under $\GF$, there may be several possibilities to embed the top partners, and thus an index $i$ should be intended in the preceding expressions. Furthermore, for the adjoint case, the right-handed top, $t^c$, may be associated with the singlet of $\GF$.
Finally, the case of models with scalars, $\mathcal{S}$, charged under HC can be recovered by replacing $XX \to \Lambda_t^2 \mathcal{S}$ in the case of the fundamental. The projectors $P_Q^\alpha$ and $P_t$ select the components of the bound state that have the same quantum numbers as the elementary SM tops.
In the following, for simplicity, we will assume that the new physics generating the four-fermion interactions will only generate mixing to a single representation of $\GF$, or equivalently that the top mass is dominantly generated by a single operator.
A more general case has been discussed at LO in Refs~\cite{Golterman:2015zwa,Golterman:2017vdj}, and it leads to the presence of a plethora of operators.

The couplings of the underlying theory in Eq.~(\ref{LtPC2}) generate, in the confined phase, linear mixing of the elementary tops to fermionic resonances (i.e. top partners). On top of this, effective operators are generated in terms of the spurions defined above: in the following we will assume that the leading contribution to the top mass is generated by the operators. This assumption is valid as long as the top partners are heavier than the NGB decay constant, $f$, and thus cannot be included as light states in the low energy chiral Lagrangian.

The LO operators contributing to the top mass, for all the choices of spurion representations, are given by the following expressions: 
\begin{equation}
\frac{y_{t_{\mathrm{L}}} y_{t_{\mathrm{R}}} f}{4 \pi} (Q_\alpha t^c)^\dagger \times
\left\{
\begin{matrix}
C_{y \mathrm{A},1}\ \mathrm {Tr}[P_Q^\alpha \Sigma^\dagger P_t \Sigma^\dagger] + C_{y \mathrm{A},2}\ \mathrm {Tr} [P_Q^\alpha \Sigma^\dagger] \mathrm {Tr} [P_t \Sigma^\dagger],
\qquad\qquad & \mathrm{A}
\\
C_{y \mathrm{S}}\ \mathrm {Tr}[P_Q^\alpha \Sigma^\dagger P_t \Sigma^\dagger], 
\qquad\qquad & \mathrm{S}
\\
C_{y \mathrm{Adj}}\ \mathrm {Tr} [P_Q^\alpha \Sigma P_t^T \Sigma^\dagger],
\qquad\qquad & \mathrm{Adj}
\\
C_{y \mathrm{F}}\ \mathrm {Tr} [(P_Q^\alpha \cdot P_t^T) \Sigma^\dagger],
\qquad\qquad & \mathrm{F}
\end{matrix}
\right.
\label{eq:PC-Yukawas}
\end{equation}
plus hermitian conjugate. The factor of $1/4\pi$ derives from applying  naive dimensional analysis (NDA) 
as explained in Refs~\cite{Georgi:1992dw,Buchalla:2013eza}. 
Note that, as expected, the above operators involve both spurions $y_{t_L}$ and $y_{t_R}$ in order to generate the top mass, and that only case A involves two independent operators.
The case of the right-handed top mixing to the singlet can be used only if the left-handed tops are in the antisymmetric representation (as that is the only case with an operator containing a single spurion; see Table~\ref{tab1}), and we do not consider it in the following because of non-minimality.

Similarly, we can construct the operators contributing to the potential for the NGBs. At leading order, there exist operators involving only two spurions only for the case of the antisymmetric and adjoint representations,
\begin{equation}
\mathcal{L}_{t, \mathrm{PC}}^{(2)} = -\frac{f^4}{4 \pi} \times
\left\{
\begin{matrix}
C_{t \mathrm{A}}\ \left( y_{t_{\mathrm{L}}}^2 \mathrm {Tr} [P_Q^\alpha \Sigma^\dagger] \mathrm {Tr} [\Sigma P^\dagger_{Q\alpha}] + y_{t_{\mathrm{R}}}^2 \mathrm {Tr} [P_t \Sigma^\dagger] \mathrm {Tr} [\Sigma P^\dagger_{t}]  \right),
\qquad\qquad & \mathrm{A}
\\
C_{t \mathrm{Adj}}\ \left( y_{t_{\mathrm{L}}}^2 \mathrm {Tr} [P_{Q \alpha}^\dagger \Sigma P_Q^{\alpha T} \Sigma^\dagger] + y_{t_{\mathrm{R}}}^2 \mathrm {Tr} [P_{t}^\dagger \Sigma P_t^T \Sigma^\dagger]  \right),
\qquad\qquad & \mathrm{Adj}
\end{matrix}
\right. \label{pot_A-Adj}
\end{equation}
where a factor of $1/4\pi$ comes from NDA. 
The only consistent chiral counting that allows for these operators to appear at LO, $\mathcal{O} (p^2)$, is that the Yukawa couplings $y_{t_{\mathrm{L/R}}}$ count as $p$. Note that this chiral counting is consistent with the appearance of the NDA factor in Eq.~(\ref{eq:PC-Yukawas}), as the top mass operator would appear at chiral order $\mathcal{O} (p^3)$.

For the spurions in the symmetric and fundamental representations, the leading operators contain at least four spurions, leading to the following expressions
\begin{equation}
\begin{split}
\left. \hspace{-0.17cm}\mathcal{L}_{t, \mathrm{PC}}^{(2)} \right|_{\rm S} =&\  - C_{t\, \mathrm{S},1} \frac{f^4}{(4 \pi)^2}\ \left( y_{t_\mathrm{L}}^4 \ \mathrm {Tr} [P_Q^\alpha \Sigma^\dagger P_Q^\beta \Sigma^\dagger] \mathrm {Tr} [\Sigma P^\dagger_{Q\alpha}\Sigma P^\dagger_{Q\beta}] +  y_{t_\mathrm{R}}^4 \ \mathrm {Tr} [P_t \Sigma^\dagger P_t \Sigma^\dagger] \mathrm {Tr} [\Sigma P^\dagger_{t}\Sigma P^\dagger_{t}] \right. \\
&~\qquad \left.  + y_{t_\mathrm{L}}^2 y_{t_\mathrm{R}}^2 \ \mathrm {Tr} [P_Q^\alpha \Sigma^\dagger P_t \Sigma^\dagger] \mathrm {Tr} [\Sigma P^\dagger_{Q\alpha}\Sigma P^\dagger_{t}] \right)\\
&- C_{t\, \mathrm{S},2} \frac{f^4}{(4 \pi)^2}\ \left( y_{t_\mathrm{L}}^4 \mathrm {Tr} [P_Q^\alpha \Sigma^\dagger P_Q^\beta P^\dagger_{Q\gamma} \Sigma P^\dagger_{Q \sigma}](\delta_\alpha^\gamma \delta_\beta^\sigma+ \delta_\alpha^\sigma \delta_\beta^\gamma) 
+   y_{t_\mathrm{R}}^4 \mathrm {Tr} [P_t \Sigma^\dagger P_t P^\dagger_t \Sigma P^\dagger_t] \right. \\
&~\qquad \left. + y_{t_\mathrm{L}}^2 y_{t_\mathrm{R}}^2  \left( \mathrm {Tr} [P_Q^\alpha \Sigma^\dagger P_t P^\dagger_{Q\alpha} \Sigma P^\dagger_t] + \mathrm {Tr} [P_Q^\alpha \Sigma^\dagger P_t P^\dagger_t \Sigma P^\dagger_{Q \alpha}]  \right)  \right)\,,
\end{split} \label{L2S}
\end{equation}
for the symmetric S, and
\begin{equation}
\begin{split}
\left. \mathcal{L}_{t, \mathrm{PC}}^{(2)} \right|_{\rm F} = &- C_{t\, \mathrm{F}} \frac{f^4}{(4 \pi)^2}\ \left( y_{t_\mathrm{L}}^4  \mathrm {Tr} [P_{Q }^\alpha \cdot P_Q^{\beta T}  \Sigma^\dagger] \mathrm {Tr} [\Sigma P_{Q \beta }^* \cdot P_{Q \alpha}^\dagger] + y_{t_\mathrm{R}}^4  \mathrm {Tr} [P_t \cdot P_t^T \Sigma^\dagger] \mathrm {Tr} [\Sigma P_t^* \cdot P_t^\dagger] \right. \\
&~\qquad \left. + y_{t_\mathrm{L}}^2 y_{t_\mathrm{R}}^2  \mathrm {Tr} [P_{Q }^\alpha \cdot P_t^{T}  \Sigma^\dagger] \mathrm {Tr} [\Sigma P_{t}^* \cdot P_{Q \alpha}^\dagger]  \right) \,,
\end{split}
\end{equation}
for the fundamental F. \footnote{For simplicity we assumed that the LECs are the same for operators that only differ on the type of spurion insertion, $P_Q^\alpha$ or $P_t$. More generally, however, differences may arise due to combinatorics of different origins in the underlying theory of the operators.}

At NLO, many more operators are generated, as listed in  App.~\ref{Operators involving mass, gauge and top bilinear spurions}. For reasons of space, we will limit ourselves here to the operators generated in the case of the symmetric representation.
For the potential, mixed operators involving two Yukawas with the mass spurion or the gauge couplings arise at the same level as the leading pure Yukawa ones listed above.
There exist only one operator with a mass insertion,
\begin{equation}
\left. \mathcal{L}_{t, \mathrm{PC}}^{(4)} \right|_{\rm S} \supset - C_{t\, \mathrm{S},3} \frac{f^4}{\Lambda_\mathrm{HC}^2}\ \left( y_{t \mathrm{L}}^2\  \mathrm{Tr} [\chi \Sigma^\dagger P_Q^\alpha P^\dagger_{Q \alpha}] +  y_{t \mathrm{R}}^2\ \mathrm{Tr} [\chi \Sigma^\dagger P_t P^\dagger_t]\right) + \mathrm{h.c.}\,,
\end{equation}
and four involving gauge couplings
\begin{equation}
\begin{split}
\left. \mathcal{L}_{t, \mathrm{PC}}^{(4)} \right|_{\rm S} \supset& \ - C_{t\, \mathrm{S},4}\frac{f^4}{\Lambda_\mathrm{HC}^2}\ \mathrm{Tr} [\XiS \Sigma^\dagger \XiAd] \mathrm{Tr} [\Sigma \XiS^\dagger \XiAd^\dagger] -  C_{t \mathrm{S},5} \frac{f^4}{\Lambda_\mathrm{HC}^2}\ \mathrm{Tr} [\XiS \Sigma^\dagger \XiAd \Sigma \XiS^\dagger \XiAd^\dagger]  \\
&-C_{t \mathrm{S},6} \frac{f^4}{\Lambda_\mathrm{HC}^2}\ \mathrm{Tr} [\XiS \Sigma^\dagger \XiAd \XiAd \Sigma \XiS^\dagger] -
\left(  C_{t\, \mathrm{S},7} \frac{f^4}{\Lambda_\mathrm{HC}^2}\ \mathrm{Tr} [\XiS \XiS^\dagger \XiAd \Sigma \XiAd^T \Sigma^\dagger] + \mathrm{h.c.} \right)\,,
\end{split}
\end{equation}
where we have left implicit all the possible combinations of Yukawas and gauge couplings.


\section{Minimal $\SU(4)/\Sp(4)$ model}
\label{Minimal Fundamental Composite (Goldstone) Higgs }

In this section, we apply the machinery developed in the previous section to the coset $\SU(4)/\Sp(4)$.
This is the minimal composite-Higgs framework with underlying four-dimensional fermionic realisations~\cite{Cacciapaglia:2014uja}.
Models based on this coset have been studied from an effective point of view in Refs~\cite{Katz:2005au,Gripaios:2009pe,Frigerio:2012uc}, and the coset has also been used to construct minimal technicolor  
models in Refs~\cite{Ryttov:2008xe,Galloway:2010bp,Barnard:2013zea,Ferretti:2013kya,Cacciapaglia:2014uja}.

The most minimal underlying fermionic model is based on a confining $\SU(2)$ gauge group with four Weyl fermions transforming under the 
fundamental representation of the new gauge group~\cite{Ryttov:2008xe,Galloway:2010bp}. Since the fundamental representation of $\SU(2) \sim \Sp(2)$ is pseudo-real, the 
fermion sector has an enhanced global symmetry, $\SU(4)$. The condensate forming due to the new strong dynamics 
then breaks this global symmetry spontaneously to $\Sp(4)$, as confirmed from lattice simulations~\cite{Lewis:2011zb,Hietanen:2014xca}.
The spectrum of this theory has also been extensively studied on the lattice~\cite{Arthur:2014lma,Arthur:2016dir,Arthur:2016ozw,Drach:2017jsh}.
Preliminary lattice studies based on a HC $\Sp(4)$\footnote{Purely fermionic underlying theories of partial compositeness need at least a $\Sp(4)$ hypercolour gauge symmetry.}
 have also been recently published~\cite{Bennett:2017kga}.

In the following, we will revisit the operator analysis that we detailed in the previous section focusing in particular on the potential generated for the NGBs of the model.

\subsection{Electroweak embedding}
  
    The full custodial symmetry of the SM, $\mathrm{SU}(2)_{\mathrm{L}}\times \SUR$, is embedded in 
    $\SU(4)$ by identifying the left and right chiral generators to be 
    \begin{equation}
	\label{eq:gensCust}
	T^A_{\mathrm{L}}=\frac{1}{2}\left(\begin{array}{cc}\sigma_A & 0 \\ 0 & 0\end{array}\right),\quad\text{and}\quad
	T^A_{\mathrm{R}}=\frac{1}{2}\left(\begin{array}{cc} 0 & 0 \\ 0 & -\sigma_A^{T}\end{array}\right),
    \end{equation}
    where $\sigma_i$ are the Pauli matrices. The generator of the hypercharge is then further identified with the diagonal generator 
    of the $\SUR$ group, $Y=T^3_{\mathrm{R}}$.
    
    As discussed in Ref.~\cite{Galloway:2010bp}, there are two inequivalent real vacua that leave the SM chiral group invariant, $E_{\pm}$,
    and we denote the one breaking the EW subgroup completely to the electromagnetic $\mathrm{U}(1)_Q$ by $E_{\mathrm{B}}$. They can be explicitly
    written as
    \begin{equation}
	E_{\pm} = \left( \begin{array}{cc}
	\ii \sigma_2 & 0 \\
	0 & \pm\ii \sigma_2
	\end{array} \right), 
	\qquad \qquad
	E_B  =\left( \begin{array}{cc}
	0 & 1 \\
	-1 & 0
	\end{array} \right) \ .
    \end{equation}
    where we chose the normalisation to be real.

    In general the vacuum can be written as the superposition of the EW preserving and breaking ones, and the physical properties of the NGBs generically do not depend on the choice of the EW preserving vacuum $E_{\pm}$. We will see later in this section that, in some cases, the choice of the EW-preserving vacuum is related to some properties of the spurions.
Following Refs~\cite{Galloway:2010bp,Cacciapaglia:2014uja}, in this paper we  use $E_{-}$ and parameterise the vacuum as
 \begin{equation}
E_\theta=U_\theta E_- U_\theta^T=\cos \theta\, E_- + \sin \theta\, E_B \,,
\qquad
U_\theta=\begin{pmatrix}
\cos \dfrac{\theta}{2} & i \sigma_2 \sin \dfrac{\theta}{2}
\\
i \sigma_2 \sin \dfrac{\theta}{2} & \cos \dfrac{\theta}{2}
\end{pmatrix}\in \SU(4)\,,
 \end{equation}
where the angle $\theta$ describes the misalignment of the unbroken $\Sp(4)$ with respect to the EW embedding and is generated by an $\SU(4)$ rotation $U_\theta$ associated with the generator of the Higgs.

    The (non-linearly-realised) scalar variable describing the dynamics of the NGBs associated with the above 
    breaking pattern and the vacuum $E_{\theta}$ can  then be written, in the unitary gauge, as a matrix \cite{Cacciapaglia:2014uja}:
    \begin{equation}
	\label{eq:sigma}
	\Sigma^\prime =U^{\prime 2} \cdot E_\theta= \exp\left[\dfrac{2 \sqrt{2} i}{f}(h {\mathrm{Y}}^{\hat{4}}+ \eta {\mathrm{Y}}^{\hat{5}}) \right]\cdot E_\theta  = \left[\cos \dfrac{x}{f} 1\!\!1 +\dfrac{2 \sqrt{2} i}{x} \sin \dfrac{x}{f}\,\, (h {\mathrm{Y}}^{\hat{4}}+ \eta {\mathrm{Y}}^{\hat{5}}) \right]\cdot E_\theta\,,
    \end{equation}
with $x=\sqrt{h^2 + \eta^2}$. The matrix $\Sigma^\prime$ transforms linearly under the flavour symmetry $\SU(4)$.
The matrix $U^\prime$ (transforming non-linearly under $\SU(4)$) contains the NGBs along the vacuum $E_\theta$, and the matrices ${\mathrm{Y}}^{\hat{4}, \hat{5}}$ are two of the broken generators associated with the Higgs and additional singlet, $\eta$ (while the remaining three generators are associated with the exact NGBs eaten by the $W$ and $Z$ bosons). Note that the normalisation we chose for the decay constant, $f$, is different from the one adopted in Refs~\cite{Ryttov:2008xe,Galloway:2010bp,Cacciapaglia:2014uja} by a factor of $2 \sqrt{2}$ as we follow the prescription defined in Eq.~(\ref{L2-general}) such that $f = v/\sin \theta$. In this way, $\theta = \pi/2$ corresponds to the technicolor limit where $v = f$.

\subsection{Explicit form of the $\SU(4)$ spurions}
\label{Explicit spurions}

We can now explicitly write the relevant spurions introduced in Sec.~\ref{Explicit breaking sources in composite Higgs models} in the case of the coset $\SU(4)/\Sp(4)$.
Let us start by the current mass:
this spurion does not explicitly break the SM gauge symmetry; thus it needs to be proportional to the EW preserving vacua,
\begin{equation}
    \chi =2B_0 \begin{pmatrix}
    m_1 i \sigma_2 & 0
    \\
    0  & m_2 (-i \sigma_2)
    \end{pmatrix}
    =
    2m B_0 ~ E_- + 2 \delta m B_0 ~ E_+ \,,
\end{equation}
where we define $m = (m_1 + m_2)/2$ and $\delta m = (m_1 - m_2)/2$. In order for the EW preserving vacuum to be aligned with $E_-$, we need to impose $\delta m \ll m$ because it is the potential generated by the mass term that will fix the preferred alignment of the vacuum. Note that both the term proportional to $m$ and the one proportional to $\delta m$ are invariant under (different) $\Sp(4)$ subgroups, while the presence of both non-zero values leaves a common $\SU(2) \times \SU(2)$ subgroup unbroken. In this sense, the parameter $\delta m$ can be thought of as a (small) explicit breaking of the $\Sp(4)$ symmetry in the confined phase. Remarkably, the signs of the mass terms (which thus decide which EW preserving vacuum is chosen) are arbitrary as they are associated with the unphysical phases of the underlying fermions: in fact, one could also choose complex masses, thus selecting a complex (but still CP conserving) vacuum.
The physics of the NGBs will be the same. This fact is very important when studying the vacuum misalignment in the model, and we will provide explicit examples at the end of this section.

The spurions corresponding to the EW gauging including the elementary fields can be written as 
$\Xi_{\mathrm{EW}}^{\mu}=g T^A_{\mathrm{L}}W^{A\,\mu}+g^{\prime}T^3_{\mathrm{R}}B^{\mu}$ 
with 
the explicit forms already given in Eq.~\eqref{eq:gensCust}.

For the top bilinear spurions, transforming as $\mathrm{A}^\dagger$, we have $\Xi_{Qt}=y_{t}P^{\alpha}(Q_{\alpha}t^c)^{\dagger}$, and there is a unique choice for the projectors $P^{1,2}$ given 
by~\cite{Galloway:2010bp,Cacciapaglia:2014uja}
\begin{equation}
    P^1= \dfrac{1}{2}
    \begin{pmatrix}
     0 & 0 & -1 & 0 
     \\
     0& 0 &0 &0 
     \\1 & 0 &0 &0
     \\
     0& 0& 0 &0
    \end{pmatrix},
    \qquad
    P^2= 
    \dfrac{1}{2}
    \begin{pmatrix}
     0 & 0 & 0 & 0 
     \\
     0& 0 &-1 &0 
     \\0 & 1 &0 &0
     \\
     0& 0& 0 &0
    \end{pmatrix}\,.
    \label{projector-bilinear}
\end{equation}
The uniqueness is due to the presence of a single (bi)doublet among the NGBs.

In the case of partial compositeness, we can write the spurions as $\Xi^Q=\sum_{i}y_{t_i L}P_{Q_i}^{\alpha}Q^{\alpha}$
and $\Xi^t=\sum_{i}y_{t_i R}P_{t_i}t^c$: the two sets of projectors, $P_{Q_i}^{\alpha}$ and $P_{t_i}$, thus select the components of the fermionic operator of the strong dynamics that match the quantum numbers of the left-handed doublet and the right-handed singlet, respectively. We recall that an additional U(1)$_X$ charge needs to be included in order to fix the hypercharge of the top partners, so that the SM hypercharge is defined as $Y = T_{\mathrm{R}}^3 + X$.
For the fundamental representation (that has $X_\mathrm{F} = 1/6$\footnote{This charge assignment refers to the partner of $Q$. For $t^c$ the charge assignment is opposite in sign, together with the colour assignment. Recall that we always refer to the left-handed components following the fact that the underlying theories are defined in terms of left-handed Weyl spinors.}) there is only one choice available as clearly seen from the decomposition of the $\SU(4)$ representation under $\SU(2)_{\mathrm{L}} \times \SU(2)_{\mathrm{R}}$, i.e. ${\bf 4} \to (2,1) \oplus (1,\bar{2})$, and the projectors $P_{Q}^{1,2}$ and $P_t$ are given by
\begin{equation}
    P_{Q_1}^1= \begin{pmatrix}
    1 \\ 0 \\ 0 \\ 0
    \end{pmatrix}\,,
    \qquad
    P_{Q_1}^2= \begin{pmatrix}
    0 \\ 1 \\ 0 \\ 0
    \end{pmatrix}\,,
    \qquad
    P_{t_1}= \begin{pmatrix}
    0 \\ 0 \\ 1 \\ 0
    \end{pmatrix}\,.
    \label{projector-PC}
\end{equation}
We recall that in the above case, $t^c$ belongs to an $\SU(2)_{\mathrm{R}}$ anti-doublet, and that the partial-compositeness couplings will violate the extended custodial symmetry needed to protect the $Z$ coupling to left-handed bottom quarks~\cite{Agashe:2006at}.
For the antisymmetric ($X_\mathrm{A} = 2/3$) the decomposition reads ${\bf 6} \to (2,\bar{2}) \oplus (1,1) \oplus (1,1)$, and thus there is a single choice for the doublet, but two for the singlet:
\begin{equation}
\begin{split}
&P_{Q}^1=\dfrac{1}{\sqrt{2}}
\begin{pmatrix}
0 & 0 & 1 &0
\\
0 & 0 &0 & 0
\\
- 1 & 0 & 0 &0
\\
0 & 0 & 0 &0
\end{pmatrix}\,,
\qquad
P_{Q}^2=\dfrac{1}{\sqrt{2}}
\begin{pmatrix}
0 & 0 & 0 &0
\\
0 & 0 & 1 & 0
\\
0 & - 1 & 0 &0
\\
0 & 0 & 0 &0
\end{pmatrix}\,, \\
& P_{t_1}=\dfrac{1}{\sqrt{2}}
\begin{pmatrix}
0 & 0 & 0 &0
\\
0 & 0 &0 & 0
\\
0 & 0 & 0 & 1
\\
0 & 0 & - 1 &0
\end{pmatrix}\,,
\qquad
P_{t_2}=\dfrac{1}{\sqrt{2}}\begin{pmatrix}
0 & 1 & 0 &0
\\
- 1 & 0 &0 & 0
\\
0 & 0 & 0 &0
\\
0 & 0 & 0 &0
\end{pmatrix}\,.
\end{split}
\end{equation}
Note that $P_{t_2} - P_{t_1}$ is aligned with the vacuum $E_-$, so it corresponds to a singlet of $\Sp(4)$ along the EW preserving vacuum, while $P_{t_1} + P_{t_2}$ is part of a 5-plet together with the doublet. 
We want to stress that this assignment is relative to the choice of vacuum, as, for instance, $P_{t_1} + P_{t_2}$ corresponds to the singlet for the $E_+$ vacuum. In general, the right-handed top will couple to a linear combination of the two spurions, i.e. with a generalised projector 
\begin{equation} \label{Pt_A}
P_{t} = A\ P_{t_1} + B\ P_{t_2}\,, \qquad \mbox{with}\;\; |A|^2 + |B|^2 = 1\,. 
\end{equation}
The relative phase of the two coefficients, however, can be rotated away by the use of an $\SU(4)$ transformation along the generator $X^{\hat{5}}$  associated with the singlet in the EW preserving vacuum $E_-$. This corresponds to a relative phase redefinition of the two hyper-fermion doublets: therefore, only if a mass term is present can this phase have physical effects, as we will see in a later section. Noteworthy, the real parts cannot be removed without affecting the gauge spurions.

For the symmetric ($X_\mathrm{S} = 2/3$), the decomposition reads ${\bf 10} \to (2,\bar{2}) + (3,1) + (1,3)$: for both doublet and singlet there is a single choice, with the  singlet associated with the neutral component of the $\SU(2)_{\mathrm{R}}$ triplet. The projectors are similar to the $P_{Q}^\alpha$ and $P_{t_1}$ of the antisymmetric by replacing $-1 \to 1$.

Finally the adjoint ($X_{\mathrm{Adj}}=2/3$) decomposes as ${\bf 15} \to (2,2) + (\bar{2},\bar{2}) + (3,1) + (1,3) + (1,1)$, and thus there are two options for both left- and right-handed tops:
\begin{equation}
\begin{split}
&P_{Q_1}^1=\begin{pmatrix}
0 & 0 & 0 &1
\\
0 & 0 & 0 & 0
\\
0 & 0 & 0 &0
\\
0 & 0 & 0 &0
\end{pmatrix}\,,
\qquad
P_{Q_1}^2=\begin{pmatrix}
0 & 0 & 0 &0
\\
0 & 0 &0 & 1
\\
0 & 0 & 0 &0
\\
0 & 0 & 0 &0
\end{pmatrix}\,,
\qquad
P_{t_1}=\frac{1}{\sqrt{2}}\begin{pmatrix}
0 & 0 & 0 &0
\\
0 & 0 &0 & 0
\\
0 & 0 & 1 &0
\\
0 & 0 & 0 &-1
\end{pmatrix}\,,\\
&P_{Q_2}^1=\begin{pmatrix}
0 & 0 & 0 &0
\\
0 & 0 &0 & 0
\\
0 & -1 & 0 &0
\\
0 & 0 & 0 &0
\end{pmatrix}\,,
\qquad
P_{Q_2}^2=\begin{pmatrix}
0 & 0 & 0 &0
\\
0 & 0 &0 & 0
\\
1 & 0 & 0 &0
\\
0 & 0 & 0 &0
\end{pmatrix}\,,
\qquad
P_{t_2}=\frac{1}{2}\begin{pmatrix}
1 & 0 & 0 &0
\\
0 & 1 &0 & 0
\\
0 & 0 & -1 &0
\\
0 & 0 & 0 &-1
\end{pmatrix}\,.
\end{split}
\end{equation}
Note that in terms of $\Sp(4)$, the adjoint decomposes into one symmetric and one antisymmetric: we find that $P_{Q_1} + P_{Q_2}$ and $P_{t_1}$ project states in the symmetric (with $t^c$ in an $\SU(2)_{\mathrm{R}}$ triplet), while $P_{Q_2} -P_{Q_1}$ and $P_{t_2}$ in the antisymmetric.
For both left- and right-handed tops, the projector is a superposition of the two: 
\begin{equation} \label{Pt_Adj}
P_Q = A_Q\ P_{Q_1} + B_Q\ P_{Q_2}\,, \qquad  P_t = A_t\ P_{t_1} + B_t\ P_{t_2}\,, \qquad  \mbox{with}\;\; |A_{Q/t}|^2 + |B_{Q/t}|^2 = 1\,.
\end{equation}
For the doublet combination $P_Q$ the relative phase of the two coefficients can be removed by the same $\SU(4)$ rotation (along $X^5$). For the right-handed top,  the two coefficients are always physical as they mix a singlet and a triplet of $\SU(2)_{\mathrm{R}}$.
Note also that along the other EW-preserving vacuum $E_+$, the role of the two combinations of doublet embeddings, $P_{Q_1} \pm P_{Q_2}$, are reversed.

\subsection{Vacuum alignment}
\label{Vacuum alignment in the minimal model}

We study the vacuum alignment induced by the breaking terms that have been discussed previously.
The purpose is to isolate cases where the misalignment angle, $\theta$, is sufficiently small, but non-zero, to comply with composite-Higgs models.
The most general form (up to NLO) of the potential can be inferred  from the tables in App.~\ref{Operators involving mass, gauge and top bilinear spurions} and takes the following form:
\begin{equation}
V(\theta)=c_1 s_{\theta}^2 + c_2 s_{\theta}^4 + c_3 c_{\theta} +c_4 c_{\theta} s_{\theta}^2.
\label{pot-theta}
\end{equation}
We use here and in the following the short-hand notations $s_x\equiv \sin x$ and $c_x\equiv \cos x,$.
Note that, only a non-zero current mass may induce the coefficients $c_{3}$ and $c_4$.
Moreover, $c_3$ is generated by LO and NLO operators (including mixed contributions), while $c_4$ arises only at NLO.
The coefficients $c_2$ comes only from NLO operators containing gauge and/or top spurions, while the remaining coefficient $c_1$ may be generated by all the three sources of explicit breaking starting from LO gauge and top operators.

For simplicity, let us discuss first the LO effects of each explicit breaking source independently.
As discussed in Sec.~\ref{Gauging of the EW symmetry}, the gauge contributions alone are not able to break the EW symmetry. 
In particular,  the LO gauge operators in Eq.~(\ref{gauge-operators}) correspond to $c_1>0$, $c_2=c_3=c_4=0$, such that the minimum of the potential is at $\theta=0$.
For the LO mass contribution, $\mathrm{Tr}[\chi\Sigma^{\dagger}+\Sigma\chi^{\dagger}]$,  we have $c_3 \neq 0$ positive or negative,
while $c_1=c_2=c_4=0$, and the minimum is, thus, either at $\theta=0$ or at  $\pi/2$ depending on the sign of $c_3$.
Finally, the LO top contribution, Eq.~\eqref{top bilinear operators},  corresponds to $c_1<0$, $c_2=c_3=c_4=0$ such that the minimum is at $\theta=\pi/2$.

The challenge in  composite-Higgs models is to generate a small misalignment ($\theta \ll 1$) in order to have a small hierarchy between the EW and the compositeness scale, $v \ll f$.
To depart from the EW preserving vacuum ($\theta=0$) and from the technicolor limit ($\theta=\pi/2$), one needs to consider several 
explicit breaking sources at the same time.
To this end, let us focus on two simplified scenarios:
\begin{itemize}
\item[(i)] A potential generated only by the gauge and top explicit breaking interactions such that the current masses are set to zero,
and we have $c_{3}=c_4=0$. In that case, the breaking of the EW symmetry is driven by the coefficient $c_2$,  and one needs to 
include the NLO contributions to the potential. This scenario is commonly used in composite models with partially composite tops based on holography~\cite{Agashe:2004rs,Contino:2010rs}.
\item[(ii)] A potential generated by gauge and top spurions as well as a non-zero current mass.
In this case, it is enough to restrict to the LO contributions, and we thus  assume $c_{2}=c_4=0$. This scenario is well known, and we 
refer to Ref.~\cite{Cacciapaglia:2014uja} for details. Here we just briefly outline this scenario for comparison.
\end{itemize}

In case $(i)$, the minimisation of the potential in Eq.~(\ref{pot-theta}) leads to: 
\begin{equation}
 \dfrac{\partial V}{\partial \theta}  = 2 c_{\theta} s_{\theta} ~ ( c_1 +2 c_2 s_{\theta}^2)=0.
\end{equation}
Setting aside the limit where the EW symmetry remains unbroken ($\theta=0$) as well as the technicolor limit ($\theta=\pi/2$), 
the third extremum corresponds to $s_{\theta}^2=- \dfrac{c_1}{2 c_2}$ for which $V(\theta)=-c_1^2/(4 c_2)$.
This extremum is the global minimum of the potential only if $c_2>0$ and $c_1<0$ (as expected, see Ref.~\cite{Cacciapaglia:2014uja}).
Moreover, a small misalignment angle requires $|c_1|\ll |c_2|$.
As we will see, this requirement can be obtained in several ways depending on the top coupling representation.

For the case $(ii)$, the minimisation of the potential leads to 
\begin{equation}
\dfrac{\partial V}{\partial \theta}=s_{\theta} [2 c_1 c_{\theta} -c_3]=0.
\end{equation}
Focusing again on the EW breaking vacuum alignment ($\theta\ll 1$), the potential is extremised for $c_{\theta}=\dfrac{c_3}{2 c_1}$ where $V(\theta)=(4c_1^2 +c_3^2)/4c_1$.
A small misalignment implies $|c_3| \simeq |2 c_1|$.
Moreover, for the extremum to be the global minimum, one needs $|c_3| \lesssim |2 c_1|$, where $c_{1,3}<0$, or $c_1<0$ and $c_3>0$. 

Let us now explore in details how the scenario $(i)$ could be realised when NLO contributions are taken into account.
In practice this requires obtaining $|c_1|\ll |c_2|$ 
in a natural way. \\

$\bullet$ \textbf{Hierarchy between the LECs} ($| C_t/ C_t^\prime| \ll 1$) 
\\
This case relies on the usual hypothesis that the top loops are the dominant contributions to the coefficients $c_1$ and $c_2$ and, for some reason, 
the strong dynamics leads to $|c_1/c_2|\ll 1$.
In other words, the LECs associated with the operators generating $c_1$ need to be suppressed.
For simplicity, one can neglect the gauging of the SM as its effect is negligible in comparison to the top quark contributions.
Moreover,  let us consider a bilinear coupling as an example.
The potential takes the following form:
\begin{equation}
V(\theta)=- C_t y_t^2  f^4  s_{\theta}^2  + C_t^\prime y_t^4  f^4 s_{\theta}^4.
\end{equation}
where the positive coefficients $C_t$ and $C_t^\prime$ are functions of the different LECs associated with the operators in 
Tab.~\ref{tab-bilinear-potential}.
Note that the discussion can also be applied to all the linear couplings as they also generate the coefficients $C_t$ and $C_{t^\prime}$ (for reference to the vast literature on this topic we refer the reader to the reviews in Refs~\cite{Contino:2010rs,Bellazzini:2014yua,Panico:2015jxa}).
To get a small misalignment requires $|C_t/y_t^2 C_t^\prime| \ll 1$;
i.e. some cancellation should happen at LO making that contribution comparable to if not smaller than the NLO one. 
In models inspired by holography this is achieved by assuming that the main contribution to the LECs comes from top and top partner 
loops and that other UV effects are negligible~\cite{Contino:2006qr}. We remark, however, that this is a very specific assumption, 
and not all models (especially with an underlying gauge-fermion theory) will respect it.\\

$\bullet$ \textbf{Linear coupling in the symmetric representation ($y_{t_L} \lesssim y_{t_R}$)}
\\
Choosing a symmetric representation for the left- and right-handed top couplings, one finds that the LO  contributions generate $c_1$ and $c_2$ at the same order in the chiral expansion.
This is due to the fact that the Goldstone matrix is antisymmetric (pseudo-real case) such that the LO operators involve four top spurions 
(see Tab.~\ref{tab-Sym-potential}).

For simplicity, let us first consider operators of the general form ${\rm Tr}[\Xi_S\Sigma^\dagger \Xi_S \Sigma^\dagger]{\rm Tr}[\Sigma \Xi_S^\dagger \Sigma \Xi_S^\dagger]$.
The corresponding potential is given by the operators in Eq.~(\ref{L2S})
\begin{eqnarray}
V_1(\theta)&=& C_{t\mathrm{S},1} \frac{f^4}{(4 \pi)^2} \left( y_{t_L}^4 {\rm Tr} [P_Q^\alpha \Sigma^\dagger P_Q^\beta
\Sigma^\dagger]{\rm Tr}[\Sigma P^\dagger_{Q \alpha} \Sigma P^\dagger_{Q\beta}]
+
 y_{t_R}^4 ~{\rm Tr} [P_t \Sigma^\dagger P_t
\Sigma^\dagger]{\rm Tr}[\Sigma P^\dagger_{t} \Sigma P^\dagger_{t}] \right.
\nonumber
\\
&&+
\left.   y_{t_L}^2 y_{t_R}^2 ~{\rm Tr} [P_Q^\alpha \Sigma^\dagger P_t
\Sigma^\dagger]{\rm Tr}[\Sigma P^\dagger_{Q \alpha} \Sigma P^\dagger_t] \right)
\nonumber
\\
&=& \dfrac{C_{t\mathrm{S},1}  f^4}{(4 \pi)^2} ~ [ y_{t_L}^4 s_{\theta}^4 + y_{t_R}^4 c_{\theta}^4 + y_{t_L}^2 y_{t_R}^2 c_{\theta}^2  s_{\theta}^2 ] ,
\end{eqnarray}
such that $c_1= C_{t\mathrm{S},1} f^4 (y_{t_R}^2 y_{t_L}^2-2 y_{t_R}^4)/(16 \pi^2)$ and $c_2= C_{t\mathrm{S},1} f^4 (y_{t_L}^4+y_{t_R}^4 - y_{t_R}^2 y_{t_L}^2)/(16 \pi^2)$.
Achieving $c_1<0$ and $c_2>0$ is fairly easy as long as $y_{t_L}\lesssim \sqrt{2}y_{t_R}$. Note that a small misalignment angle is achieved by tuning the value of $y_{t_L}$ close to the upper bound.
Using the constraint on the top mass coming from Eq.~(\ref{eq:PC-Yukawas}), we can express the Higgs mass as a function of the two relevant LECs and the misalignment angle as follows:
\begin{equation}
m_h^2  = 48 \frac{C_{t\mathrm{S},1}}{C_{yS}^2} \frac{m_\mathrm{top}^2}{(\sqrt{10-6 c_{4\theta}} + 2)}   \sim \frac{12 C_{t\mathrm{S},1}}{C_{yS}^2} m_\mathrm{top}^2 + \mathcal{O} (\theta^4)
\end{equation}
while the singlet remains massless~\cite{Gripaios:2009pe} (a mass can easily be generated by adding current masses). We see that a small enhancement in $C_{yS}$, or an order $1/10$ suppression in $C_{t\mathrm{S},1}$, is sufficient to achieve the measured value of the Higgs mass.
The second type of operators that follow the template ${\rm Tr}[\Xi_S\Sigma^\dagger \Xi_S \Xi_S^\dagger \Sigma \Xi_S^\dagger]$ provide an additional term in the potential proportional to  $s_{\theta}^2$,
\begin{equation}
\dfrac{C_{t\mathrm{S},2}  f^4}{(4 \pi)^2} ~ \frac{6 y_{t_L}^4 - y_{t_L}^2 y_{t_R}^2 - 2 y_{t_R}^2}{4}  s_{\theta}^2\, ,
\end{equation}
which adds up to $c_1$ and might help relieve the tension in the alignment and Higgs mass if $C_{t\mathrm{S},2}<0$.

For completeness, we also report the expression for the top mass and linear couplings to the NGBs,
\begin{equation}
 (Q_1 t^c)^\dagger \left( m_\mathrm{top} + \frac{m_\mathrm{top}}{v} \left( \frac{c_{2\theta}}{c_\theta}\ h - i \frac{s_\theta}{c_\theta}\ \eta \right) + \dots \right)\,, \qquad m_{\mathrm{top}} = C_{yS} \frac{y_{tL} y_{tR} c_\theta s_\theta}{4 \pi} f\,,  \label{mtop_S}
\end{equation}
where we remark the presence of a coupling of the pseudo-scalar singlet $\eta$ to tops.

\subsection{Masses and couplings of the pNGBs}
\label{Masses and couplings}

The general potential presented in Eq.~(\ref{pot-theta}) can be further expanded to obtain the masses for the pNGBs (Higgs and $\eta$) as well as the couplings among them. We find that, if all the coefficients and couplings are real, the four terms correspond to universal functions of the fields:
\begin{equation}
V(\theta, h, \eta) = \sum_{i=1}^4 c_i\ f_i (\theta, h, \eta) + \sum_{i=1}^3 c'_i\ f'_i (\theta, h, \eta)\,.
\end{equation}
The four functions $f_i$ correspond to the four basic functions appearing in the potential in Eq.(\ref{pot-theta}), while the three functions $f'_i$ contain additional contributions to the mass and couplings of the singlet $\eta$ that arise in special cases.\footnote{$f'_1$ arises in cases where the potential contains constant pieces, $f'_2$ when the potential consists of $s_\theta^2$ or $s_\theta^4$ terms,  and both are only present when the top spurions are embedded into two different $\Sp(4)$ representations of a given $\SU(4)$ spurion; $f'_3$ corresponds to potential terms $c_\theta$ and $c_\theta s_\theta^2$, and it receives contributions from NLO operators containing the mass spurion in Tab.~\ref{tab-mass-gauge-spurions}.} For simplicity, in the following we will neglect these special contributions and set $c'_i=0$.
Up to trilinear couplings, the functions $f_i$ read:
\begin{equation}
\begin{split}
f_1(\theta, h,\eta) =& s_\theta^2 + 2 c_\theta s_\theta \dfrac{h}{f}+ c_{2 \theta}\dfrac{h^2}{f^2}-s_\theta^2 \dfrac{\eta^2}{f^2}
    -\dfrac{4}{3} c_\theta s_\theta \dfrac{h}{f} \dfrac{(h^2+ \eta^2)}{f^2} +\dots
\\
f_2(\theta, h,\eta) =& s_\theta^4 + 4 c_\theta s_\theta^3 \dfrac{h}{f}+ 2 s_{ \theta}^2(1+2 c_{2\theta}) \dfrac{h^2}{f^2}-2 s_\theta^4 \dfrac{\eta^2}{f^2}
    -\dfrac{4}{3} c_\theta s_\theta (1-4c_{2\theta}) \dfrac{h^3}{f^3}-\dfrac{20}{3} c_\theta s_\theta^3 \dfrac{h \eta^2}{f^3} + \dots
\\
f_3(\theta, h,\eta) =& c_\theta-s_\theta \dfrac{h}{f} -\dfrac{1}{2} c_{\theta}\dfrac{(h^2+\eta^2)}{f^2}
    +\dfrac{1}{6} s_\theta \dfrac{h}{f} \dfrac{(h^2+ \eta^2)}{f^2} + \dots
\\
f_4(\theta, h,\eta) =& c_\theta s_\theta^2 +\dfrac{1}{2} s_\theta(1+3 c_{2 \theta}) \dfrac{h}{f} + c_{ \theta}(1-\dfrac{9}{2}s^2_\theta)\dfrac{h^2}{f^2}
    -\dfrac{3}{2} c_\theta s_\theta^2 \dfrac{\eta^2}{f^2}- s_\theta (\dfrac{13}{12}+\dfrac{27}{12} c_{2 \theta}) \dfrac{h^3}{f^3}\\
&-\dfrac{7}{12} s_\theta (1+3 c_{2 \theta}) \dfrac{h \eta^2}{f^3} +\dots
\end{split}
\end{equation}
We can thus trade three of the coefficients, say  $c_1$, $c_2$ and $c_3$, for the value of the misalignment angle at the minimum, $\theta$,
and the masses of the Higgs, $m_h$, and of the singlet, $m_\eta$:
\begin{eqnarray}
c_1 &=& - \frac{m_h^2 f^2}{4 c_\theta^2}  - \frac{f^2 m_\eta^2}{8 c_\theta^2} (1+3 c_{2\theta}) - \frac{c_4}{c_\theta}\,, \\
c_2 &=& \frac{f^2 m_h^2}{2 s_{2\theta}^2} - \frac{f^2 m_\eta^2}{8 c_\theta^2} + \frac{c_4}{2 c_\theta}\,, \\
c_3 &=& - f^2 m_\eta^2 c_\theta - c_4 s_\theta^2\,.
\end{eqnarray}
With these, we can thus predict the value of the trilinear coupling of the Higgs bosons and the coupling between the singlet and the Higgs:
\begin{eqnarray}
g_{h^3} &=& \frac{m_h^2}{2 v} \frac{c_{2\theta}}{c_\theta} + \frac{m_\eta^2}{2 v} \frac{s_\theta^4}{c_\theta} - \frac{c_4}{f^3} s_\theta^3\,, \label{eq:h3}\\
g_{h \eta^2} &=& \frac{(m_\eta^2 - m_h^2)}{2 v} \frac{s_\theta^2}{c_\theta} - \frac{c_4}{f^3} s_\theta\,.  \label{eq:heta2}
\end{eqnarray}
The case $(i)$, which includes the results for the symmetric top partner representation, is recovered for $c_4=0$ and $m_\eta=0$. The fact that the singlet remains massless is to be expected, and a mass can be generated by adding a current mass that will generate a non-zero $c_3$. In the case $(ii)$ obtained for $c_2 = c_4 = 0$, the condition $c_2=0$ imposes the well-known relation between the masses $m_h = m_\eta s_\theta$: this was already shown at LO for a bilinear top coupling agreeing with Ref.~\cite{Arbey:2015exa}.
Here we show that the relation also holds for any linear coupling up to two-index representations.

\begin{figure}[tb!]
\begin{center}
\includegraphics[width=6.5cm]{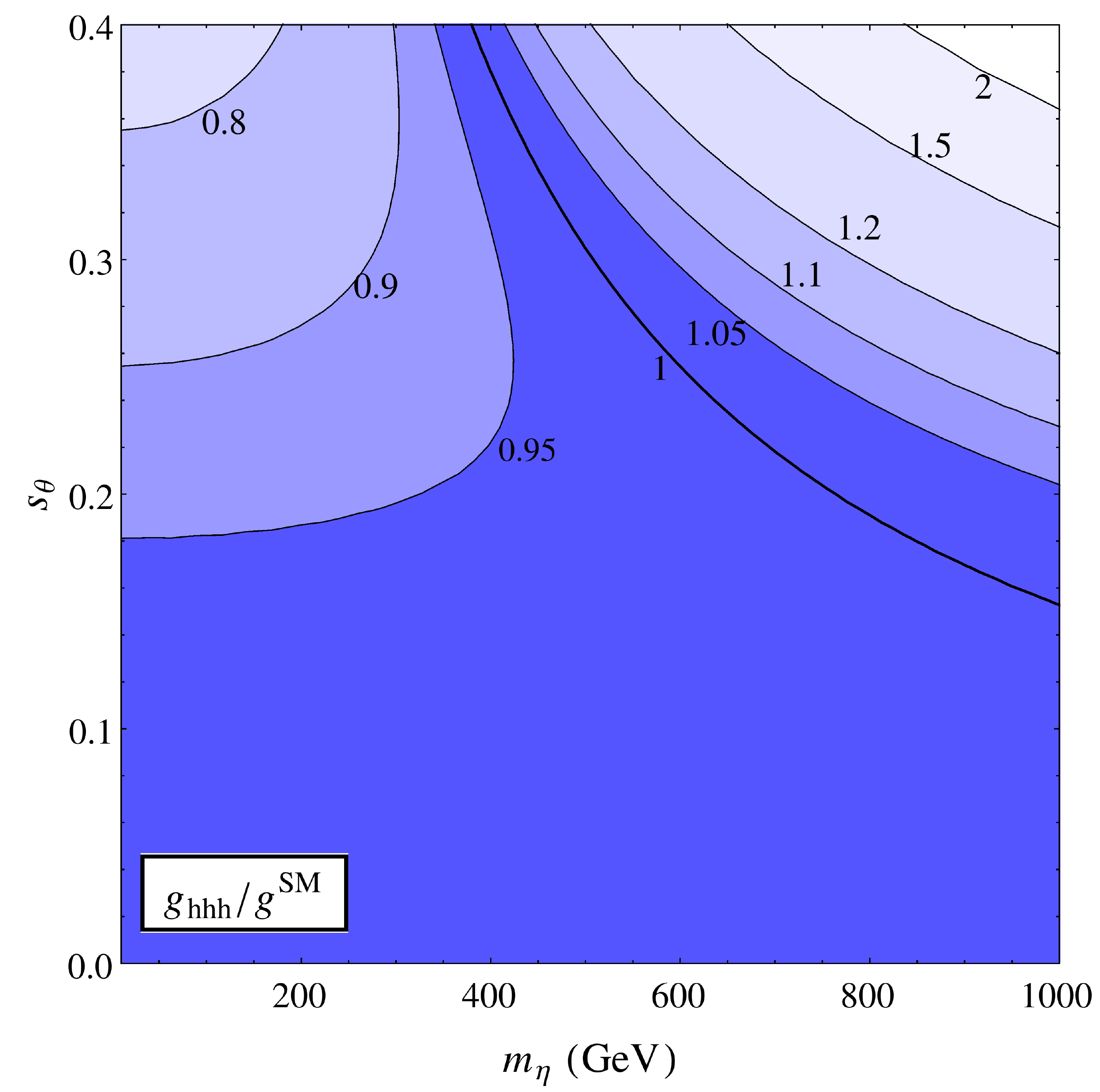} \hspace{1cm}
\includegraphics[width=6.5cm]{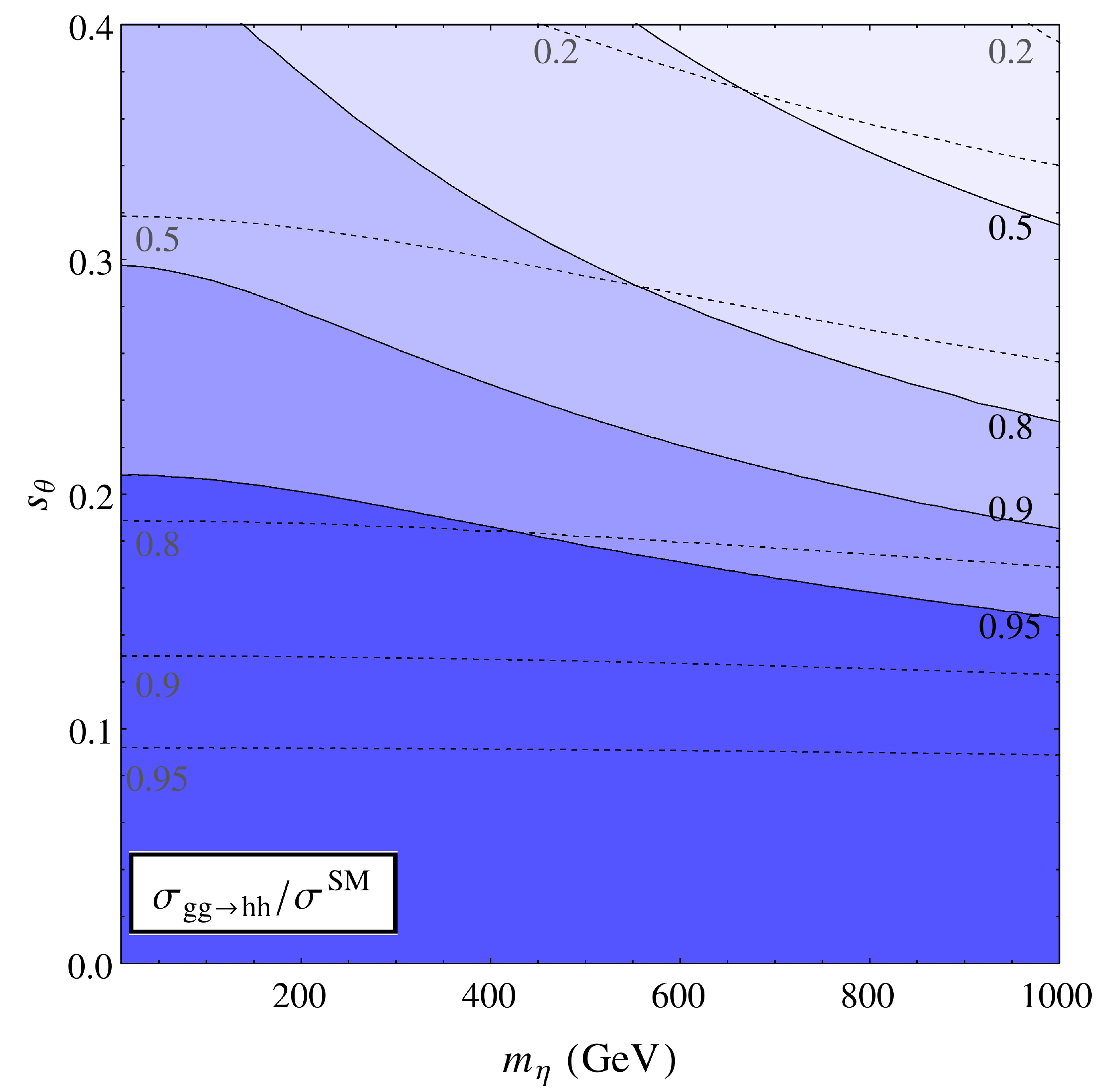}
\end{center}
\caption{Right panel: trilinear Higgs coupling normalised to the SM value as a function of $m_\eta$ and $s_\theta$ (for $c_4 = 0$). Left panel: contours of the di-Higgs cross section via gluon fusion at the LHC@14 TeV, normalised to the SM one. The continuous lines correspond to bilinear top couplings and linear couplings in the fundamental representation, and the dotted lines correspond to linear couplings in two-index representations.} \label{fig:h3}
\end{figure}

\begin{figure}[tb!]
\begin{center}
\includegraphics[width=6.5cm]{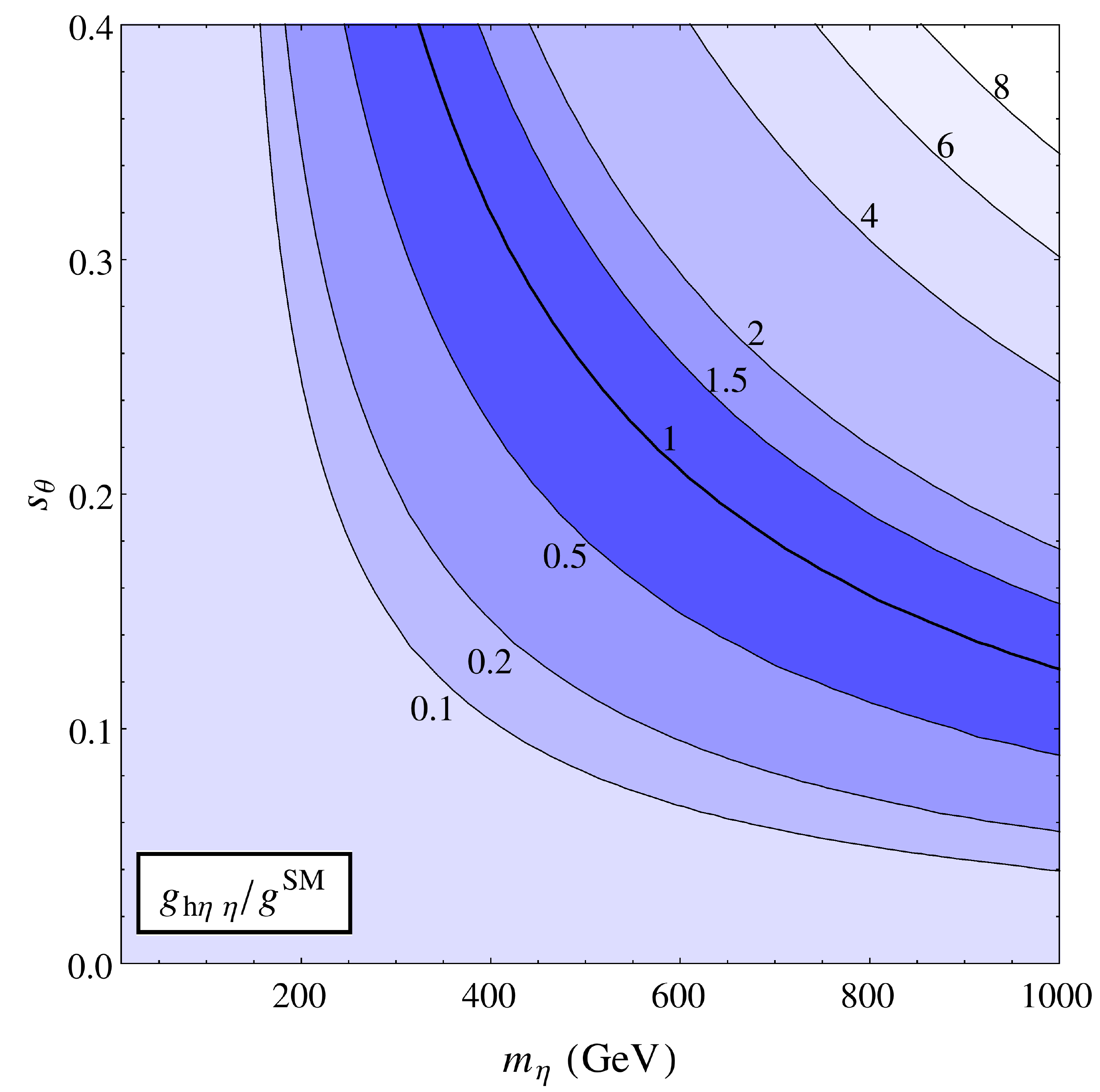} \hspace{1cm}
\includegraphics[width=6.5cm]{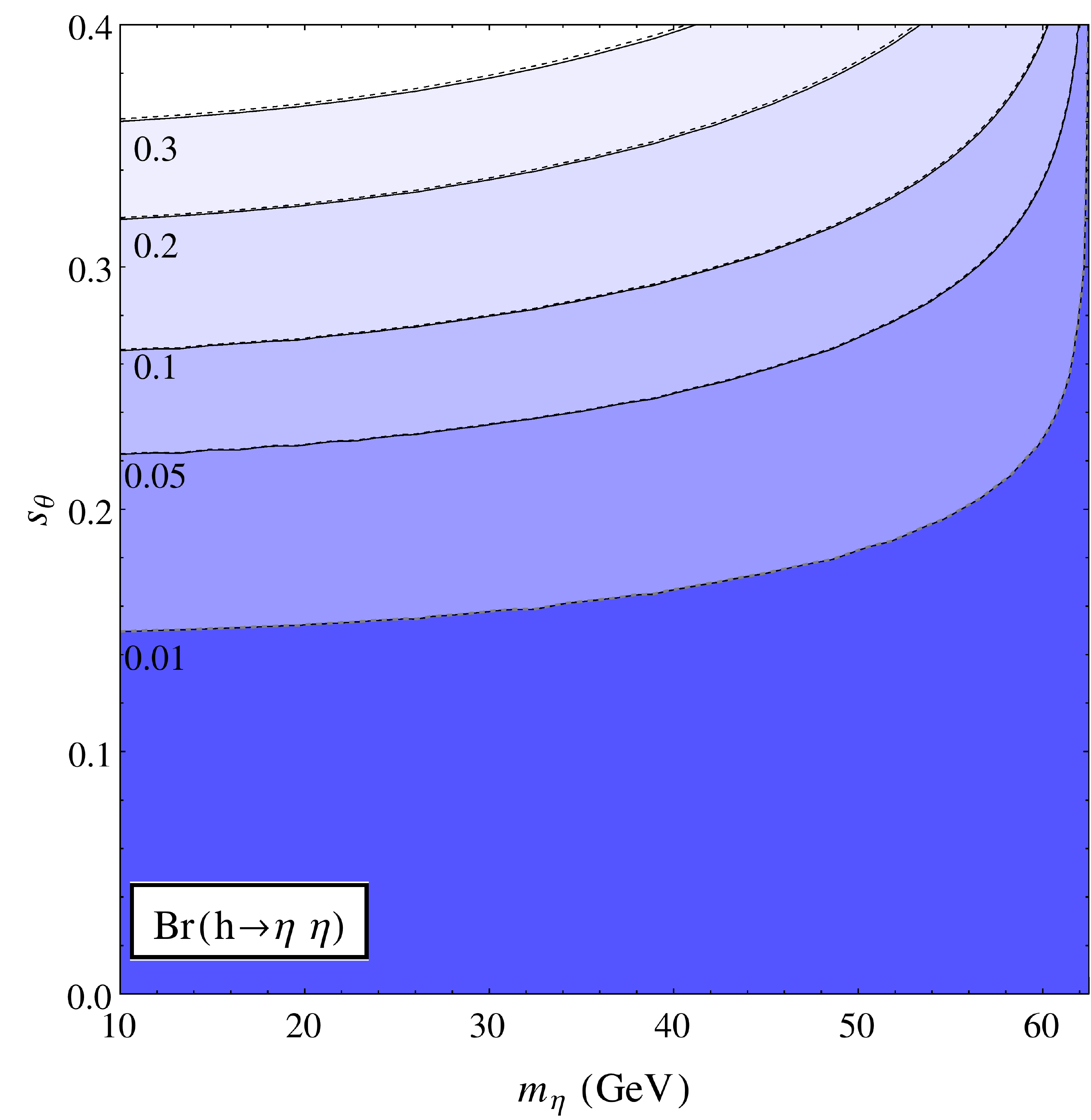}
\end{center}
\caption{Right panel: Higgs coupling to $\eta$ normalised to the SM Higgs trilinear coupling as a function of $m_\eta$ and $s_\theta$ (for $c_4 = 0$). Left panel: contours of the branching ratio $BR (h \to \eta \eta)$ for singlet mass below threshold. The continuous lines correspond to bilinear top couplings and linear couplings in the fundamental representation, and the dotted lines correspond to linear couplings in two-index representations.} \label{fig:heta2}
\end{figure}

The trilinear couplings in Eqs~(\ref{eq:h3}) and (\ref{eq:heta2}) are relevant for the phenomenology of the composite Higgs. It is well known that the modifications of the Higgs couplings to fermions and gauge bosons with respect to the SM ones are small, as corrections scale with $s_\theta^2 \sim v^2/f^2$. On the other hand, the Higgs trilinear coupling may receive larger corrections. In Figure~\ref{fig:h3} we show contours of the trilinear coupling normalised to the SM value as a function of $m_\eta$ and $s_\theta$ (for $c_4 = 0$). Sizeable modifications are only present for large $s_\theta \geq 0.2$, with an increase of the couplings for large singlet masses. The effect of the trilinear coupling on di-Higgs production via gluon fusion is shown in the left panel, where we plot contours of the cross section at the LHC with a centre of mass energy of 14 TeV~\cite{Azatov:2015oxa}. Interestingly, the cross section is always reduced with respect to the SM ones: the solid (dashed) contours correspond to bilinear top couplings (linear top couplings) for which the top coupling to the Higgs is rescaled by a factor $c_\theta$ ($c_{2 \theta}/c_\theta$) with respect to the SM value. The case of linear couplings to the fundamental representation follows the bi-linear case. Note that we do not consider here operators generated by the strong dynamics that couple the composite Higgs directly to gluons. The coupling of the Higgs to two singlets in Eq.~(\ref{eq:heta2}) is shown in Figure~\ref{fig:heta2}. If the mass of the singlet $\eta$ is smaller than half the Higgs mass, this coupling will contribute to non-standard decays of the Higgs, as shown in the left panel of Figure~\ref{fig:heta2}. Bounds on this branching ratio are obtained from global fits of the Higgs properties, independently on the decay modes of the singlet: the current bound from the Higgs data combination after Run-I is at 34\%~\cite{Khachatryan:2016vau}, and thus is unable to probe the parameter space, while projections for the high luminosity phase with a data set of 3 ab$^{-1}$ estimate the reach to 10\%~\cite{ATL-PHYS-PUB-2014-016}. We remark that dedicated searches for $h \to \eta \eta$ may give stronger bounds, but depend on the final states the singlets decay into.\\

\textbf{$\bullet$ $\eta t \overline{t}$ coupling:}\\
A coupling of the singlet $\eta$ to tops may be generated from the same operator that generates the top mass, as we have seen in Eq.~(\ref{mtop_S}).
This coupling is phenomenologically very important as it opens new decay modes for the singlet, besides the di-boson final states from the Wess--Zumino--Witten anomaly~\cite{Galloway:2010bp,Cacciapaglia:2014uja}, and induces gluon fusion at one loop thus enhancing its production at hadron colliders~\cite{Arbey:2015exa}.
The $\eta t\bar{t}$ coupling is not present at LO for bilinear top couplings nor in the case of linear coupling with fundamental top-partner 
representation. However, it appears at NLO in mixed operators involving mass and top spurions.
For instance, for the bilinear case we have \cite{Galloway:2010bp}
\begin{equation}
y_t~ \mathrm{Tr}[P^\alpha \Sigma] \mathrm{Tr}[\Sigma \chi^\dagger] (\overline{q}_{\mathrm{L} \alpha} t_{\mathrm{R}})+\mathrm{h.c.}
=-4 y_tB_0 s_\theta [c_\theta (m_1+m_2)+ i(m_1-m_2)\dfrac{\eta}{f}+\cdots](\overline{t}_{\mathrm{L} } t_{\mathrm{R}})+\mathrm{h.c.}
\end{equation} 
and similarly for the other mixed operators. Note that the coupling is proportional to the $\Sp(4)$ violating current mass. Interestingly, couplings that do not need such violation are generated by higher-order operators containing all types of spurions~\cite{Arbey:2015exa}.

The situation is different for linear couplings to two-index representations. For the symmetric, we already found in Eq.~(\ref{mtop_S}) that a coupling of the singlet proportional to the top mass is generated. For the other two representations the situation is more complex due to the fact that the embedding of the elementary top fields within the top partner representation is not unique. 
For the antisymmetric, two possible embeddings of the singlet are allowed. Using the spurion defined in Eq.~(\ref{Pt_A}), the mass of the top is given by the first line in Eq.~(\ref{eq:PC-Yukawas}),
\begin{equation}
m_\mathrm{top,A} = \frac{y_{t_L} y_{t_R}}{4 \pi} (C_{y\mathrm{A},1} + 2 C_{y\mathrm{A},2}) (B-A) f c_\theta s_\theta\,,
\end{equation}
where we see that only the component of the spurion aligned with the $\Sp(4)$ singlet (along the vacuum $E_-$) contributes.
From the same operator, we derive the couplings,
\begin{equation}
\mathcal{L} \supset - m_\mathrm{top} \bar{t} t \left( 1 + \frac{c_{2\theta}}{c_\theta} \frac{h}{v} + \dots \right) - i \frac{m_\mathrm{top}}{f c_\theta} \frac{B+A}{B-A} \eta \bar{t} \gamma^5 t + \dots\,,
\end{equation}
where we see that the coupling of the singlet is only generated by the component of the right-handed top aligned with the singlet inside the $\Sp(4)$ 5-plet.
Via the same mechanism, couplings of the single to a top partner and a top are also generated \cite{Serra:2015xfa}. 
A similar situation occurs for the adjoint, where for both doublet and singlet two possible embeddings are allowed. Using the spurions in Eq.~(\ref{Pt_Adj}), the top mass given by the third line in Eq.~(\ref{eq:PC-Yukawas}) reads:
\begin{equation}
m_\mathrm{top,Adj} = \frac{y_{t_L} y_{t_R}}{4 \pi} C_{y\mathrm{Adj}} \left(  (A_Q + B_Q) B_t + \frac{1}{\sqrt{2}} (B_Q - A_Q) A_t \right) f c_\theta s_\theta\,.
\end{equation}
The above result can easily be interpreted: when the right-handed top is aligned with the singlet of $\SU(2)_{\mathrm{R}}$ (i.e. in the 5-plet of $\Sp(4)$ of $E_-$) the doublet is projected on the 10-plet, while when the right-handed top is in the $\SU(2)_{\mathrm{R}}$ triplet (which is part of the 10-plet) the doublet is projected on the 5-plet.
The couplings acquire the form:
 \begin{equation}
\mathcal{L} \supset - m_\mathrm{top} \bar{t} t \left( 1 + \frac{c_{2\theta}}{c_\theta} \frac{h}{v} + \dots \right) - i \frac{m_\mathrm{top}}{f c_\theta} \frac{\sqrt{2} (B_Q - A_Q) B_t +  (B_Q + A_Q) A_t }{\sqrt{2} (A_Q + B_Q) B_t +  (B_Q - A_Q) A_t } \eta \bar{t} \gamma^5 t + \dots\,,
\end{equation}
where we see that a coupling to the singlet $\eta$ is not generated only when  the left- and right-handed tops are in different $\Sp(4)$ representations.\\

$\bullet$ \textbf{Vacuum expectation value of $\eta$, CP-violation, and the choice of vacuum:}
\\
So far we have only considered a vacuum misaligned along the direction of the Higgs. However, in general, we should also consider a misalignment along the direction of the singlet $\eta$. This can be done by rotating the vacuum with an $\SU(4)$ transformation along $X^{\hat{5}}$:
\begin{equation}
U_\alpha = e^{\sqrt{2} i \alpha X^{\hat{5}}} = \begin{pmatrix}
e^{i \alpha/2} & 0 \\
0 & e^{-i \alpha/2}
\end{pmatrix} \in \SU(4)\,,
\end{equation}
where $\alpha$ is related to the vacuum expectation value of the singlet. Remarkably, it appears as a phase, in accordance to the fact that $\eta$ is a pseudo-scalar.
This corresponds to a change in the relative phase of the two hyper-fermion doublets, and it will affect the phase associated with the current mass, if present. Thus, the presence of the phase in the vacuum is correlated to a phase in the current mass. One can always make the simplifying assumption of real masses and thus start with a real vacuum.
As a consistency check, one can verify that a tadpole for $\eta$ is generated by the current mass spurion if a phase is present:
\begin{equation}
{\rm Tr}[\chi \Sigma^\dagger]+ \mathrm{h.c.}=8 B_0 ~{\rm Im} (m_1-m_2) \dfrac{\eta}{f}+\cdots
\end{equation}
Once other spurions are included, it is always the phase of the current mass term that generates a tadpole for $\eta$: this is clearly seen as the gauge couplings are real, while the two Yukawas $y_{t_L}$ and $y_{tR}$ can be made real\footnote{Note that the phase appearing in the  Cabibbo--Kobayashi--Maskawa matrix does not play any role here, as we are dealing with overall phases carried by the Yukawas.} by choosing the phase of the elementary quark fields.

The situation is different in cases, such as partial compositeness with tops in the antisymmetric or adjoint representations, where more than one embedding is possible for the same SM elementary field: physical phases may remain as not all couplings can be made real by a phase shift of the fermion fields.
We will first consider in detail the case of the antisymmetric. As before, we parametrise the spurion for the right-handed top following Eq.~(\ref{Pt_A}), allowing for a phase between the two coefficients.  The potential generated by the LO operator in Eq.~(\ref{pot_A-Adj}) gives, up to linear terms in the fields,
\begin{equation}
\begin{split}
V_\mathrm{A} = &2 C_{t\mathrm{A}} f^4 \left( |B-A|^2 y_{t_R}^2 c_\theta^2 + y_{t_L}^2 s_\theta^2 \right) + 4 C_{t\mathrm{A}} f^3 \left( y_{t_L}^2 - |B-A|^2 y_{t_r}^2 \right) c_\theta s_\theta h \\
&+8 C_{t\mathrm{A}} f^3 \mathrm{Im} [A B^\ast] y_{t_R}^2 c_\theta\ \eta + \dots
\end{split}
\end{equation}
From the equation above, we clearly see that a tadpole for the singlet is present only if a relative phase between the two coefficients $A$ and $B$ is present. As already commented above, such a phase can be removed by the $\SU(4)$ rotation $U_\alpha$, and in the absence of a current mass one can use this to remove it from the Lagrangian. In other words, the vacuum expectation value of the singlet $\eta$ is not physical as it is associated with an arbitrary phase that can be removed from the theory (this point was missed in the discussion in Ref.~\cite{Gripaios:2009pe}). The only situation where a tadpole for $\eta$ could be physical is when both a current mass and a phase in the right-handed top spurion are present. As a misalignment of the vacuum along the singlet would imply the presence of a CP violating phase in the vacuum, this result shows that the only way to achieve this is to add a CP-violating phase in the underlying theory. Thus, no spontaneous CP-violation via the vacuum misalignment, or pNGB vacuum expectation value, is possible. 
We checked that the same conclusion can be drawn for the adjoint representation: the tadpole reads 
\begin{equation}
V_\mathrm{Adj} =  4 C_{t\mathrm{Adj}} f^3 \mathrm{Im} [A_Q B_Q^\ast] y_{t_L}^2 c_\theta\ \eta + \dots
\end{equation}
thus it is again proportional to the only phase that can be removed by $U_\alpha$. A mass mixing is also present and proportional to the same phase.
Another related point is the presence of a mixing between the Higgs boson, $h$, and the singlet, $\eta$, in the potential: we checked that the mixing is also proportional to the same phase generating the tadpole.
This mixing, which is only physical in theories with explicit CP-violation, has been used in Ref.~\cite{Banerjee:2017qod} to reduce the fine-tuning in the Higgs mass.

Another case where a misalignment along the singlet direction is needed is when the potential generates a negative mass squared for $\eta$ in the absence of a tadpole. This situation can occur for real coefficients; however, it is a diagnostics that the initial choice of the EW preserving vacuum is not correct. As an example, we reconsider the case of partial compositeness with the antisymmetric representation.
As mentioned in Ref.~\cite{Serra:2015xfa}, if the right-handed top is mostly aligned with the $\Sp(4)$ 5-plet, i.e. $B+A > B-A$ in our notation, the singlet may develop a vacuum expectation value via a negative squared mass (for real coefficients). However, this situation can be inverted by changing the EW preserving vacuum from $E_-$ to $E_+$ by use of a $U_\alpha$ transformation with $\alpha = \pi/2$ (plus an overall phase shift). This shows that the vacuum expectation value of the singlet (that generates $\alpha$) is unphysical in this case too, as it corresponds to an inappropriate choice of the vacuum. 

It is well known from QCD~\cite{Dashen:1970et,Witten:1980sp} that CP violation can also occur spontaneously via a phase generated by the strong dynamics, resulting in a CP violating interactions of the pNGBs~\cite{DiVecchia:2013swa}. This also applies to the models under discussion. The presence of a physical effect depends, however, on the number of non-vanishing spurions in the theory: the phases of the pre-Yukawas can always be removed by redefining the phases of the elementary spinors if only one embedding for the left-handed and right-handed top is present. Thus, as in QCD, the strong $\theta$-phase can be removed from the theory only if at least one underlying fermion is massless, i.e. $m_1=0$ or $m_2=0$ in our case.


\section{Conclusion} 
\label{Conclusion}

To date, composite-Higgs models remain a valid alternative to the SM and to supersymmetric models in describing the physics of the discovered Higgs boson. One of the tools we have to explore the physics of composite Higgses is the construction of effective theories.
In this work, we offer an exhaustive classification of template operators that can be used to construct effective Lagrangians, up to NLO in the chiral expansion, for models based on the symmetry breaking patterns $\SU(\NF)/\Sp(\NF)$ (with $\NF$ even) and $\SU(\NF)/\SO(\NF)$. The main interest of these two patterns is that they can be generated by simple underlying theories based on gauge interactions and fermionic matter. Such theories are being studied on the lattice; thus the exercise we perform in this work is essential for lattice studies to calculate the relevant low energy constants that impact the physics of the composite Higgs. 
This is, in our view, the constructive way to endow this class of models with predictive power.
Furthermore, the templates we provide, together with a discussion of the counting of each class of spurions, i.e. fermion mass terms, gauge couplings and top Yukawas, are the relevant building blocks for the extension of effective theories up to NLO.
The utility of this tool goes beyond composite-Higgs models, but can be applied to any class of models based on composite dynamics.

After a general discussion, we specialised our results to the simplest case based on $\SU(4)/\Sp(4)$. 
We discuss in detail the issue of the vacuum misalignment, which is generated by operators containing the spurions associated with SM interactions.
One of our main results has been to find a general set of functions of the fields that allow us to study in a model-independent way the vacuum alignment together with the masses and couplings involving the Higgs and the additional singlet.
We also defined the most general embedding of the elementary tops in representations of $\SU(4)$ with up to two indices (as they are generated in underlying theories). This allowed us to clarify misunderstandings present in the literature about the properties of the singlet $\eta$.
Our general results show that a vacuum expectation value for the singlet is not physical, unless explicit sources of CP-violation are present in the theory. Any apparent misalignment in the singlet direction can be removed either by removing unphysical phases in the underlying theory, or by redefining the EW-preserving vacuum around which the theory is constructed.

\section*{Acknowledgments} 
TA acknowledges partial funding from a Villum foundation grant when part of this article was being completed. 
NB and GC acknowledge partial support from the Labex-LIO (Lyon Institute of Origins) under grant ANR-10-LABX-66 and FRAMA (FR3127, F\`ed\`eration de Recherche ``Andr\'e Marie Amp\`ere''), and support from the ``Institut Franco Danois''.
The CP$^3$-Origins centre is partially funded by the Danish National Research Foundation, grant number DNRF90.

\clearpage

\appendix


\section{Spurionic operators involved in the  composite-Higgs potential}
\label{Operators involving mass, gauge and top bilinear spurions}

In this appendix, we classify all the operators, up to NLO, that contribute to the NGB potential at tree level.\footnote{Sticking to the spirit of our analysis,  we do not consider operators containing elementary SM fields that may contribute to the NGB potential at one-loop level.}
We specialise the  classification of the generic spurionic operators (see Sec.~\ref{Introduction of spurions} and App.~\ref{General classification of the spurionic operators})  to the three main sources of explicit breaking relevant for composite-Higgs models: a current mass for the underlying fermions, the gauging of the EW symmetry and bilinear or linear couplings ({\it \`{a} la} partial compositeness) generating the top quark Yukawa coupling.
As we only consider tree-level contributions, we use all spurions in Tab.~\ref{tab-spurions} except those containing the elementary SM fermions and gauge bosons (the latter appearing only in covariant derivatives due to gauge invariance).
Finally, specialising to the minimal $\SU(4)/\Sp(4)$ coset discussed in Sec.~\ref{Minimal Fundamental Composite (Goldstone) Higgs }, we expand\footnote{For simplicity, we assume that no explicit CP-violation is present in the underlying theory (see Sec.~\ref{Masses and couplings}), i.e. that all couplings are real.} these operators and extract the field-independent part relevant to determine the vacuum alignment.

Several points are worthwhile to remember at this point:
\begin{itemize}
    \item[(i)] Once the spurions are specified, their chiral counting is fixed such that the general basis of non-derivative operators involving up to four spurions contains operators that appear beyond NLO. 
For instance all of the operators involving three or four mass spurions $\XiA=\chi$ are subleading and should not be added to our NLO analysis.
    \item[(ii)] The underlying fundamental theory  (see 
Sec.~\ref{Explicit breaking sources in composite Higgs models}) dictates the properties of the spurions.
For instance, the gauge spurion  $\XiAd^A=g T^A_{\mathrm{L}}$ ($\XiAd=g^\prime T_Y$) as well as the linear spurions can appear only in pairs in order to respect the EW gauge symmetry.
    \item[(iii)] Some  operators contain traces made only with spurionic fields (no NGB matrix, $\Sigma$).
    We will neglect in general these subleading effects as we focus in this analysis on the general form of the couplings.
A simple example is provided by the two following operators associated with the bilinear spurion:
	\begin{equation}
	     \dfrac{y_t^2 f^2 \Lambda_{\mathrm{HC}}^2}{16 \pi^2} \mathrm{Tr}[P^\alpha \Sigma^\dagger]\mathrm{Tr}[\Sigma P_\alpha^\dagger]
	    \qquad\qquad
	    \dfrac{y_t^4 f^4}{16 \pi^2} \mathrm{Tr}[P^\alpha \Sigma^\dagger]\mathrm{Tr}[\Sigma P_\alpha^\dagger]\mathrm{Tr}[P^\beta P^\dagger_\beta] 
	\end{equation}
where $\Lambda_{\mathrm{HC}}^2/(16 \pi^2) \simeq f^2$.
Thus, despite the large number of operators present in the general classification 
of App.~\ref{General classification of the spurionic operators}, only a smaller set that depends on the spurions under consideration is relevant in practice.
On the other hand, several spurions may transform in the same representation of flavour increasing the number of operators compared to the ones listed in App.~\ref{General classification of the spurionic operators}.

\end{itemize}

\subsection{Current mass and gauge spurions}
\label{Operators involving mass spurion}

Let us start with the operators containing only the mass and gauge spurions.
These operators are already well known in QCD and can then be used to check the completeness of our classification.

$\bullet$ \textbf{Current mass}
\\
We first consider the operators  containing only the mass spurion, $\XiA=\chi$.
The latter enters in the chiral expansion at ${\cal O}(p^2)$, resulting in the three following classes of operators up to NLO: 
\begin{itemize}
    \item[(i)] Only one spurion $\XiA$ or $\XiA^\dagger$ (class $\chi$).
    \item[(ii)] Two spurions  $\XiA^2$, $\XiA^{\dagger 2}$ or $\XiA \XiA^\dagger$ (class $\chi^2$).
    \item[(iii)] One spurion $\XiA$ or $\XiA^\dagger$ and two derivatives (class $\chi D^2 $).
\end{itemize}
%

\begin{table}[t]
\footnotesize 
    \renewcommand{\arraystretch}{0.6}
    \begin{center}
    \scalebox{0.7}{
	\begin{tabular}{ c c  c c c}
\hline\hline
Class & General form & Operator & Associated LEC & $\SU(4)/\Sp(4)$ \\
\hline 
$\chi$ 
& $\mathrm{Tr}[\XiA \Sigma^\dagger]+\mathrm{h.c.}$  
& $\mathrm{Tr}[\chi \Sigma^\dagger+ \Sigma \chi^\dagger]$ 
& $B_0$ 
& $8 B_0 (m_1 + m_2) \cos \theta + \dots$
\\
\hline
$\chi^2$ 
&  $\mathrm{Tr}[\XiA \XiA^\dagger]$  
&  $\mathrm{Tr}[\chi \chi^\dagger]$  
& $H_2$ 
& $-$
\\
& $\mathrm{Tr}[\XiA \Sigma^\dagger \XiA \Sigma^\dagger]+\mathrm{h.c.}$  
&   $\mathrm{Tr}[\chi \Sigma^\dagger \chi \Sigma^\dagger + \Sigma \chi^\dagger \Sigma \chi^\dagger]$ 
&  $L_8$ 
& $8 B_0^2 [(m_1-m_2)^2+ (m_1 + m_2)^2 \cos (2\theta) + \dots]$
\\
& $\mathrm{Tr}[\XiA \Sigma^\dagger] \mathrm{Tr}[\Sigma \XiA^\dagger]$  
& $\mathrm{Tr}[\chi \Sigma^\dagger] \mathrm{Tr}[\Sigma \chi^\dagger]$ 
& $L_{6,7}$ 
& $16 B_0^2 (m_1+m_2)^2 \cos^2 \theta +\dots$
\\
& $\mathrm{Tr}[\XiA \Sigma^\dagger]^2 +\mathrm{h.c.}$  
& $\mathrm{Tr}[\chi \Sigma^\dagger]^2 + \mathrm{Tr}[ \Sigma \chi^\dagger]^2$ 
& $L_{6,7}$ 
& $32 B_0^2 (m_1+m_2)^2 \cos^2 \theta +\dots$
\\
\hline
$g^2, g^{\prime 2}$ 
& $\mathrm{Tr}[\XiAd\Sigma \XiAd^T \Sigma^\dagger]$  
& $g^2  \mathrm{Tr}[T^A_{\mathrm{L}} \Sigma (T^A_{\mathrm{L}})^T \Sigma^\dagger]$ 
& $K_0^g$  
& $-\dfrac{3}{2} g^2 \cos^2 \theta+\dots$
\\
\hline
$g^4, g^{\prime 4}, g^2 g^{\prime 2}$
& $\Tr[\XiAd\Sigma\XiAd^T\Sigma^{\dagger}]^2$     
&	$g^4~ \Tr[T^A_{\mathrm{L}}\Sigma (T^A_{\mathrm{L}})^T \Sigma^{\dagger}]^2$ 
& $-$
& $\dfrac{9}{4} g^4 \cos^4 \theta+\dots$
\\
& $\Tr[\XiAd^3 \Sigma\XiAd^T\Sigma^{\dagger}]$     
&	$g^4~ \Tr[T^A_{\mathrm{L}} T^A_{\mathrm{L}} T^B_{\mathrm{L}}\Sigma (T^B_{\mathrm{L}})^T \Sigma^{\dagger}]$ 
& $-$
& $-\dfrac{9}{8} g^4 \cos^2 \theta+\dots$
\\
& $\Tr[\XiAd^2 \Sigma (\XiAd^T)^2\Sigma^{\dagger}]$     
&	$g^4~ \Tr[T^A_{\mathrm{L}} T^A_{\mathrm{L}} \Sigma (T^B_{\mathrm{L}})^T (T^B_{\mathrm{L}})^T \Sigma^{\dagger}]$ 
& $-$
& $\dfrac{9}{8} g^4 \cos^2 \theta+\dots$
\\
& $\Tr[\XiAd \Sigma\XiAd^T\Sigma^{\dagger}\XiAd \Sigma\XiAd^T\Sigma^{\dagger}]$     
&	$g^4~ \Tr[T^A_{\mathrm{L}}  \Sigma (T^A_{\mathrm{L}})^T \Sigma^{\dagger} T^B_{\mathrm{L}}  \Sigma (T^B_{\mathrm{L}})^T \Sigma^{\dagger}]$ 
& $-$
& $\dfrac{9}{8} g^4 \cos^4 \theta+\dots$
\\
\hline
$g^2 \chi, g^{\prime 2} \chi$ 
& $\mathrm{Tr}[\XiA \Sigma^\dagger] \mathrm{Tr}[\XiAd^2]+\mathrm{h.c.}$  
& $g^2~ \mathrm{Tr}[\chi \Sigma^\dagger+ \Sigma \chi^\dagger] \mathrm{Tr}[T^A_{\mathrm{L}} T^A_{\mathrm{L}}]$ 
& $K_7^g$ 
& $12  B_0 g^2 (m_1+m_2) \cos \theta+\dots$
\\
& $\mathrm{Tr}[\XiA \Sigma^\dagger] \mathrm{Tr}[\XiAd\Sigma \XiAd^T \Sigma^\dagger]+\mathrm{h.c.}$  
& $g^2~ \mathrm{Tr}[\chi \Sigma^\dagger+ \Sigma \chi^\dagger] \mathrm{Tr}[T^A_{\mathrm{L}} \Sigma (T^A_{\mathrm{L}})^T \Sigma^\dagger]$ 
& $K_8^g$ 
& $ 12  B_0 g^2 (m_1+m_2) \cos \theta \sin^2\theta+\dots$
\\
& $\mathrm{Tr}[\XiA \Sigma^\dagger \XiAd^2]+\mathrm{h.c.}$  
& $g^2~\mathrm{Tr}[(\chi \Sigma^\dagger + \Sigma \chi^\dagger) T^A_{\mathrm{L}}  T^A_{\mathrm{L}}]$ 
& $K_9^g$ 
& $6 B_0 g^2 m_1 \cos \theta+\dots$
\\
& 
& $g^{\prime 2}~\mathrm{Tr}[(\chi \Sigma^\dagger + \Sigma \chi^\dagger) T_Y  T_Y]$ 
& $K_9^{g^\prime}$ 
& $2 B_0 g^{\prime 2} m_2 \cos \theta+\dots$
\\
& $ \mathrm{Tr}[\XiA \XiAd^T \Sigma^\dagger \XiAd]+\mathrm{h.c.}$  
& $g^2~\mathrm{Tr}[\chi (T^A_{\mathrm{L}})^T \Sigma^\dagger  T^A_{\mathrm{L}}]+\mathrm{h.c.}$ 
& $K_{10,11}^g$ 
& $-6 B_0 g^{ 2} m_1 \cos \theta+\dots$
\\
& 
& $g^{\prime 2}~\mathrm{Tr}[\chi T_Y^T \Sigma^\dagger  T_Y]+\mathrm{h.c.}$ 
& $K_{10,11}^{g^\prime}$ 
& $-2 B_0 g^{\prime 2} m_2 \cos \theta+\dots$
\\
& $\mathrm{Tr}[\XiA \Sigma^\dagger \XiAd \Sigma \XiAd^T \Sigma^\dagger]+\mathrm{h.c.}$  
& $g^2~ \mathrm{Tr}[\chi \Sigma^\dagger T^A_{\mathrm{L}} \Sigma   (T^A_{\mathrm{L}})^T \Sigma^\dagger]+\mathrm{h.c.}$ 
& $K_{10,11}^g$ 
&  $-3 B_0  \cos \theta [(m_1-m_2) +(m_1+m_2) \cos (2\theta) + \dots]$
\\
\hline\hline
	\end{tabular}
	}
\end{center}
\caption{ \footnotesize Non-derivative operators up to NLO that contain the mass spurion $\XiA=\chi$ and/or  the gauge spurions $\XiAd=g T^A_{\mathrm{L}}, ~g^\prime T_Y$ in the pseudo-real case.
The operators can be grouped according to different classes depending on whether they contain only the mass spurion, only the gauge spurions, or both kinds of spurions.
The gauging of $\mathrm{U}(1)_Y$ can be taken into account through the replacements $g\rightarrow g^\prime$ and $T^A_{\mathrm{L}}\rightarrow T_Y$ in the third column.
Similarly, the same replacements hold in the last column with an additional factor of three less for each factor $g^{\prime 2}$ (operators of class $g^{\prime 4}$, $g^2 g^{\prime 2}$ and $g^{\prime 2} \chi$).
When several orderings of the spurions lead to different operators, only one  is shown for each general form of operators as the others can easily be inferred from the table.
}
\label{tab-mass-gauge-spurions}
\end{table}

The corresponding operators are displayed in Tabs.~\ref{tab-mass-gauge-spurions} and~\ref{tab-mass-gauge-spurions-derivative}.
Note that the derivative operators (see App.~\ref{General classification of the spurionic operators}) as well as the contact terms (with traces made of spurions only) have been included  in order to check the completeness of our basis with Eq.~(\ref{Gasser-Leutwyler-Lagrangian}).

$\bullet$ \textbf{Gauge spurion}
\\
We now include both the mass and  gauge spurions,  $\XiAd^A=gT^A_{\mathrm{L}}$ and
$\XiAd= g^\prime T_Y$.
The latter enter in the chiral expansion at order ${\cal O}(p)$ and the resulting four classes (including the mixed operators involving both mass and gauge spurions) of operators correspond to:
\begin{itemize}
    \item[(i)] Two  spurions $\XiAd^2$ (classes $g^2$ and $g^{\prime 2}$).
    \item[(ii)] Four spurions  $\XiAd^4$ (classes $g^4$, $g^{\prime 4}$ and $g^2 g^{\prime 2}$).
    \item[(iii)] Two spurions $\XiAd^2$ and two derivatives (classes $g^2 D^2 $ and $g^{\prime 2} D^2$).
    \item[(iv)] One  spurion $\XiA$ or $\XiA^\dagger$ and two spurions $\XiAd^2$ (classes $g^2 \chi$ and $g^{\prime 2} \chi$).
\end{itemize}
As already mentioned, in order to check the consistency of our classification, the derivative operators as well as contact terms (see Refs~\cite{Urech:1994hd,Knecht:1999ag}) have also been included.
All of the operators corresponding to the above classes are reported in Tabs.~\ref{tab-mass-gauge-spurions} and \ref{tab-mass-gauge-spurions-derivative}, while those associated with the gauging of $\mathrm{U}(1)_Y$ can be obtained from $g\rightarrow g^\prime$ and $T^A_{\mathrm{L}}\rightarrow T_Y$.
Furthermore, the expansion of the operator gives the same result as for the $\SU(2)$ spurion but with an additional factor of $1/3$, except when explicitly listed in the tables.
%

\begin{table}[t]
\footnotesize 
    \renewcommand{\arraystretch}{0.6}
    \begin{center}
    \scalebox{0.7}{
	\begin{tabular}{ c c  c c }
\hline\hline
Class & General form & Operator & Associated LEC 
\\
\hline 
$\chi D^2 $~~ 
& $\mathrm{Tr}[\XiA \Sigma^\dagger] \mathrm{Tr}[D_\mu \Sigma (D^\mu \Sigma)^\dagger] +\mathrm{h.c.} $~~~~ 
&$\mathrm{Tr}[\chi \Sigma^\dagger+ \Sigma \chi^\dagger] \mathrm{Tr}[D_\mu \Sigma (D^\mu \Sigma)^\dagger] $
& $L_4$ 
\\
& $\mathrm{Tr}[\XiA \Sigma^\dagger D_\mu \Sigma (D^\mu \Sigma)^\dagger] +\mathrm{h.c.}$  
&$\mathrm{Tr}[(\chi \Sigma^\dagger+ \Sigma \chi^\dagger) D_\mu \Sigma (D^\mu \Sigma)^\dagger] $
& $L_5$ 
\\
& $\mathrm{Tr}[\XiA (D^2 \Sigma)^\dagger] +\mathrm{h.c.}$ 
& $\mathrm{Tr}[ \chi (D^2 \Sigma)^\dagger+ (D^2 \Sigma) \chi^\dagger]+\mathrm{h.c.}$ 
& e.o.m 
\\
\hline
$g^2 D^2$ 
& $\mathrm{Tr}[(D_\mu \Sigma)^\dagger (D^\mu \Sigma)] \mathrm{Tr}[\XiAd^2]$  
& $\frac{f^2 g^2}{16 \pi^2}  \mathrm{Tr}[(D_\mu \Sigma)^\dagger (D^\mu \Sigma)] \mathrm{Tr}[T^A_{\mathrm{L}} T^A_{\mathrm{L}}]$ 
& $K_1^g$ 
\\
& $\mathrm{Tr}[(D_\mu \Sigma)^\dagger (D^\mu \Sigma)] \mathrm{Tr}[\XiAd \Sigma \XiAd^T \Sigma^\dagger]$  
& $\frac{f^2 g^2}{16 \pi^2}  \mathrm{Tr}[(D_\mu \Sigma)^\dagger (D^\mu \Sigma)] \mathrm{Tr}[T^A_{\mathrm{L}} \Sigma (T^A_{\mathrm{L}})^T \Sigma^\dagger]$ 
& $K_2^g$ 
\\
& $\mathrm{Tr}[\XiAd (D_\mu \Sigma) \Sigma^\dagger] \mathrm{Tr}[\XiAd (D^\mu \Sigma) \Sigma^\dagger]$ 
& $ \frac{f^2 g^2}{16 \pi^2} \mathrm{Tr}[T^A_{\mathrm{L}} (D_\mu \Sigma) \Sigma^\dagger] \mathrm{Tr}[T^A_{\mathrm{L}} (D^\mu \Sigma) \Sigma^\dagger]$ 
& $K_{3,4}^g$ 
\\
& $\mathrm{Tr}[\XiAd^2 (D_\mu \Sigma)(D^\mu \Sigma)^\dagger]$ 
& $\frac{f^2 g^2}{16 \pi^2}  \mathrm{Tr}[T^A_{\mathrm{L}} T^A_{\mathrm{L}} (D_\mu \Sigma)(D^\mu \Sigma)^\dagger]$  
& $K_5^g$ 
\\
& $ \mathrm{Tr}[\XiAd (D^\mu \Sigma) (D_\mu \Sigma)^\dagger  \Sigma \XiAd^T \Sigma^\dagger]+\mathrm{h.c.}$ 
& $\frac{f^2 g^2}{16 \pi^2} \mathrm{Tr}[T^A_{\mathrm{L}} (D^\mu \Sigma) (D_\mu \Sigma)^\dagger  \Sigma (T^A_{\mathrm{L}})^T \Sigma^\dagger]+\mathrm{h.c.}$   
& $K_6^g$ 
\\
& $\mathrm{Tr}[(D_\mu \XiAd)(D^\mu \Sigma) \XiAd^T \Sigma^\dagger]+\mathrm{h.c.}$ 
& $ \frac{f^2 g^2}{16 \pi^2} \mathrm{Tr}[(D_\mu T^A_{\mathrm{L}})(D^\mu \Sigma) (T^A_{\mathrm{L}})^T \Sigma^\dagger]+\mathrm{h.c.}$ 
& $K_{12}^g$ 
\\
& $\mathrm{Tr}[(D_\mu \XiAd) \Sigma(D^\mu \XiAd)^T \Sigma^\dagger]$ 
&$\frac{f^2 g^2}{16 \pi^2} \mathrm{Tr}[(D_\mu T^A_{\mathrm{L}}) \Sigma(D^\mu T^A_{\mathrm{L}})^T \Sigma^\dagger]$  
& $K_{13}^g$ 
\\ 
& $\mathrm{Tr}[(D_\mu \XiAd) (D^\mu \XiAd)]$ 
&  $\frac{f^2 g^2}{16 \pi^2} \mathrm{Tr}[(D_\mu T^A_{\mathrm{L}}) (D^\mu T^A_{\mathrm{L}})]$ 
& $K_{14}^g$ 
\\
\hline\hline
\end{tabular}
}
\end{center}
\caption{\footnotesize Same as in Tab.~\ref{tab-mass-gauge-spurions} but for the derivative operators. 
The covariant derivatives involve the gauge spurions $\XiAd^\mu=g T^A_{\mathrm{L}} W_\mu^A+g^\prime T_Y B_\mu$ such that mixed operators belong to the class $D^2 \chi$.}
\label{tab-mass-gauge-spurions-derivative}
\end{table}

\newpage

\subsection{Top quark spurions}

We now discuss the spurions generating the top mass:
in the following, we consider a bilinear coupling  as well as linear couplings {\it \`{a} la} partial compositeness.
For the linear top coupling cases with antisymmetric and adjoint spurions there
are more than one possible spurion embedding, and we use the general linear combinations defined in Eqs~\eqref{Pt_A} and~\eqref{Pt_Adj}.

$\bullet$ \textbf{Bilinear coupling}

The four classes of non-derivative operators involving the top bilinear spurion ${\XiA^{\alpha,\dagger}=y_t P^\alpha}$ correspond to:
\begin{itemize}
    \item[(i)] Two top spurions $(\XiA \XiA^\dagger)$ (class $y_t^2$).
    \item[(ii)] Four top spurions  $(\XiA \XiA^\dagger)^2$ (class $y_t^4$).
    \item[(iii)] Two top spurions $(\XiA \XiA^\dagger)$ and one mass spurion $\XiA$ or $\XiA^\dagger$   (class $y_t^2 \chi$).
    \item[(iv)] Two top spurions $(\XiA \XiA^\dagger)$ and two gauge spurions $\XiAd^2$ (classes $y_t^2 g^2 $ and $y_t^2 g^{\prime 2}$)   .
\end{itemize}
where the two last classes involved mixed operators with two different spurions.
All of the operators, up to NLO, that contribute to the NGB potential at tree level and involve the top bilinear spurion are listed in Tab.~\ref{tab-bilinear-potential}.

\begin{table}[tb]
\footnotesize 
\renewcommand{\arraystretch}{0.6}
\begin{center}
\scalebox{0.7}{
\begin{tabular}{ c c  c c }
\hline\hline
Class & General form & Operator & $\SU(4)/\Sp(4)$ 
\\
\hline
$y_t^2$ 
&  $\mathrm{Tr}[\XiA \Sigma^\dagger] \mathrm{Tr}[\Sigma \XiA^\dagger]$  
& $y_t^2~ \mathrm{Tr}[P^\alpha \Sigma^\dagger] \mathrm{Tr}[\Sigma P_{\alpha}^\dagger] $ 
& $y_t^2~ \sin^2 \theta +\dots$
\\
\hline
$y_t^4$ 
& $\left(\mathrm{Tr}[\XiA \Sigma^\dagger] \mathrm{Tr}[\Sigma \XiA^\dagger]\right)^2$  
& $y_t^4~ \left(\mathrm{Tr}[P^\alpha \Sigma^\dagger]  \mathrm{Tr}[\Sigma P_{\alpha}^\dagger]\right)^2 $ 
&  $y_t^4~ \sin^4 \theta +\dots$
\\
& $\mathrm{Tr}[\XiA \Sigma^\dagger]^2 \mathrm{Tr}[\Sigma \XiA^\dagger \Sigma \XiA^\dagger]+\mathrm{h.c.}$  
&$y_t^4~\mathrm{Tr}[P^\alpha \Sigma^\dagger] \mathrm{Tr}[P^\beta \Sigma^\dagger] \mathrm{Tr}[\Sigma P_{\alpha}^\dagger \Sigma P_{\beta}^\dagger] +\mathrm{h.c.} $ 
& $y_t^4~ \sin^4 \theta +\dots$
\\
& $\mathrm{Tr}[\XiA \Sigma^\dagger] \mathrm{Tr}[\XiA \XiA^\dagger \Sigma \XiA^\dagger]+\mathrm{h.c.}$  
&$y_t^4~ \mathrm{Tr}[P^\alpha \Sigma^\dagger] \mathrm{Tr}[P^\beta P_\alpha^\dagger \Sigma P_{\beta}^\dagger]  +\mathrm{h.c.} $ 
& $\dfrac{3}{4} y_t^4~ \sin^2 \theta +\dots$
\\
& $\mathrm{Tr}[\XiA \Sigma^\dagger \XiA \Sigma^\dagger] \mathrm{Tr}[\Sigma \XiA^\dagger \Sigma \XiA^\dagger]$  
& $y_t^4~\mathrm{Tr}[P^\alpha \Sigma^\dagger P^\beta \Sigma^\dagger]  
\mathrm{Tr}[\Sigma P_{\alpha}^\dagger \Sigma P_{\beta}^\dagger]$ 
& $\dfrac{1}{4} y_t^4~ \sin^4 \theta +\dots$
\\
& $\mathrm{Tr}[\XiA \Sigma^\dagger \XiA \XiA^\dagger \Sigma \XiA^\dagger]$  
& $y_t^4~\mathrm{Tr}[P^\alpha \Sigma^\dagger P^\beta P_\alpha^\dagger \Sigma P_\beta^\dagger]$ 
& $- \dfrac{1}{8} y_t^4~ \cos  (2\theta) +\dots$
\\
\hline\hline
$y_t^2~ \chi$ 
& $\mathrm{Tr}[\Xi_{A_{1}} \Sigma^\dagger \Xi_{A_{2}} \Xi_{A_2}^\dagger]+\mathrm{h.c.}$ 
& $y_t^2~ \mathrm{Tr}[\chi\Sigma^\dagger P^\alpha P_\alpha^\dagger]+\mathrm{h.c.}$   
&  $2 y_t^2 B_0 (m_1+m_2) \cos \theta + \dots$
\\
& $\mathrm{Tr}[\Xi_{A_1} \Sigma^\dagger] \mathrm{Tr}[\Xi_{A_2} \Sigma^\dagger] \mathrm{Tr}[\Sigma \Xi_{A_2}^\dagger] +\mathrm{h.c.}$ 
&$y_t^2~ \mathrm{Tr}[\chi \Sigma^\dagger] \mathrm{Tr}[P^\alpha \Sigma^\dagger] \mathrm{Tr}[\Sigma P_{\alpha}^\dagger] +\mathrm{h.c.}$ 
&  $8  y_t^2 B_0 (m_1+m_2) \cos \theta \sin^2 \theta+\dots $
\\
& $\mathrm{Tr}[\Xi_{A_2} \Sigma^\dagger] \mathrm{Tr}[\Sigma \Xi_{A_1}^\dagger \Sigma \Xi_{A_2}^\dagger] +\mathrm{h.c.}$ 
&$y_t^2~ \mathrm{Tr}[P^\alpha \Sigma^\dagger] \mathrm{Tr}[\Sigma \chi^\dagger \Sigma P_{\alpha}^\dagger] +\mathrm{h.c.}$ 
&  $4  y_t^2 B_0 (m_1+m_2) \cos \theta \sin^2 \theta+\dots $
\\
\hline\hline
$y_t^2 g^2$ 
& $\mathrm{Tr}[\XiA \Sigma^\dagger] \mathrm{Tr}[\Sigma \XiA^\dagger] \mathrm{Tr}[\XiAd \Sigma \XiAd^T \Sigma^\dagger ]$ 
&$ y_t^2 g^2 ~\mathrm{Tr}[P^\alpha \Sigma^\dagger] \mathrm{Tr}[\Sigma P_{\alpha}^\dagger] \mathrm{Tr}[T^A_{\mathrm{L}} \Sigma (T^A_{\mathrm{L}})^T \Sigma^\dagger]$ 
& $-\dfrac{3}{8} y_t^2 g^2 \sin^2 (2\theta)+ \dots$
\\
& $\mathrm{Tr}[\XiA \Sigma^\dagger \XiAd] \mathrm{Tr}[\Sigma \XiA^\dagger \XiAd]$ 
&$y_t^2 g^2 ~ \mathrm{Tr}[P^\alpha \Sigma^\dagger T^A_{\mathrm{L}}]  \mathrm{Tr}[\Sigma P_{\alpha}^\dagger  T^A_{\mathrm{L}}]$
& $\dfrac{3}{16} y_t^2 g^2 \sin^2 \theta + \dots$
\\
& $\mathrm{Tr}[\XiA \Sigma^\dagger] \mathrm{Tr}[\Sigma \XiA^\dagger \XiAd^2]+\mathrm{h.c.}$ 
&$y_t^2 g^2 ~\mathrm{Tr}[P^\alpha \Sigma^\dagger]  \mathrm{Tr}[\Sigma P_{\alpha}^\dagger  T^A_{\mathrm{L}} T^A_{\mathrm{L}}]+\mathrm{h.c.}$ 
& $\dfrac{3}{4} y_t^2 g^2 \sin^2 \theta + \dots$
\\
& $\mathrm{Tr}[\XiA \Sigma^\dagger] \mathrm{Tr}[ \XiA^\dagger \XiAd \Sigma \XiAd^T ]+\mathrm{h.c.}$ 
&$y_t^2 g^2 ~\mathrm{Tr}[P^\alpha \Sigma^\dagger]  \mathrm{Tr}[ P_{\alpha}^\dagger  T^A_{\mathrm{L}} \Sigma (T^A_{\mathrm{L}})^T]+\mathrm{h.c.}$ 
& $0$
\\
& $\mathrm{Tr}[\XiA \Sigma^\dagger] \mathrm{Tr}[ \Sigma \XiA^\dagger \Sigma \XiAd^T \Sigma^\dagger \XiAd]+\mathrm{h.c.}$ 
&$y_t^2 g^2 ~\mathrm{Tr}[P^\alpha \Sigma^\dagger]  \mathrm{Tr}[\Sigma P_{\alpha}^\dagger \Sigma (T^A_{\mathrm{L}})^T \Sigma^\dagger T^A_{\mathrm{L}}]+\mathrm{h.c.}$ 
& $-\dfrac{3}{8} y_t^2 g^2 \sin^2 (2\theta) + \dots$
\\
& $\mathrm{Tr}[\XiA \XiAd^T \XiA^\dagger \XiAd]$ 
&$y_t^2 g^2 ~\mathrm{Tr}[P^\alpha (T^A_{\mathrm{L}})^T  P_{\alpha}^\dagger T^A_{\mathrm{L}}]$ 
& $0$
\\
& $\mathrm{Tr}[\XiA \XiA^\dagger \XiAd \Sigma  \XiAd^T \Sigma^\dagger]+\mathrm{h.c.}$ 
&$y_t^2 g^2 ~\mathrm{Tr}[P^\alpha  P_{\alpha}^\dagger T^A_{\mathrm{L}} \Sigma (T^A_{\mathrm{L}})^T \Sigma^\dagger]+\mathrm{h.c.}$ 
& $-\dfrac{3}{4} y_t^2 g^2 \cos^2 \theta+\dots$
\\
& $\mathrm{Tr}[\XiA  \Sigma^\dagger \XiAd \Sigma \XiA^\dagger \XiAd ]$ 
&$y_t^2 g^2 ~\mathrm{Tr}[P^\alpha \Sigma^\dagger T^A_{\mathrm{L}} \Sigma P_{\alpha}^\dagger  T^A_{\mathrm{L}}]$
& $0$
\\
& $\mathrm{Tr}[\XiA  \Sigma^\dagger \XiAd^2 \Sigma \XiA^\dagger ]$ 
&$y_t^2 g^2 ~\mathrm{Tr}[P^\alpha \Sigma^\dagger T^A_{\mathrm{L}} T^A_{\mathrm{L}} \Sigma P_{\alpha}^\dagger ]$
& $\dfrac{3}{8} y_t^2 g^2 +\dots$
\\
& $\mathrm{Tr}[\XiA \Sigma^\dagger \XiAd \Sigma \XiA^\dagger \Sigma \XiAd^T \Sigma^\dagger]$ 
&$y_t^2 g^2 ~\mathrm{Tr}[P^\alpha \Sigma^\dagger T^A_{\mathrm{L}}\Sigma P_{\alpha}^\dagger   \Sigma (T^A_{\mathrm{L}})^T \Sigma^\dagger]$ 
& $-\dfrac{3}{32} y_t^2 g^2 \sin^2(2 \theta)+\dots$
\\
\hline\hline
\end{tabular}
}
\end{center}
\caption{\footnotesize Non-derivative operators up to NLO involving the top bilinear spurion $\XiA^{\alpha,\dagger}=y_t P^\alpha$ and contributing to the scalar potential.
Also shown are the mixed operators involving the top bilinear spurion and the gauge or the mass spurion.
When not explicitly written, the $\mathrm{U}(1)_Y$ contributions are obtained by the following replacements  $g\rightarrow g^\prime$ and $T^A_L \rightarrow T_Y$ in the third column and similarly in the last column with a factor of three less. 
When several orderings of the spurions lead to different operators, only one  is shown for each general form of operators as the others can easily be inferred from the table.}
\label{tab-bilinear-potential}
\end{table}

\newpage

$\bullet$ \textbf{Linear coupling in the fundamental representation}

The three classes of operators involving the linear spurions in the fundamental representation $\XiF^\alpha=y_{t_L} P_q^\alpha$ and $\XiF=y_{t_R}P_t$ correspond to:

\begin{itemize}
    \item[(i)] Four top spurions  $(\XiF \XiF^\dagger)^2$ (classes $y_{t_L}^2 y_{t_R}^2$ and $y_{t_{L,R}}^4$).
    \item[(ii)] Two top spurions $(\XiF \XiF^\dagger)$ and one mass spurion $\XiA$ or $\XiA^\dagger$   (classes $y_{t_{L,R}}^2 \chi$).
    \item[(iii)] Two top spurions $(\XiF \XiF^\dagger)$ and two gauge spurions $\XiAd^2$ (classes $y_{t_{L,R}}^2 g^2 $ and $y_{t_{L,R}}^2 g^{\prime 2}$).   
\end{itemize}
The operators, belonging to the three above classes are listed in Tab.~\ref{tab-fund-potential}.

\begin{table}[h!]
\footnotesize 
\renewcommand{\arraystretch}{0.6}
\begin{center}
\scalebox{0.7}{
\begin{tabular}{ c c  c c }
\hline\hline
Class & General form & Operator & $\SU(4)/\Sp(4)$  \\
\hline 
$y_{t_{L}}^2 y_{t_R}^2,~ y_{t_{L,R}}^4 $  ~~
&  $\mathrm{Tr}[\XiF\cdot \XiF^T \Sigma^\dagger]\mathrm{Tr}[\Sigma \XiF^* \cdot \XiF^\dagger]$ & ~~~~~~$y_{t_L}^2 y_{t_R}^2 ~\mathrm{Tr}[P_Q^\alpha \cdot P_t^T \Sigma^\dagger] \mathrm{Tr}[\Sigma P_{Q \alpha}^* \cdot P_t^\dagger]$
& $- y_{t_L}^2 y_{t_R}^2 \sin^2 \theta +\dots$
\\ 
& 
& $y_{t_L}^4 ~\mathrm{Tr}[P_Q^\alpha \cdot P_Q^{\beta T} \Sigma^\dagger] \mathrm{Tr}[\Sigma P_{Q \beta}^* \cdot P_{Q \alpha}^\dagger]$ 
& $2 y_{t_L}^4 \cos^2 \theta + \dots$
\\
& 
& $y_{t_R}^4 ~\mathrm{Tr}[P_t \cdot P_t^{ T} \Sigma^\dagger] \mathrm{Tr}[\Sigma P_t^* \cdot P_t^\dagger]$ 
& $0$
\\
\hline
$y_{t_{L,R}}^2 \chi$ 
& $\mathrm{Tr}[\XiA \Sigma^\dagger  \XiF \cdot \XiF^\dagger]+\mathrm{h.c.} $ 
& $y_{t_L}^2 ~\mathrm{Tr}[\chi \Sigma^\dagger  P_Q^\alpha \cdot P_{Q \alpha}^\dagger]+\mathrm{h.c.} $ 
& $8 y_{t_L}^2 m_1 B_0 \cos \theta +\dots$
\\
& 
& $y_{t_R}^2 ~\mathrm{Tr}[\chi \Sigma^\dagger  P_t \cdot P_t^\dagger]+\mathrm{h.c.} $ 
&  $4 y_{t_R}^2 m_2 B_0 \cos \theta +\dots$
\\
\hline
$y_{t_{L,R}}^2 g^2$ 
& $\mathrm{Tr}[\Sigma \XiAd^T \XiAd^T  \Sigma^\dagger  \XiF \cdot \XiF^\dagger]$   
& $ y_{t_L}^2 g^2 ~\mathrm{Tr}[\Sigma (T^A_{\mathrm{L}})^T (T^A_{\mathrm{L}})^T  \Sigma^\dagger  P_Q^\alpha \cdot P_{Q\alpha}^\dagger]$ 
& $ \dfrac{3}{2} y_{t_L}^2 g^2 \cos^2 \theta+\dots$
\\
&  
& $ y_{t_L}^2 g^{\prime 2} ~\mathrm{Tr}[\Sigma T_Y^T T_Y^T  \Sigma^\dagger  P_Q^\alpha \cdot P_{Q\alpha}^\dagger]$ 
& $ \dfrac{1}{2} y_{t_L}^2 g^{\prime 2} \sin^2 \theta+\dots$
\\
 &   
 & $ y_{t_R}^2 g^2~ \mathrm{Tr}[\Sigma (T^A_{\mathrm{L}})^T (T^A_{\mathrm{L}})^T  \Sigma^\dagger  P_t \cdot P_{t}^\dagger]$ 
 & $ \dfrac{3}{4} y_{t_R}^2 g^2 \sin^2 \theta+\dots$
\\
 &   
 & $ y_{t_R}^2 g^{\prime 2}~ \mathrm{Tr}[\Sigma T_Y^T T_Y^T  \Sigma^\dagger  P_t \cdot P_{t}^\dagger]$ 
 & $ \dfrac{1}{4} y_{t_R}^2 g^{\prime 2} \cos^2 \theta+\dots$
\\
 & $\mathrm{Tr}[\XiAd \Sigma \XiAd^T \Sigma^\dagger    \XiF \cdot \XiF^\dagger]+\mathrm{h.c.}$ 
 & $ y_{t_L}^2 g^2 ~\mathrm{Tr}[T^A_L\Sigma (T^A_{\mathrm{L}})^T \Sigma^\dagger    P_Q^\alpha \cdot P_{Q\alpha}^\dagger]+\mathrm{h.c.}$ 
 & $ -3 y_{t_L}^2 g^2 \cos^2 \theta+\dots$
\\
 &
 & $ y_{t_L}^2 g^{\prime 2} ~\mathrm{Tr}[T_Y \Sigma T_Y^T \Sigma^\dagger    P_Q^\alpha \cdot P_{Q\alpha}^\dagger]+\mathrm{h.c.}$ 
 & $ 0$
\\
 & 
 & $ y_{t_R}^2 g^2 ~\mathrm{Tr}[T^A_L\Sigma (T^A_{\mathrm{L}})^T \Sigma^\dagger    P_t \cdot P_{t}^\dagger]+\mathrm{h.c.}$ 
 &  $0$
\\
 & 
 & $ y_{t_R}^2 g^{\prime 2} ~\mathrm{Tr}[T_Y \Sigma T_Y^T \Sigma^\dagger    P_t \cdot P_{t}^\dagger]+\mathrm{h.c.}$ 
 &  $-\dfrac{1}{2}y_{t_R}^2 g^{\prime 2} \cos^2 \theta + \dots$
\\
\hline\hline
\end{tabular}
}
\end{center}
\caption{\footnotesize Same as in Tab.~\ref{tab-bilinear-potential} but for  linear spurions in the fundamental representation, namely $\XiF^\alpha =y_{t_L} P_Q^\alpha$ and $\XiF=y_{t_R} P_t$.}
\label{tab-fund-potential}
\end{table}

\newpage

$\bullet$ \textbf{Linear coupling in the adjoint representation}

The four classes of operators involving the linear spurions in the adjoint representation $\XiAd^\alpha=y_{t_L} P_q^\alpha$ and $\XiAd=y_{t_R}P_t$ correspond to:

\begin{itemize}
    \item[(i)] Two top spurions  $\XiAd^2$ (classes  $y_{t_{L,R}}^2$).
    \item[(ii)] Four top spurions  $\XiAd^4$ (classes $y_{t_L}^2 y_{t_R}^2$ and $y_{t_{L,R}}^4$).
    \item[(iii)] Two top spurions $\XiAd^2$ and one mass spurion $\XiA$ or $\XiA^\dagger$   (classes $y_{t_{L,R}}^2 \chi$).
    \item[(iv)] Two top spurions $\XiAd^2$ and two gauge spurions $\XiAd^2$ (classes $y_{t_{L,R}}^2 g^2 $ and $y_{t_{L,R}}^2 g^{\prime 2}$).   
\end{itemize}
The operators, belonging to the above classes are listed in Tab.~\ref{tab-Adj-potential}.

\begin{table}[h!]
\footnotesize 
\renewcommand{\arraystretch}{0.6}
\begin{center}
\scalebox{0.7}{
\begin{tabular}{ c c  c c }
\hline\hline
Class & General form & Operator & $\SU(4)/\Sp(4)$  \\
\hline 
$y_{t_{L,R}}^2$  ~~
&  $\mathrm{Tr}[\XiAd \Sigma \XiAd^T \Sigma^\dagger]$ 
& $y_{t_{L}}^2 \mathrm{Tr}[P_{Q\alpha}^\dagger \Sigma P_{Q}^{\alpha T} \Sigma^\dagger]$
& $ -y_{t_{L}}^2\left((A_Q^2+B_Q^2)\sin^2\theta+4A_Q B_Q\cos^2\theta\right)+\dots$ 
\\
&  
& $y_{t_{R}}^2~ \mathrm{Tr}[P_{t}^\dagger \Sigma P_{t}^T \Sigma^\dagger]$
& $-\frac{1}{2}y_{t_{R}}^2(A_t^2-2B_t^2)\cos (2\theta)+ \dots$
\\
\hline
$y_{t_{L,R}}^4,~ y_{t_L}^2 y_{t_R}^2$  ~~
&  $\mathrm{Tr}[\XiAd \Sigma \XiAd^T \Sigma^\dagger]^2$ 
& $y_{t_{L}}^4~\mathrm{Tr}[P_{Q\alpha}^\dagger \Sigma P_{Q}^{\alpha T} \Sigma^\dagger]
    \mathrm{Tr}[P_{Q\beta}^\dagger \Sigma P_{Q}^{\beta T} \Sigma^\dagger]$
& $\frac{1}{4}y_{t_{L}}^4\left(-(A_Q^2-4A_Q B_Q+B_Q^2)\cos(2 \theta)+ (A_Q^2+4A_Q B_Q+B_Q^2)\right)^2+ \dots$
\\
&  
& $y_{t_{L}}^4~\mathrm{Tr}[P_{Q\alpha}^\dagger \Sigma P_{Q}^{\beta T} \Sigma^\dagger]
    \mathrm{Tr}[P_{Q\beta}^\dagger \Sigma P_{Q}^{\alpha T} \Sigma^\dagger]$
& $y_{t_{L}}^4\left[(A_Q^2+B_Q^2)\left((A_Q^2+B_Q^2)\sin^4 \theta + A_Q B_Q\sin^2(2\theta)\right)+8A_Q^2 B_Q^2\cos^4\theta\right]+\dots$
\\
&  
& $y_{t_{R}}^4~\mathrm{Tr}[P_{t}^\dagger \Sigma P_{t}^T \Sigma^\dagger]
    \mathrm{Tr}[P_{t}^\dagger \Sigma P_{t}^T \Sigma^\dagger]$
& $\frac{1}{4}y_{t_{R}}^4\left((A_t^2-2B_t^2)\cos(2\theta)+A_t^2\right)^2+\dots$
\\
%
&  
& $y_{t_{L}}^2 y_{t_{R}}^2~\mathrm{Tr}[P_{Q\alpha}^\dagger \Sigma P_{Q}^{\alpha T} \Sigma^\dagger] \mathrm{Tr}[P_{t}^\dagger \Sigma P_{t}^T \Sigma^\dagger]$
& $-\frac{1}{4}y_{t_{L}}^2y_{t_{R}}^2\left((A_Q^2-4A_Q B_Q+B_Q^2)\cos(2\theta)-(A_Q^2+4A_Q B_Q+B_Q^2)\right)$
\\
&
&
&\hspace{-1cm}$\cdot\left((A_t^2-2B_t^2)\cos(2\theta)+A_t^2\right)+\dots$
\\
&  
& $y_{t_{L}}^2 y_{t_{R}}^2~\mathrm{Tr}[P_{Q\alpha}^\dagger \Sigma P_{t}^{T} \Sigma^\dagger] \mathrm{Tr}[P_{t}^\dagger \Sigma P_{Q}^{\alpha T} \Sigma^\dagger]$
& $\frac{1}{8}y_{t_{L}}^2 y_{t_{R}}^2\sin^2(2\theta)\left(A_Q^2(A_t^2-2\sqrt{2}A_tB_t+2B_t^2)-2A_Q B_Q(A_t^2-2B_t^2)\right.$
\\
&
&
&\quad$\left.+B_Q^2 (A_t^2+2\sqrt{2}A_t B_t+2B_t^2)\right)+\dots$
\\
&  $\mathrm{Tr}[\XiAd^3 \Sigma \XiAd^T \Sigma^\dagger] + \mathrm{h.c.}$ 
& $y_{t_{L}}^4 \mathrm{Tr}[P_{Q\alpha}^{\dagger}P_{Q}^{\alpha}P_{Q\beta}^{\dagger}\Sigma P_{Q}^{\beta T}\Sigma^{\dagger}]+ \mathrm{h.c.}$
& $-2 y_{t_{L}}^4\left( (2A_Q^4+B_Q^4) \sin^2\theta+2A_Q B_Q(2A_Q^2+B_Q^2)\cos^2\theta\right)+\dots$
\\
&
& $y_{t_{R}}^4 \mathrm{Tr}[P_{t}^{\dagger}P_{t} P_{t}^{\dagger}\Sigma P_{t}^{T}\Sigma^{\dagger}]+ \mathrm{h.c.}$
& $-\frac{1}{4}y_{t_R}^4(A_t^2-2B_t^2)(2A_t^2+B_t^2)\cos(2\theta)+\dots$
\\
&
& $y_{t_{L}}^2y_{t_{R}}^2 \mathrm{Tr}[P_{t}^{\dagger}P_{t} P_{Q\alpha}^{\dagger}\Sigma P_{Q}^{\alpha T}\Sigma^{\dagger}] + \mathrm{h.c.}$
& $-\frac{1}{2}y_{t_{L}}^2y_{t_{R}}^2\left[\left(A_Q^2(2A_t^2+2\sqrt{2}A_tB_t+B_t^2)+B_Q^2B_t^2\right)\sin^2\theta+4A_QB_Q(A_t^2+\sqrt{2}A_tB_t+B_t^2)\cos^2\theta
\right]+\dots$
\\
&
& $y_{t_{L}}^2y_{t_{R}}^2 \mathrm{Tr}[P_{Q\alpha}^{\dagger}P_{Q}^{\alpha} P_{t}^{\dagger}\Sigma P_{t}^{T}\Sigma^{\dagger}] + \mathrm{h.c.}$
& 
 $y_{t_{L}}^2 y_{t_{R}}^2\left(B_t^2(A_Q^2+B_Q^2) \cos(2\theta) -A_Q^2A_t(2A_t\cos^2\theta+\sqrt{2}B_t\sin^2\theta)\right)+\dots $ 
\\
&  $\mathrm{Tr}[\XiAd^2 \Sigma \XiAd^T \XiAd^T \Sigma^\dagger]$ 
& $y_{t_{L}}^4 \mathrm{Tr}[P_{Q\alpha}^{\dagger}P_{Q}^{\alpha}\Sigma P_{Q\beta}^{*}P_{Q}^{\beta T}\Sigma^{\dagger}]$
& $2y_{t_L}^4\left((A_Q^4+B_Q^4)\sin^2\theta+3A_Q^2B_Q^2\cos^2\theta\right)+\dots$
\\
&
& $y_{t_{R}}^4\mathrm{Tr}[P_{t}^{\dagger}P_{t}\Sigma P_{t}^{*}P_{t}^{T}\Sigma^{\dagger}]$
& $\frac{1}{4}y_{t_{R}}^4A_t^2(A_t^2-2B_t^2)\cos(2\theta)+\dots$
\\
&
& $y_{t_{L}}^2y_{t_{R}}^2 \mathrm{Tr}[P_{t}^{\dagger}P_{t}\Sigma P_{Q\alpha}^{*}P_{Q}^{\alpha T}\Sigma^{\dagger}]$
&
 $-\frac{1}{2} y_{t_{L}}^2y_{t_{R}}^2A_t\left(A_Q^2A_t-B_Q^2(A_t+\sqrt{2}B_t)\right)\cos(2\theta)+\dots$
\\
&
& $y_{t_{L}}^2y_{t_{R}}^2 \mathrm{Tr}[P_{Q\alpha}^{\dagger}P_{Q}\Sigma P_{t}^{*}P_{t}^{T}\Sigma^{\dagger}]$
& $  \dfrac{1}{2}  y_{t_{L}}^2y_{t_{R}}^2\left(2A_Q^2A_t(A_t-\sqrt{2}B_t)\cos^2\theta-B_Q^2A_t^2\cos(2\theta)\right)+ \dots$

\\
&  $\mathrm{Tr}[\XiAd \Sigma \XiAd^T \Sigma^\dagger \XiAd \Sigma \XiAd^T \Sigma^\dagger] + \mathrm{h.c.}$ 
& $y_{t_{L}}^4 \mathrm{Tr}[P_{Q\alpha}^{\dagger}\Sigma P_{Q}^{\alpha T}\Sigma^{\dagger} P_{Q\beta}^{\dagger}\Sigma P_{Q}^{\beta T}\Sigma^{\dagger}]+ \mathrm{h.c.}$
& $2 y_{t_{L}}^4\left((A_Q^4+B_Q^4)\sin^4\theta+6A_Q^2B_Q^2\cos^4\theta+6A_QB_Q(A_Q^2-A_QB_Q+B_Q^2)\sin^2\theta\cos^2\theta\right)+\dots$
\\
&
& $y_{t_{R}}^4 \mathrm{Tr}[P_{t}^{\dagger}\Sigma P_{t}^{T}\Sigma^{\dagger} P_{t}^{\dagger}\Sigma P_{t}^{T}\Sigma^{\dagger}]+ \mathrm{h.c.}$
& $\frac{1}{8}y_{t_R}^4\left(4A_t^2(A_t^2-2B_t^2)\cos(2\theta)+(A_t^2-2B_t^2)^2\cos(4\theta)\right)+\dots$
\\
&
& $y_{t_{L}}^2y_{t_{R}}^2 \mathrm{Tr}[P_{t}^{\dagger}\Sigma P_{t}^{T}\Sigma^{\dagger} P_{Q\alpha}^{\dagger}\Sigma P_{Q}^{\alpha T}\Sigma^{\dagger}]+ \mathrm{h.c.}$
& $ - \frac{1}{8}y_{t_{L}}^2y_{t_{R}}^2\left[2\left(A_Q^2B_t(\sqrt{2}A_t+B_t)+4A_QB_Q(B_t^2-A_t^2)+B_Q^2B_t(B_t-\sqrt{2}A_t)\right)\cos(2\theta)\right.$
\\
&
&
&\hspace{-1.5cm}$\left.+(A_Q^2-4A_QB_Q+B_Q^2)(A_t^2-2B_t^2)\cos(4\theta)\right]+\dots$
\\
\hline
$y_{t_{L,R}}^2 \chi$ 
& $\mathrm{Tr}[\XiA \Sigma^\dagger  \XiAd^2]+\mathrm{h.c.} $ 
& $y_{t_{L}}^2 ~\mathrm{Tr}[\chi \Sigma^\dagger  P_{Q}^\alpha P_{Q\alpha}^\dagger]+\mathrm{h.c.} $
& $ 8 B_0 y_{t_L}^2 \left(  m_1 A_Q^2+  m_2 B_Q^2\right)\cos\theta + \dots$
\\
& 
&$y_{t_{R}}^2 ~\mathrm{Tr}[\chi \Sigma^\dagger  P_{t} P_{t}^\dagger]+\mathrm{h.c.} $
& $ 2B_0y_{t_R}^2\left(2A_t^2 m_2 +B_t^2(m_1+ m_2)\right)\cos\theta + \dots$
\\
& $\mathrm{Tr}[\XiA \XiAd^T \Sigma^\dagger  \XiAd]+\mathrm{h.c.} $ 
& $y_{t_{L}}^2~\mathrm{Tr}[\chi P_{Q}^{\alpha T} \Sigma^\dagger  P_{Q\alpha}^\dagger]+\mathrm{h.c.} $
& $-8B_0y_{t_L}^2A_QB_Q(m_1+m_2)\cos\theta+\dots$
\\
& 
& $y_{t_{R}}^2~\mathrm{Tr}[\chi P_{t}^T \Sigma^\dagger  P_{t}^\dagger]+\mathrm{h.c.} $
& $2B_0y_{t_R}^2\left(-2A_t^2 m_2+B_t^2(m_1+m_2)\right) \cos \theta + \dots$
\\
& $\mathrm{Tr}[\XiA \Sigma^\dagger]  \mathrm{Tr}[\XiAd \Sigma  \XiAd^T \Sigma^\dagger]+\mathrm{h.c.} $ 
& $y_{t_{L}}^2~\mathrm{Tr}[\chi\Sigma^\dagger]	    
    \mathrm{Tr}[P_{Q\alpha}^\dagger \Sigma  P_{Q}^{\alpha T} \Sigma^\dagger]+\mathrm{h.c.} $
& $4B_0 y_{t_{L}}^2 (m_1+m_2)\cos\theta\left((A_Q^2-4A_QB_Q+B_Q^2)\cos(2\theta)-(A_Q^2+4A_QB_Q+B_Q^2)\right) + \dots$
\\
& 
& $y_{t_{R}}^2~\mathrm{Tr}[\chi\Sigma^\dagger]  \mathrm{Tr}[P_{t}^\dagger \Sigma  P_{t}^{ T} \Sigma^\dagger]+\mathrm{h.c.} $
& $-4B_0 y_{t_R}^2 (m_1+m_2) \cos \theta \left(A_t^2+(A_t^2-2B_t^2)\cos(2\theta)\right) \dots$
\\
& $\mathrm{Tr}[\XiA \Sigma^\dagger \XiAd \Sigma  \XiAd^T \Sigma^\dagger]+\mathrm{h.c.} $ 
& $y_{t_{L}}^2 ~\mathrm{Tr}[\chi \Sigma^\dagger P_{Q \alpha}^\dagger \Sigma  P_{Q}^{\alpha T} \Sigma^\dagger]+\mathrm{h.c.} $
& $2B_0 y_{t_{L}}^2 (m_1+m_2) \cos \theta\left((A_Q^2-4A_QB_Q+B_Q^2) \cos(2\theta)-(A_Q^2+B_Q^2)\right) + \dots$
\\
&  
& $y_{t_{R}}^2 ~\mathrm{Tr}[\chi \Sigma^\dagger P_{t}^\dagger \Sigma  P_{t}^{ T} \Sigma^\dagger]+\mathrm{h.c.} $
& $ 2B_0y_{t_R}^2\cos\theta\left((2A_t^2m_1-3B_t^2(m_1+m_2))\sin^2\theta+(B_t^2(m_1+m_2)-2A_t^2m_2)\cos^2\theta\right)+\dots$ \\
\hline
$y_{t_{L,R}}^2 g^2$ 
& $\mathrm{Tr}[\XiAd \Sigma \XiAd^T \Sigma^\dagger]^2$ 
& $y_{t_{L}}^2 g^2 \mathrm{Tr}[T^A_L \Sigma T_L^{A T} \Sigma^\dagger] 
    \mathrm{Tr}[P_{Q\alpha}^\dagger \Sigma P_{Q}^{\alpha T} \Sigma^\dagger]$
& $\dfrac{3}{2}g^2 y_{t_{L}}^2 \cos^2\theta\left((A_Q^2+B_Q^2)\sin^2\theta+4A_QB_Q\cos^2\theta\right)+ \dots$
\\
& 
& $y_{t_{R}}^2 g^2 \mathrm{Tr}[T^A_L \Sigma T_L^{A T} \Sigma^\dagger] 
    \mathrm{Tr}[P_{t}^\dagger \Sigma P_{t}^{ T} \Sigma^\dagger]$ 
& $\frac{3}{4}g^2y_{t_R}^2\cos^2\theta\left((A_t^2-2B_t^2)\cos(2\theta)+A_t^2\right) + \dots$
\\
&  $\mathrm{Tr}[\XiAd^3 \Sigma \XiAd^T \Sigma^\dagger]$ 
&  $y_{t_{L}}^2 g^2 \mathrm{Tr}[T_L^A T_L^A P_{Q \alpha}^\dagger \Sigma P_{Q}^{\alpha T} \Sigma^\dagger]+ \mathrm{h.c.}$
& $ -\frac{3}{2} g^2y_{t_{L}}^2B_Q(2A_Q\cos^2\theta+B_Q\sin^2\theta)+\dots$
\\
&  
&  $y_{t_{L}}^2 g^{\prime\,2} \mathrm{Tr}[T_Y T_Y P_{Q \alpha}^\dagger \Sigma P_{Q}^{\alpha T} \Sigma^\dagger]+ \mathrm{h.c.}$
& $- \frac{1}{2} g^{\prime\,2}y_{t_{L}}^2A_Q\left(A_Q\sin^2\theta+2B_Q\cos^2\theta\right)+\dots$
\\
&  
&  $y_{t_{L}}^2 g^2 \mathrm{Tr}[ P_{Q \alpha}^\dagger  P_{Q}^{\alpha} T_L^A \Sigma T_L^{A\,T}\Sigma^\dagger]+ \mathrm{h.c.}$
& $ -3 g^2 y_{t_{L}}^2B_Q^2  \cos^2\theta+\dots$
\\
&  
&  $y_{t_{L}}^2 g^{\prime\,2} \mathrm{Tr}[ P_{Q \alpha}^\dagger  P_{Q}^{\alpha} T_Y \Sigma T_Y^{T}\Sigma^\dagger]+ \mathrm{h.c.}$
& $ - g^{\prime\,2}  y_{t_{L}}^2A_Q^2\cos^2\theta+\dots$
\\
&  
&  $y_{t_{R}}^2 g^2 \mathrm{Tr}[T_L^A T_L^A P_{t}^\dagger \Sigma P_{t}^{ T} \Sigma^\dagger]+ \mathrm{h.c.}$
& $ \frac{3}{4}  g^2y_{t_{R}}^2B_t^2\cos(2\theta)+\dots$
\\
&  
&  $y_{t_{R}}^2 g^{\prime\,2} \mathrm{Tr}[T_Y T_Y P_{t}^\dagger \Sigma P_{t}^{ T} \Sigma^\dagger]+ \mathrm{h.c.}$
& $ -\frac{1}{4}  g^{\prime\,2} \left(A_t^2-B_t^2\right)\cos(2\theta)+\dots$
\\
&  
&  $y_{t_{R}}^2 g^2 \mathrm{Tr}[ P_{t}^\dagger P_{t}T_L^A\Sigma T_L^{A\,T}\Sigma^\dagger]+ \mathrm{h.c.}$
& $ -\frac{3}{4}  g^2y_{t_{R}}^2B_t^2\cos^2\theta+\dots$
\\
&  
& $y_{t_{R}}^2 g^{\prime\,2} \mathrm{Tr}[ P_{t}^\dagger P_{t}T_Y\Sigma T_Y^{T}\Sigma^\dagger]+ \mathrm{h.c.}$
& $  -\frac{1}{4} g^{\prime\,2}y_{t_R}^2(2A_t^2+B_t^2)\cos^2\theta+\dots$
\\
&  $\mathrm{Tr}[\XiAd^2 \Sigma \XiAd^T \XiAd^T \Sigma^\dagger]$ 
& $y_{t_{L}}^2 g^2\mathrm{Tr}[T^A_L T^A_L \Sigma P_{Q \alpha}^* P_{Q}^{\alpha T} \Sigma^\dagger]$
& $  \dfrac{3}{2}  g^2y_{t_L}^2\left(A_Q^2\cos^2 \theta+  B_Q^2\sin^2 \theta\right) +\dots$
\\
& 
& $y_{t_{L}}^2 g^{\prime\,2}\mathrm{Tr}[T_Y T_Y \Sigma P_{Q \alpha}^* P_{Q}^{\alpha T} \Sigma^\dagger]$
& $  \dfrac{1}{2}  g^{\prime\,2}y_{t_L}^2\left(A_Q^2\sin^2 \theta+  B_Q^2\cos^2 \theta\right) +\dots$
\\
& 
& $y_{t_{L}}^2 g^2\mathrm{Tr}[P_{Q \alpha}^{\dagger}P_{Q}^{\alpha}\Sigma T^{A\,T}_L T^{A\,T}_L \Sigma^\dagger]$
& $  \dfrac{3}{2} g^2y_{t_L}^2\left(A_Q^2\sin^2 \theta+ B_Q^2\cos^2 \theta\right) +\dots$
\\
& 
& $y_{t_{L}}^2 g^{\prime\,2}\mathrm{Tr}[P_{Q \alpha}^{\dagger}P_{Q}^{\alpha}\Sigma T^{T}_Y T^{T}_Y \Sigma^\dagger]$
& $ \dfrac{1}{2}  g^{\prime\,2}y_{t_R}^2\left(A_Q^2\cos^2 \theta+ B_Q^2\sin^2 \theta\right) +\dots$
\\
&
& $y_{t_{R}}^2 g^2\mathrm{Tr}[T^A_L T^A_L \Sigma P_{t}^* P_{t}^{T} \Sigma^\dagger]$
& $-\dfrac{3}{8} g^2 y_{t_{R}}^2A_t^2\cos(2\theta)+ \dots$
\\
& 
& $y_{t_{R}}^2 g^{\prime\,2}\mathrm{Tr}[T_Y T_Y \Sigma P_{t}^* P_{t}^{T} \Sigma^\dagger]$ 
& $\dfrac{1}{8} g^{\prime\,2}y_{t_{R}}^2A_t^2\cos(2 \theta)+\dots$
\\
& 
& $y_{t_{R}}^2 g^2\mathrm{Tr}[P_{t}^{\dagger}P_{t}\Sigma T^{A\,T}_L T^{A\,T}_L \Sigma^\dagger]$
& $-\dfrac{3}{8} g^2y_{t_{R}}^2A_t^2\cos(2 \theta) +\dots$
\\
& 
& $y_{t_{R}}^2 g^{\prime\,2}\mathrm{Tr}[P_{t}^{\dagger}P_{t}^\Sigma T^{T}_Y T^{T}_Y \Sigma^\dagger]$ 
& $\dfrac{1}{8} g^{\prime\,2}y_{t_{R}}^2A_t^2\cos(2 \theta) +\dots$
\\
&  $\mathrm{Tr}[\XiAd \Sigma \XiAd^T \Sigma^\dagger \XiAd \Sigma \XiAd^T \Sigma^\dagger]$ 
& $y_{t_{L}}^2 g^2 \mathrm{Tr}[T^A_L \Sigma T_L^{A T} \Sigma^\dagger P_{Q\alpha}^\dagger 
    \Sigma P_{Q}^{\alpha T} \Sigma^\dagger]+ \mathrm{h.c.}$
& $-\dfrac{3}{4} g^{2} y_{t_{L}}^2 \cos^2 \theta\left((A_Q^2-4A_QB_Q+B_Q^2) \cos(2 \theta) - (A_Q^2+B_Q^2)\right) +\dots$
\\
&
& $y_{t_{L}}^2 g^{\prime\,2} \mathrm{Tr}[T_Y \Sigma T_Y^{ T} \Sigma^\dagger P_{Q\alpha}^\dagger 
    \Sigma P_{Q}^{\alpha T} \Sigma^\dagger]+ \mathrm{h.c.}$
& $-\dfrac{1}{4} g^{\prime\,2} y_{t_{L}}^2 \cos^2 \theta\left((A_Q^2-4A_QB_Q+B_Q^2) \cos(2 \theta)  - (A_Q^2+B_Q^2)\right) +\dots$  
\\
& 
& $y_{t_{R}}^2 g^2 \mathrm{Tr}[T^A_L \Sigma T_L^{A T} \Sigma^\dagger P_{t}^\dagger \Sigma P_{t}^{T} \Sigma^\dagger]
    + \mathrm{h.c.}$
& $ \frac{3}{4}  g^2y_{t_R}^2\cos^2\theta\left((A_t^2-2B_t^2)\cos(2\theta)-A_t^2+B_t^2\right) +\dots$
\\
& 
& $y_{t_{R}}^2 g^{\prime 2} \mathrm{Tr}[T_Y \Sigma T_Y^{T} \Sigma^\dagger P_{t}^\dagger \Sigma P_{t}^{T} \Sigma^\dagger]
    + \mathrm{h.c.}$
& $ \frac{1}{4}  g^{\prime\,2}y_{t_R}^2\cos^2\theta\left((A_t^2-2B_t^2)\cos(2\theta)+A_t^2+B_t^2\right) +\dots$
\\
\hline\hline
\end{tabular}
}
\end{center}
\caption{\footnotesize Same as in Tab.~\ref{tab-bilinear-potential} but for  linear spurions in the adjoint representation, namely $\XiAd^\alpha =y_{t_L} P_Q^\alpha$ and $\XiAd=y_{t_R} P_t$.\\
}
\label{tab-Adj-potential}
\end{table}


$\bullet$ \textbf{Linear coupling in the symmetric representation}

The three classes of operators involving the linear spurions in the symmetric representation $\XiS^\alpha=y_{t_L} P_q^\alpha$ and $\XiS=y_{t_R}P_t$ correspond to:

\begin{itemize}
    \item[(i)] Four top spurions  $(\XiS \XiS^\dagger)^2$ (classes $y_{t_L}^2 y_{t_R}^2$ and $y_{t_{L,R}}^4$).
    \item[(ii)] Two top spurions $(\XiS \XiS^\dagger)$ and one mass spurion $\XiA$ or $\XiA^\dagger$   (classes $y_{t_{L,R}}^2 \chi$).
    \item[(iii)] Two top spurions $(\XiS \XiS^\dagger)$ and two gauge spurions $\XiAd^2$ (classes $y_{t_{L,R}}^2 g^2 $ and $y_{t_{L,R}}^2 g^{\prime 2}$).   
\end{itemize}
The operators,  belonging to the three above classes are listed in Tab.~\ref{tab-Sym-potential}.

\begin{table}[tb]
\footnotesize 
\renewcommand{\arraystretch}{0.6}
\begin{center}
\scalebox{0.7}{
\begin{tabular}{ c c  c c }
\hline\hline
Class & General form & Operator & $\SU(4)/\Sp(4)$ \\
\hline 
$y_{t_{L,R}}^4, ~y_{t_L}^2 y_{t_R}^2 $  
&  ~~$\mathrm{Tr}[\XiS  \Sigma^\dagger \XiS \Sigma^\dagger] 
\mathrm{Tr}[ \Sigma \XiS ^\dagger  \Sigma \XiS^\dagger]$ 
& ~~~~~$y_{t_L}^4~ \mathrm{Tr}[P_Q^\alpha  \Sigma^\dagger P_Q^\beta \Sigma^\dagger] 
\mathrm{Tr}[ \Sigma P_{Q\alpha}^\dagger  \Sigma P_{Q\beta}^\dagger]$
& $ y_{t_L}^4~ \sin^4 \theta + \dots $
\\
& 
& $y_{t_R}^4 ~\mathrm{Tr}[P_t \Sigma^\dagger P_t\Sigma^\dagger] 
\mathrm{Tr}[ \Sigma P_t^\dagger  \Sigma P_t^\dagger]$
& $ y_{t_R}^4~ \cos^4 \theta + \dots $
\\
& 
& $y_{t_L}^2 y_{t_R}^2 ~\mathrm{Tr}[P_Q^\alpha \Sigma^\dagger P_t\Sigma^\dagger] 
\mathrm{Tr}[ \Sigma P_{Q \alpha}^\dagger  \Sigma P_t^\dagger]$
& $y_{t_L}^2 y_{t_R}^2~ \sin^2 \theta \cos^2 \theta + \dots $
\\
&  $\mathrm{Tr}[\XiS  \Sigma^\dagger \XiS  \XiS^\dagger \Sigma\XiS^\dagger ] +\mathrm{h.c.} $ 
& $y_{t_L}^4~ \mathrm{Tr}[P_Q^\alpha  \Sigma^\dagger P_Q^\beta  P_{Q\alpha}^\dagger \Sigma P_{Q\beta}^\dagger ] +\mathrm{h.c.} $
& $- y_{t_L}^4 ~ \cos (2\theta) + \dots $
\\
&  
& $y_{t_L}^4~ \mathrm{Tr}[P_Q^\alpha  \Sigma^\dagger P_Q^\beta  P_{Q\beta}^\dagger \Sigma P_{Q\alpha}^\dagger ] +\mathrm{h.c.} $
& $ y_{t_L}^4 ~ (1+ \sin^2 \theta ) + \dots $
\\
&  
& $y_{t_R}^4~ \mathrm{Tr}[P_t  \Sigma^\dagger P_t  P_{t}^\dagger \Sigma P_{t}^\dagger ] +\mathrm{h.c.} $
& $ y_{t_R}^4 ~  \cos^2 \theta  + \dots $
\\
&  
& $y_{t_L}^2 y_{t_R}^2 ~\mathrm{Tr}[P_Q^\alpha  \Sigma^\dagger P_t  P_{Q\alpha}^\dagger \Sigma P_{t}^\dagger ] +\mathrm{h.c.} $
& $-\dfrac{1}{2} y_{t_L}^2 y_{t_R}^2 ~  \sin^2 \theta  + \dots $
\\
&  
& $y_{t_L}^2 y_{t_R}^2~ \mathrm{Tr}[P_Q^\alpha  \Sigma^\dagger P_t  P_{t}^\dagger \Sigma P_{Q\alpha}^\dagger ] +\mathrm{h.c.} $
& $y_{t_L}^2 y_{t_R}^2  + \dots $
\\
\hline
$y_{t_{L,R}}^2 \chi$ 
&  $\mathrm{Tr}[\XiA \Sigma^\dagger  \XiS \XiS^\dagger]+\mathrm{h.c.} $ 
& $ y_{t_L}^2~\mathrm{Tr}[\chi \Sigma^\dagger  P_Q^\alpha P_{Q\alpha}^\dagger]+\mathrm{h.c.} $ 
& $4 y_{t_L}^2 B_0 (m_1+m_2) \cos \theta + \dots $
\\
& 
& $y_{t_R}^2 ~\mathrm{Tr}[\chi \Sigma^\dagger  P_t P_{t}^\dagger]+\mathrm{h.c.} $ 
& $4 y_{t_R}^2 B_0 m_2 \cos \theta + \dots $
\\
\hline
$y_{t_{L,R}}^2 g^2$, $y_{t_{L,R}}^2 g^{\prime 2}$  
& $\mathrm{Tr}[\XiS \Sigma^\dagger \XiAd ] \mathrm{Tr}[\Sigma \XiS^\dagger \XiAd]$ 
& $y_{t_L}^2 g^2 ~\mathrm{Tr}[P_Q^\alpha \Sigma^\dagger T^A_L ] \mathrm{Tr}[\Sigma P_{Q\alpha}^\dagger T^A_L]$ 
& $\dfrac{3}{8} y_{t_L}^2 g^2 ~\sin^2 \theta +\dots$
\\
& 
& $y_{t_R}^2 g^2 ~\mathrm{Tr}[P_t \Sigma^\dagger T^A_L ] \mathrm{Tr}[\Sigma P_t^\dagger T^A_L]$ 
& $0$
\\
& 
& $y_{t_R}^2 g^{\prime 2} ~\mathrm{Tr}[P_t \Sigma^\dagger T_Y ] \mathrm{Tr}[\Sigma P_t^\dagger T_Y]$ 
& $\dfrac{1}{2} y_{t_R}^2 g^{\prime 2} ~\cos^2 \theta +\dots$
\\
& $\mathrm{Tr}[\XiS \XiS^\dagger \XiAd\Sigma \XiAd^T\Sigma^\dagger ] +\mathrm{h.c.}$ 
& $y_{t_L}^2 g^2  ~\mathrm{Tr}[P_Q^\alpha P_{Q\alpha}^\dagger T^A_L \Sigma T^{A T}_L \Sigma^\dagger ] +\mathrm{h.c.}$ 
& $-\dfrac{3}{2} y_{t_L}^2 g^2 \cos^2\theta+\dots$
\\
& 
& $y_{t_R}^2 g^2  ~\mathrm{Tr}[P_t P_t^\dagger T^A_L \Sigma T^{A T}_L \Sigma^\dagger ] +\mathrm{h.c.}$ 
& $0$
\\
& 
& $y_{t_R}^2 g^{\prime 2}  ~\mathrm{Tr}[P_t P_t^\dagger T_Y\Sigma T_Y^T \Sigma^\dagger ] +\mathrm{h.c.}$ 
& $-\dfrac{1}{2} y_{t_R}^2 g^{\prime 2} \cos^2 \theta + \dots$
\\
& $\mathrm{Tr}[\XiS \Sigma^\dagger \XiAd \Sigma \XiS^\dagger  \XiAd ]$ 
& $y_{t_L}^2 g^2~ \mathrm{Tr}[P_Q^\alpha \Sigma^\dagger T^A_L \Sigma P_{Q\alpha}^\dagger T^A_L]$ 
& $0$
\\
& 
& $y_{t_R}^2 g^2 ~\mathrm{Tr}[P_t \Sigma^\dagger T^A_L \Sigma P_t^\dagger T^A_L]$ 
& $0$
\\
& 
& $y_{t_R}^2 g^{\prime 2} ~\mathrm{Tr}[P_t \Sigma^\dagger T_Y \Sigma P_t^\dagger T_Y]$ 
& $\dfrac{1}{4} y_{t_R}^2 g^{\prime 2} \cos^2 \theta + \dots$
\\
& $\mathrm{Tr}[\XiS \Sigma^\dagger \XiAd^2 \Sigma \XiS^\dagger ]$ 
& $y_{t_L}^2 g^2~ \mathrm{Tr}[P_Q^\alpha \Sigma^\dagger T_L^A T_L^A  \Sigma P_{Q\alpha}^\dagger]$ 
& $\dfrac{3}{4} y_{t_L}^2 g^2~ +\dots$
\\
& 
& $y_{t_R}^2 g^2~ \mathrm{Tr}[P_t \Sigma^\dagger T_L^A T_L^A  \Sigma P_{t}^\dagger]$ 
& $\dfrac{3}{4} y_{t_R}^2 g^2~\sin^2 \theta +\dots$
\\
& 
& $y_{t_R}^2 g^{\prime 2}~ \mathrm{Tr}[P_t \Sigma^\dagger T_Y T_Y  \Sigma P_{t}^\dagger]$ 
& $\dfrac{1}{4} y_{t_R}^2 g^{\prime 2}~\cos^2 \theta +\dots$
\\
\hline\hline
\end{tabular}
}
\end{center}
\caption{\footnotesize Same as in Tab.~\ref{tab-bilinear-potential} but for  linear spurions in the symmetric representation, namely $\XiS^\alpha =y_{t_L} P_Q^\alpha$ and $\XiS=y_{t_R} P_t$.}
\label{tab-Sym-potential}
\end{table}


$\bullet$ \textbf{Linear coupling in the antisymmetric representation}

Finally, the four classes of operators involving the linear spurions in the antisymmetric representation $\XiA^\alpha=y_{t_L} P_q^\alpha$ and 
$\XiA=y_{t_R}P_t$ correspond to:
\begin{itemize}
    \item[(i)] Two top spurions  $(\XiA \XiA^\dagger)$ (classes  $y_{t_{L,R}}^2$).
    \item[(ii)] Four top spurions  $(\XiA \XiA^\dagger)^2$ (classes $y_{t_L}^2 y_{t_R}^2$ and $y_{t_{L,R}}^4$).
    \item[(iii)] Two top spurions $(\XiA \XiA^\dagger)$ and one mass spurion $\XiA$ or $\XiA^\dagger$   (classes $y_{t_{L,R}}^2 \chi$).
    \item[(iv)] Two top spurions $(\XiA \XiA^\dagger)$ and two gauge spurions $\XiAd^2$ (classes $y_{t_{L,R}}^2 g^2 $ and $y_{t_{L,R}}^2 g^{\prime 2}$).   
\end{itemize}
The operators, belonging to the above classes are listed in Tab.~\ref{tab-Anti-potential}.

\begin{table}[h!]
\footnotesize 
\renewcommand{\arraystretch}{0.6}
\begin{center}
\scalebox{0.6}{
\begin{tabular}{ c c  c c }
\hline\hline
Class & General form & Operator & $\SU(4)/\Sp(4)$  \\
\hline 
$y_{t_{L,R}}^2$  ~~
&  $\mathrm{Tr}[\XiA \Sigma^\dagger]\mathrm{Tr}[\Sigma \XiA^\dagger]$ 
& $y_{t_L}^2 \mathrm{Tr}[P_Q^\alpha \Sigma^\dagger]\mathrm{Tr}[\Sigma P_{Q\alpha}^\dagger]$ 
& $2y_{t_L}^2  \sin^2 \theta +\dots$
\\
& 
& $y_{t_{R}}^2 \mathrm{Tr}[P_{t} \Sigma^\dagger]\mathrm{Tr}[\Sigma P_{t}^\dagger]$ 
& $2y_{t_{R}}^2(A-B)^2  \cos^2 \theta +\dots $
\\
\hline
$y_{t_{L,R}}^4,~ y_{t_L}^2 y_{t_R}^2$  
&  $\mathrm{Tr}[\XiA \Sigma^\dagger]^2 \mathrm{Tr}[\Sigma \XiA ^\dagger]^2$ 
& $y_{t_L}^4 \left(\mathrm{Tr}[P_Q^\alpha \Sigma^\dagger] \mathrm{Tr}[\Sigma P_{Q\alpha}^\dagger]\right)^2$ 
& $4y_{t_L}^4 \sin^4\theta +\dots$
\\ 
&  
& $y_{t_{R}}^4 \left(\mathrm{Tr}[P_{t} \Sigma^\dagger] \mathrm{Tr}[\Sigma P_{t}^\dagger] \right)
    \left(\mathrm{Tr}[P_{t} \Sigma^\dagger] \mathrm{Tr}[\Sigma P_{t}^\dagger] \right)$ 
& $4y_{t_{R}}^4(A-B)^4 \cos^4\theta +\dots$
\\ 
&  
& $y_{t_L}^2 y_{t_{R}}^2 \mathrm{Tr}[P_Q^\alpha \Sigma^\dagger] \mathrm{Tr}[\Sigma P_{Q\alpha}^\dagger] \mathrm{Tr}[P_{t} \Sigma^\dagger] \mathrm{Tr}[\Sigma P_{t}^\dagger]$ 
& $y_{t_L}^2 y_{t_{R}}^2(A-B)^2 \sin^2(2\theta) +\dots$
\\
&  $\mathrm{Tr}[\XiA \Sigma^\dagger]^2 \mathrm{Tr}[\Sigma \XiA ^\dagger \Sigma \XiA ^\dagger] +\mathrm{h.c.} $ 
& $y_{t_L}^4 \mathrm{Tr}[P_Q^\alpha \Sigma^\dagger] \mathrm{Tr}[P_Q^\beta \Sigma^\dagger]  \mathrm{Tr}[\Sigma P_{Q\alpha}^\dagger \Sigma P_{Q\beta}^\dagger] +\mathrm{h.c.} $ 
& $4y_{t_L}^4 \sin^4\theta +\dots$
\\
&  
& $y_{t_{R}}^4 \mathrm{Tr}[P_{t} \Sigma^\dagger]\mathrm{Tr}[P_{t} \Sigma^\dagger]  \mathrm{Tr}[\Sigma P_{t}^\dagger \Sigma P_{t}^\dagger] +\mathrm{h.c.} $ 
& $2y_{t_{R}}^4(A-B)^2 \cos^2\theta\left((A-B)^2\cos(2\theta)+(A+B)^2\right) +\dots$
\\
&   
& $y_{t_L}^2 y_{t_{R}}^2 \mathrm{Tr}[P_Q^\alpha \Sigma^\dagger] \mathrm{Tr}[P_{t} \Sigma^\dagger]  \mathrm{Tr}[\Sigma P_{Q\alpha}^\dagger \Sigma P_{t}^\dagger] +\mathrm{h.c.} $ 
& $y_{t_L}^2  y_{t_{R}}^2(A-B)^2 \sin^2(2\theta) +\dots$
\\
&  $\mathrm{Tr}[\XiA \Sigma^\dagger]   \mathrm{Tr}[\XiA \XiA^\dagger \Sigma \XiA^\dagger] +\mathrm{h.c.} $ 
& $y_{t_L}^4 \mathrm{Tr}[P_Q^\alpha \Sigma^\dagger]   \mathrm{Tr}[P_Q^\beta P_{Q\alpha}^\dagger \Sigma P_{Q\beta}^\dagger] +\mathrm{h.c.} $ 
& $ 3 y_{t_L}^4 \sin^2\theta +\dots$
\\
& 
& $y_{t_{R}}^4 \mathrm{Tr}[P_{t} \Sigma^\dagger]   \mathrm{Tr}[P_{t} P_{t}^\dagger \Sigma P_{t}^\dagger] +\mathrm{h.c.} $ 
& $2 y_{t_{R}}^4(A-B)^2(A^2+A B + B^2) \cos^2\theta +\dots$
\\
& 
& $y_{t_L}^2 y_{t_{R}}^2 \mathrm{Tr}[P_Q^\alpha \Sigma^\dagger]   \mathrm{Tr}[P_{t} P_{Q\alpha}^\dagger \Sigma P_{t}^\dagger] +\mathrm{h.c.} $
& $  y_{t_L}^2 y_{t_{R}}^2(A^2+B^2)\sin^2\theta +\dots$
\\
& 
& $y_{t_L}^2 y_{t_{R}}^2 \mathrm{Tr}[P_{t} \Sigma^\dagger]   \mathrm{Tr}[P_Q^\alpha P_{Q\alpha}^\dagger \Sigma P_{t}^\dagger] +\mathrm{h.c.} $ 
& $ 2 y_{t_L}^2 y_{t_{R}}^2(A-B)^2\cos^2\theta +\dots$
\\
&  $\mathrm{Tr}[\XiA \Sigma^\dagger \XiA \Sigma^\dagger]   \mathrm{Tr}[ \Sigma \XiA^\dagger \Sigma \XiA^\dagger] $ 
& $y_{t_L}^4 \mathrm{Tr}[P_Q^\alpha \Sigma^\dagger P_Q^\beta \Sigma^\dagger]   \mathrm{Tr}[ \Sigma P_{Q\alpha}^\dagger \Sigma P_{Q\beta}^\dagger] $ 
& $ y_{t_L}^4 \sin^4 \theta +\dots$
\\
&  
& $y_{t_{R}}^4 \mathrm{Tr}[P_{t} \Sigma^\dagger P_{t} \Sigma^\dagger]   \mathrm{Tr}[ \Sigma P_{t}^\dagger \Sigma P_{t}^\dagger] $ 
& $ \dfrac{1}{4} y_{t_{R}}^4 \left((A-B)^2 \cos(2 \theta)+(A+B)^2\right)^2 +\dots$
\\
&  
& $y_{t_L}^2 y_{t_{R}}^2 \mathrm{Tr}[P_Q^\alpha \Sigma^\dagger P_{t}\Sigma^\dagger]   \mathrm{Tr}[ \Sigma P_{Q\alpha}^\dagger \Sigma P_{t}^\dagger] $ 
& $  y_{t_L}^2 y_{t_{R}}^2(A-B)^2 \cos^2 \theta \sin^2 \theta  +\dots$
\\
\hline
$y_{t_{L,R}}^2 \chi$ 
& $\mathrm{Tr}[\XiA \Sigma^\dagger  \XiA \XiA^\dagger]+\mathrm{h.c.} $ 
& $y_{t_L}^2~ \mathrm{Tr}[P_Q^\alpha \Sigma^\dagger  \chi P_{Q\alpha}^\dagger]+\mathrm{h.c.} $ 
& $4 y_{t_L}^2 B_0 (m_1+m_2)\cos \theta+ \dots$
\\
& 
& $y_{t_{R}}^2~ \mathrm{Tr}[P_{t} \Sigma^\dagger  \chi P_{t}^\dagger]+\mathrm{h.c.} $ 
& $4 y_{t_{R}}^2B_0 (A^2m_2 +B^2m_1)\cos \theta+ \dots$
\\
& $\mathrm{Tr}[\XiA \Sigma^\dagger]^2 \mathrm{Tr}[ \Sigma \XiA^\dagger]+\mathrm{h.c.} $ 
& $y_{t_L}^2 \mathrm{Tr}[P_Q^\alpha \Sigma^\dagger] \mathrm{Tr}[ \Sigma P_{Q\alpha}^\dagger] \mathrm{Tr}[\chi \Sigma^\dagger]+\mathrm{h.c.}$ 
& $8 y_{t_L}^2 B_0 (m_1+m_2) \sin^2(2\theta)+ \dots$
\\
&
& $ y_{t_{R}}^2 \mathrm{Tr}[P_{t} \Sigma^\dagger] \mathrm{Tr}[ \Sigma P_{t}^\dagger] \mathrm{Tr}[\chi \Sigma^\dagger]+\mathrm{h.c.}$ 
& $16 y_{t_{R}}^2 B_0(A-B)^2 (m_1+m_2) \cos^3 \theta + \dots$
\\
& $\mathrm{Tr}[\XiA \Sigma^\dagger] \mathrm{Tr}[\Sigma \XiA^\dagger \Sigma \XiA^\dagger]+\mathrm{h.c.} $ 
& $y_{t_L}^2 ~ \mathrm{Tr}[P_Q^\alpha \Sigma^\dagger] \mathrm{Tr}[\Sigma P_{Q\alpha}^\dagger \Sigma \chi^\dagger]+\mathrm{h.c.} $ 
& $8 y_{t_L}^2 B_0 (m_1+m_2)\cos \theta \sin^2 \theta+ \dots$
\\
& 
& $y_{t_{R}}^2 ~ \mathrm{Tr}[P_{t} \Sigma^\dagger] \mathrm{Tr}[\Sigma P_{t}^\dagger \Sigma \chi^\dagger]+\mathrm{h.c.} $ 
& $4 B_0y_{t_R}^2(A-B) \cos\theta \left((A-B) (m_1 +m_2)\cos(2\theta)-(A+B)(m_1-m_2)\right)+ \dots$
\\
\hline
$y_{t_{L,R}}^2 g^2$ 
&  $\mathrm{Tr}[\XiA \Sigma^\dagger] \mathrm{Tr}[\Sigma \XiA^\dagger] \mathrm{Tr}[\XiAd \Sigma \XiAd^T \Sigma^\dagger] $ 
& $y_{t_L}^2 g^2~\mathrm{Tr}[P_Q^\alpha \Sigma^\dagger] \mathrm{Tr}[\Sigma P_{Q\alpha}^\dagger] \mathrm{Tr}[T^A_L\Sigma T_L^{AT} \Sigma^\dagger] $ 
& $-3 y_{t_L}^2 g^2 \cos^2 \theta \sin^2 \theta+ \dots$
\\
&  
& $y_{t_{R}}^2 g^2~\mathrm{Tr}[P_{t}\Sigma^\dagger] \mathrm{Tr}[\Sigma P_{t}^\dagger] \mathrm{Tr}[T^A_L\Sigma T_L^{AT} \Sigma^\dagger] $ 
& $-3 y_{t_{R}}^2(A-B)^2 g^2 \cos^4 \theta + \dots$
\\
& $\mathrm{Tr}[\XiA \Sigma^\dagger \XiAd] \mathrm{Tr}[\Sigma \XiA^\dagger\XiAd ]$ 
& $y_{t_L}^2 g^2~ \mathrm{Tr}[P_Q^\alpha \Sigma^\dagger T^A_L] \mathrm{Tr}[\Sigma P_{Q\alpha}^\dagger T^A_L ]$ 
& $\dfrac{3}{8} y_{t_L}^2 g^2 \sin^2 \theta+ \dots$
\\
& 
& $y_{t_{R}}^2 g^2~ \mathrm{Tr}[P_{t} \Sigma^\dagger T^A_L] \mathrm{Tr}[\Sigma P_{t}^\dagger T^A_L ]$ 
& $0$
\\
& $\mathrm{Tr}[\XiA \Sigma^\dagger]  \mathrm{Tr}[\Sigma \XiA^\dagger \XiAd^2]+\mathrm{h.c.}$ 
& $y_{t_L}^2 g^2~ \mathrm{Tr}[P_Q^\alpha \Sigma^\dagger]  \mathrm{Tr}[\Sigma P_{Q\alpha}^\dagger T^A_L T^A_L]+\mathrm{h.c.}$ 
& $\dfrac{3}{2} y_{t_L}^2 g^2 \sin^2 \theta+ \dots$
\\
& 
& $y_{t_{R}}^2 g^2~ \mathrm{Tr}[P_{t} \Sigma^\dagger]  \mathrm{Tr}[\Sigma P_{t}^\dagger T^A_L T^A_L]+\mathrm{h.c.}$ 
& $ 3 y_{t_{R}}^2g^2B(B-A)\cos^2\theta$+\dots
\\
& 
& $y_{t_{R}}^2 g^{\prime 2}~ \mathrm{Tr}[P_{t} \Sigma^\dagger]  \mathrm{Tr}[\Sigma P_{t}^\dagger T_Y T_Y]+\mathrm{h.c.}$ 
& $ y_{t_{R}}^2 g^{\prime 2}A(A-B) \cos^2 \theta+ \dots$
\\
& $\mathrm{Tr}[\XiA \Sigma^\dagger]  \mathrm{Tr}[\XiA^\dagger \XiAd \Sigma \XiAd^T]+\mathrm{h.c.}$ 
& $y_{t_L}^2 g^2~ \mathrm{Tr}[P_Q^\alpha \Sigma^\dagger]  \mathrm{Tr}[P_{Q\alpha}^\dagger T^A_L \Sigma T_L^{AT}]+\mathrm{h.c.}$ 
& $0$
\\
& 
& $y_{t_{R}}^2 g^2~ \mathrm{Tr}[P_{t}\Sigma^\dagger]  \mathrm{Tr}[P_{t}^\dagger T^A_L \Sigma T_L^{AT}]+\mathrm{h.c.}$
& $3y_{t_{R}}^2g^2B(A-B)\cos^2\theta+\dots$
\\
& 
& $y_{t_{R}}^2 g^{\prime 2}~ \mathrm{Tr}[P_{t}\Sigma^\dagger]  \mathrm{Tr}[P_{t}^\dagger T_Y \Sigma T_Y^{T}]+\mathrm{h.c.}$ 
& $ y_{t_{R}}^2 g^{\prime 2}A(B-A) \cos^2 \theta + \dots$
\\
& $\mathrm{Tr}[\XiA \Sigma^\dagger]  \mathrm{Tr}[\Sigma \XiA^\dagger \Sigma \XiAd^T \Sigma^\dagger \XiAd]+\mathrm{h.c.}$ 
& $y_{t_L}^2 g^2~\mathrm{Tr}[P_Q^\alpha \Sigma^\dagger]  \mathrm{Tr}[\Sigma P_{Q\alpha}^\dagger \Sigma T_L^{AT} \Sigma^\dagger T^A_L]+\mathrm{h.c.}$ 
& $-3 y_{t_L}^2 g^{ 2} \cos^2 \theta \sin^2 \theta + \dots$
\\
& 
&$y_{t_{R}}^2 g^2~\mathrm{Tr}[P_{t} \Sigma^\dagger]  \mathrm{Tr}[\Sigma P_{t}^\dagger \Sigma T_L^{AT} \Sigma^\dagger T^A_L]+\mathrm{h.c.}$ 
& $\dfrac{3}{2} g^2 y_{t_R}^2(B-A)\cos^2\theta\left((A-B)\cos(2\theta)-A-B\right)+ \dots$
\\
& 
&$y_{t_{R}}^2 g^{\prime 2}~\mathrm{Tr}[P_{t} \Sigma^\dagger]  \mathrm{Tr}[\Sigma P_{t}^\dagger \Sigma T_Y^{T} \Sigma^\dagger T_Y]+\mathrm{h.c.}$ 
& $\dfrac{1}{2}  g^{\prime 2}y_{t_R}^2(B-A) \cos^2 \theta\left((A-B)\cos(2\theta)+A+B\right)  + \dots$
\\
& $\mathrm{Tr}[\XiA \XiA^\dagger \XiAd \Sigma \XiAd^T \Sigma^\dagger]+\mathrm{h.c.}$ 
& $y_{t_L}^2 g^2~ \mathrm{Tr}[P_Q^\alpha P_{Q\alpha}^\dagger T^A_L \Sigma T_L^{AT}\Sigma^\dagger]+\mathrm{h.c.}$ 
& $-\dfrac{3}{2} y_{t_L}^2 g^{ 2}  \cos^2 \theta  + \dots$
\\
& 
& $y_{t_{R}}^2 g^2~ \mathrm{Tr}[P_{t} P_{t}^\dagger T^A_L \Sigma T_L^{AT}\Sigma^\dagger]+\mathrm{h.c.}$ 
& $-\frac{3}{2}y_{t_{R}}^2g^2B^2\cos^2\theta+\dots$
\\
& 
& $y_{t_{R}}^2 g^{\prime 2}~ \mathrm{Tr}[P_{t} P_{t}^\dagger T_Y \Sigma T_Y^{T}\Sigma^\dagger]+\mathrm{h.c.}$ 
& $-\dfrac{1}{2} y_{t_{R}}^2 g^{\prime 2}A^2 \cos^2 \theta +\dots$
\\
& $\mathrm{Tr}[\XiA \Sigma^\dagger  \XiAd \Sigma \XiA^\dagger  \XiAd]$ 
& $y_{t_L}^2 g^2~ \mathrm{Tr}[P_Q^\alpha \Sigma^\dagger  T^A_L \Sigma P_{Q\alpha}^\dagger  T^A_L]$ 
&  $0$
\\
& 
& $y_{t_{R}}^2 g^2~ \mathrm{Tr}[P_{t} \Sigma^\dagger  T^A_L \Sigma P_{t}^\dagger  T^A_L]$ 
& $\frac{3}{4}y_{t_{R}}g^2B^2\cos^2\theta+\dots$
\\
& 
& $y_{t_{R}}^2 g^{\prime 2}~ \mathrm{Tr}[P_{t} \Sigma^\dagger  T_Y \Sigma P_{t}^\dagger  T_Y]$ 
& $\dfrac{1}{4} y_{t_{R}}^2 g^{\prime 2}A^2 \cos^2 \theta+\dots$
\\
& $\mathrm{Tr}[\XiA \Sigma^\dagger  \XiAd^2 \Sigma \XiA^\dagger]$ 
& $y_{t_L}^2 g^2~ \mathrm{Tr}[P_Q^\alpha \Sigma^\dagger  T^A_L T^A_L \Sigma P_{Q\alpha}^\dagger]$ 
& $\dfrac{3}{4} y_{t_L}^2 g^2+\dots$
\\
&
& $y_{t_{R}}^2 g^2~ \mathrm{Tr}[P_{t} \Sigma^\dagger  T^A_L T^A_L \Sigma P_{t}^\dagger]$ 
& $\dfrac{3}{4}  g^2 y_{t_{R}}^2\left(A^2\sin^2 \theta +B^2\cos^2 \theta\right)+\dots$
\\
&
& $y_{t_{R}}^2 g^{\prime 2}~ \mathrm{Tr}[P_{t} \Sigma^\dagger  T_Y T_Y \Sigma P_{t}^\dagger]$ 
& $\dfrac{1}{4}  g^{\prime 2}y_{t_R}^2\left( A^2\cos^2 \theta+B^2\sin^2 \theta\right)+\dots$
\\
& $\mathrm{Tr}[\XiA \Sigma^\dagger \XiAd \Sigma \XiA^\dagger \Sigma \XiAd^T \Sigma^\dagger]$ 
& $y_{t_L}^2 g^2~ \mathrm{Tr}[P_Q^\alpha \Sigma^\dagger T^A_L \Sigma P_{Q\alpha}^\dagger \Sigma T_L^{AT} \Sigma^\dagger]$ 
& $-\dfrac{3}{4} y_{t_L}^2 g^2 \cos^2 \theta \sin^2 \theta+\dots$
\\
& 
& $y_{t_{R}}^2 g^2~ \mathrm{Tr}[P_{t} \Sigma^\dagger T^A_L \Sigma P_{t}^\dagger \Sigma T_L^{AT} \Sigma^\dagger]$ 
& $-\dfrac{3}{16}  g^2 y_{t_R}^2\left((B-A)\cos(2 \theta)+A+B\right)^2+\dots$
\\
& 
& $y_{t_{R}}^2 g^{\prime 2}~ \mathrm{Tr}[P_{t} \Sigma^\dagger T_Y \Sigma P_{t}^\dagger \Sigma T_Y^{T} \Sigma^\dagger]$ 
& $-\dfrac{1}{16}  g^{\prime 2}y_{t_R}^2\left((A-B)\cos(2 \theta)+A+B\right)^2+\dots$
\\
\hline\hline
\end{tabular}
}
\end{center}
\caption{\footnotesize Same as in Tab.~\ref{tab-bilinear-potential} but for  linear spurions in the antisymmetric representation, namely $\XiA^\alpha =y_{t_L} P_Q^\alpha$ and $\XiA=y_{t_R} P_t$.\\
}
\label{tab-Anti-potential}
\end{table}

\clearpage


\section{Spurionic operators generating the top-quark mass}
\label{Spurionic operators generating the top quark mass}

In this appendix we list the operators up to NLO that contribute to the top mass at tree level.
We consider a bilinear top coupling as well as linear couplings in the fundamental, adjoint, symmetric, or antisymmetric representations.
These operators also generate the top quark couplings to the pNGBs and in particular, the $\eta t \overline{t}$ coupling as discussed in Sec.~\ref{Masses and couplings} for the $\SU(4)/\Sp(4)$ case.
Let us remind the reader that our classification corresponds to a pseudo-real coset while  the real case can  easily be obtained in a similar way as explained in Sec.~\ref{Chiral perturbation theory}.
In order to isolate the contributions to the top mass, we use all spurions in Tab.~\ref{tab-spurions} except those involving SM gauge bosons.
Indeed, the latter  can only appear in covariant derivatives.

In addition to the points outlined in App.~\ref{Operators involving mass, gauge and top bilinear spurions}, it is worthwhile to notice that
\begin{itemize}
\item[(i)]  The top quark spurions containing elementary fermions generate corrections to their kinetic term.
For instance, the generic operator $\mathrm{Tr}[\XiA \Sigma^\dagger] \mathrm{Tr}[\Sigma \XiA^\dagger]$ cannot contribute to the top mass.
However,  in the case of a bilinear coupling it leads to the two following operators:
\begin{equation}
     \dfrac{y_t^2 f^2}{\Lambda_{HC}^2} \mathrm{Tr}[P^\alpha \Sigma^\dagger] \mathrm{Tr}[\Sigma P_{\beta}^\dagger]  (\overline{q}_{\mathrm{L} \alpha} \slashed{D} q_{\mathrm{L}}^\beta),
     \qquad
     \dfrac{y_t^2 f^2}{\Lambda_{HC}^2} \mathrm{Tr}[P^\alpha \Sigma^\dagger] \mathrm{Tr}[\Sigma P_{\alpha}^\dagger]  (\overline{t}_{R} \slashed{D} t_{\mathrm{R}})~.
     \label{correction-kinetic-term}
\end{equation}
We do not include these kinds of operators in our analysis.

\item[(ii)] In the same way, four-fermion operators are in general generated.
Using the same generic operator as before, we obtain in the bilinear case the following operator:
\begin{equation}
    \dfrac{y_t^2 f^2}{\Lambda_{HC}^4}\mathrm{Tr}[P^\alpha \Sigma^\dagger] \mathrm{Tr}[\Sigma P_{\beta}^\dagger] (\overline{q}_{\mathrm{L} \alpha} t_{\mathrm{R}})(\overline{t}_{\mathrm{R}} q_{\mathrm{L}}^\beta)~.
    \label{four-fermion-int}
\end{equation}
Again, we do not include these kinds of operators in our analysis.
\end{itemize}

The number of operators is again drastically reduced compared to those present in the generic classification of App.~\ref{General classification of the spurionic operators}.

\newpage

$\bullet$ \textbf{Bilinear coupling }

For a bilinear top spurion, we get four different classes of operators that contribute at tree level to the top mass:
\begin{itemize}
    \item[(i)] Only one top spurion $\XiA$  (class $y_t$).
    \item[(ii)] Three top spurions   $\XiA (\XiA \XiA^\dagger)$ (class $y_t^3$).
    \item[(iii)] One top spurion $\XiA$ and one mass spurion $\XiA$ or $\XiA^\dagger$ (class  $y_t \chi$).
    \item[(iv)] One top spurion $\XiA$ and two gauge spurions $\XiAd^2$ (classes  $y_t g^2$, $y_t g^{\prime 2}$).
\end{itemize}
The operators belonging to the above classes are displayed in Tab.~\ref{tab-bilinear-mass}.

\begin{table}[b]
\renewcommand{\arraystretch}{1.1}
\begin{center}
\scalebox{0.8}{
\begin{tabular}{ c c  c  }
\hline\hline
Class & General form & Operator 
\\
\hline
$y_t$ 
& $ \mathrm{Tr}[\XiA \Sigma^\dagger]+\mathrm{h.c.}$ 
& $y_t~\mathrm{Tr}[P^\alpha \Sigma^\dagger](Q_\alpha t^c)^\dagger+h.c$ 
\\
\hline
$y_t^3$ 
& $ \mathrm{Tr}[\XiA \Sigma^\dagger \XiA \XiA^\dagger]+\mathrm{h.c.}$ 
& $y_t^3~\mathrm{Tr}[P^\alpha \Sigma^\dagger P^\beta P_\beta^\dagger](Q_\alpha t^c)^\dagger+\mathrm{h.c.}$ 
\\
& $\mathrm{Tr}[\XiA \Sigma^\dagger]^2 \mathrm{Tr}[\Sigma \XiA^\dagger]+\mathrm{h.c.}$ 
& $y_t^3~ \mathrm{Tr}[P^\alpha \Sigma^\dagger] \mathrm{Tr}[P^\beta \Sigma^\dagger] \mathrm{Tr}[\Sigma P_{\beta}^\dagger](Q_\alpha t^c)^\dagger+\mathrm{h.c.}$ 
\\
& $ \mathrm{Tr}[\Sigma \XiA^\dagger] \mathrm{Tr}[\XiA \Sigma^\dagger\XiA \Sigma^\dagger ]+\mathrm{h.c.}$ 
&$y_t^3~ \mathrm{Tr}[\Sigma P_{\beta}^\dagger] 
		    \mathrm{Tr}[P^\beta \Sigma^\dagger P^\alpha \Sigma^\dagger] (Q_\alpha t^c)^\dagger+\mathrm{h.c.}$ 
\\
\hline\hline
$y_t ~\chi$ 
& $\mathrm{Tr}[\Xi_{A_2} \Sigma^\dagger] \mathrm{Tr}[\Sigma \Xi_{A_1}^\dagger]$ 
& $y_t~ \mathrm{Tr}[P^\alpha \Sigma^\dagger] \mathrm{Tr}[\Sigma \chi^\dagger] (Q_\alpha t^c)^\dagger+\mathrm{h.c.}$ 
\\
& $\mathrm{Tr}[\Xi_{A_1} \Sigma^\dagger]\mathrm{Tr}[\Xi_{A_2} \Sigma^\dagger]+\mathrm{h.c.}$  
& $y_t~ \mathrm{Tr}[\chi \Sigma^\dagger]\mathrm{Tr}[P^\alpha \Sigma^\dagger] (Q_\alpha t^c)^\dagger+\mathrm{h.c.}$ 
\\
& $\mathrm{Tr}[\Xi_{A_1} \Sigma^\dagger \Xi_{A_2} \Sigma^\dagger]+\mathrm{h.c.}$ 
& $y_t~ \mathrm{Tr}[\chi \Sigma^\dagger P^\alpha \Sigma^\dagger] (Q_\alpha t^c)^\dagger+\mathrm{h.c.}$
\\
\hline\hline 
$y_t ~g^2$, $y_t g^{\prime 2}$  
&  $\mathrm{Tr}[\XiA  \Sigma^\dagger \XiAd^2]+\mathrm{h.c.}$
& $y_t g^2~\mathrm{Tr}[P^\alpha  \Sigma^\dagger T^A_{\mathrm{L}} T^A_{\mathrm{L}}] (Q_\alpha t^c)^\dagger+\mathrm{h.c.}$
\\
& $\mathrm{Tr}[\XiA \XiAd^T \Sigma^\dagger \XiAd]+\mathrm{h.c.}$ 
& $y_t g^2~ \mathrm{Tr}[P^\alpha (T^A_{\mathrm{L}})^T \Sigma^\dagger T^A_{\mathrm{L}}] (Q_\alpha t^c)^\dagger+\mathrm{h.c.}$
\\ 
& $\mathrm{Tr}[\XiA \Sigma^\dagger] \mathrm{Tr}[\XiAd \Sigma \XiAd^T \Sigma^\dagger]+\mathrm{h.c.}$ 
& $y_t g^2~\mathrm{Tr}[P^\alpha \Sigma^\dagger] \mathrm{Tr}[T^A_{\mathrm{L}} \Sigma (T^A_{\mathrm{L}})^T \Sigma^\dagger](Q_\alpha t^c)^\dagger +\mathrm{h.c.}$ 
\\
& $\mathrm{Tr}[\XiA \Sigma^\dagger \XiAd \Sigma \XiAd^T \Sigma^\dagger]+\mathrm{h.c.}$ 
& $ y_t g^2~ \mathrm{Tr}[P^\alpha \Sigma^\dagger T^A_{\mathrm{L}} \Sigma (T^A_{\mathrm{L}})^T \Sigma^\dagger] (Q_\alpha t^c)^\dagger+\mathrm{h.c.} $ 
\\
\hline\hline
\end{tabular}
}
\end{center}
\caption{\footnotesize Non-derivative operators up to NLO involving the top bilinear spurions $\XiA^\dagger=y_t P^\alpha (Q_\alpha t^c)^\dagger$ and possibly $\XiA^{\alpha,\dagger}=y_t P^\alpha$  and contributing to the tree-level top mass.
Also shown are the mixed operators involving the top bilinear spurion and the gauge spurions or the mass spurion.
The $\mathrm{U}(1)_Y$ contributions are obtained by the following replacements  $g\rightarrow g^\prime$ and $T^A_{\mathrm{L}} \rightarrow T_Y$ in the third column. }
\label{tab-bilinear-mass}
\end{table}

\newpage

$\bullet$ \textbf{Linear coupling in the fundamental representation}

For a linear top coupling transforming in the fundamental representation, the operators contributing at tree level to the top mass organise as follows:
\begin{itemize}
    \item[(i)] Two top spurions $\XiF^2$  (class $y_{t_L} y_{t_R}$).
    \item[(ii)] Four top spurions   $\XiF^2 (\XiF \XiF^\dagger)$ (classes $y_{t_L}^3 y_{t_R}$, $y_{t_L} y_{t_R}^3$).
    \item[(iii)] Two top spurions $\XiF^2$ and one mass spurion $\XiA$ or $\XiA^\dagger$ (class  $y_{t_L} y_{t_R} \chi$).
    \item[(iv)] Two top spurions $\XiF^2$ and two gauge spurions $\XiAd^2$ (classes  $y_{t_L} y_{t_R} g^2$, $y_{t_L} y_{t_R} g^{\prime 2}$).
\end{itemize}
They are listed in Tab.~\ref{tab-fund-mass}.

\begin{table}[h!]
\renewcommand{\arraystretch}{1.1}
\begin{center}
\scalebox{0.8}{
\begin{tabular}{ c c  c }
\hline\hline
Class & General form & Operator  \\
\hline 
$y_{t_{L}} y_{t_R}$  
&  $\mathrm{Tr}[\XiF\cdot \XiF^T \Sigma^\dagger]+\mathrm{h.c.}$
& $y_{t_L} y_{t_R} ~\mathrm{Tr}[P_Q^\alpha \cdot P_t^T \Sigma^\dagger] (Q_\alpha t^c)^\dagger +\mathrm{h.c.}$
\\
\hline
 $y_{t_{L}}^3 y_{t_R},~ y_{t_{L}} y_{t_R}^3$ 
&  $\mathrm{Tr}[\XiF\cdot \XiF^\dagger \XiF\cdot \XiF^T \Sigma^\dagger]+\mathrm{h.c.}$ 
& $y_{t_L}^3 y_{t_R} ~\mathrm{Tr}[P_Q^\beta \cdot P_{Q\beta}^\dagger P_Q^\alpha \cdot P_t^T \Sigma^\dagger](Q_\alpha t^c)^\dagger +\mathrm{h.c.}$
\\
&
& $y_{t_L} y_{t_R}^3 ~\mathrm{Tr}[P_t \cdot P_t^\dagger P_Q^\alpha \cdot P_t^T \Sigma^\dagger](Q_\alpha t^c)^\dagger+\mathrm{h.c.}$
\\
\hline
$y_{t_{L}} y_{t_R} \chi$ 
&  $\mathrm{Tr}[\XiA \Sigma^\dagger  \XiF \cdot \XiF^T \Sigma^\dagger]+\mathrm{h.c.} $ 
 & $y_{t_L} y_{t_R} ~\mathrm{Tr}[\chi \Sigma^\dagger  P_Q^\alpha \cdot P_t^T \Sigma^\dagger](Q_\alpha t^c)^\dagger+\mathrm{h.c.} $  
\\
 & $\mathrm{Tr}[\XiA \Sigma^\dagger]  \mathrm{Tr}[\XiF \cdot \XiF^T \Sigma^\dagger]+\mathrm{h.c.} $ 
 & $y_{t_L} y_{t_R} ~\mathrm{Tr}[\chi \Sigma^\dagger]  \mathrm{Tr}[P_Q^\alpha \cdot P_t^T \Sigma^\dagger] (Q_\alpha t^c)^\dagger +\mathrm{h.c.} $ 
\\
 & $\mathrm{Tr}[ \Sigma \XiA^\dagger]  \mathrm{Tr}[\XiF \cdot \XiF^T \Sigma^\dagger]+\mathrm{h.c.} $ 
 & $y_{t_L} y_{t_R} ~\mathrm{Tr}[ \Sigma \chi^\dagger]  \mathrm{Tr}[P_Q^\alpha \cdot P_t^T \Sigma^\dagger] (Q_\alpha t^c)^\dagger +\mathrm{h.c.} $ 
\\
\hline
$y_{t_{L}} y_{t_R} g^2$, $y_{t_{L}} y_{t_R} g^{\prime 2}$  
&$\mathrm{Tr}[\XiAd^T \Sigma^\dagger \XiAd   \XiF \cdot \XiF^T]+\mathrm{h.c.} $ 
 & $y_{t_L} y_{t_R} g^2~ \mathrm{Tr}[(T^A_{\mathrm{L}})^T  \Sigma^\dagger T_L^A   P_Q^\alpha \cdot P_t^T](Q_\alpha t^c)^\dagger+\mathrm{h.c.} $ 
\\
 & $\mathrm{Tr}[\XiAd^2   \XiF \cdot \XiF^T \Sigma^\dagger]+\mathrm{h.c.} $ 
 & $y_{t_L} y_{t_R} g^2~ \mathrm{Tr}[T^A_L T^A_L  P_Q^\alpha \cdot P_t^T \Sigma^\dagger](Q_\alpha t^c)^\dagger+\mathrm{h.c.} $ 
\\
 & $\mathrm{Tr}[\XiF \cdot \XiF^T \Sigma^\dagger] \mathrm{Tr}[\XiAd \Sigma \XiAd^T \Sigma^\dagger]+\mathrm{h.c.} $ 
 & ~~~$ y_{t_L} y_{t_R} g^2~\mathrm{Tr}[P_Q^\alpha\cdot P_t^T \Sigma^\dagger] \mathrm{Tr}[T^A_L \Sigma (T^A_{\mathrm{L}})^T  \Sigma^\dagger] (Q_\alpha t^c)^\dagger +\mathrm{h.c.} $ 
\\
 & $\mathrm{Tr}[\XiAd \Sigma \XiAd^T \Sigma^\dagger   \XiF \cdot \XiF^T \Sigma^\dagger]+\mathrm{h.c.} $ 
 & $y_{t_L} y_{t_R} g^2~ \mathrm{Tr}[T^A_L \Sigma (T^A_{\mathrm{L}})^T  \Sigma^\dagger   P_Q^\alpha \cdot P_t^T \Sigma^\dagger](Q_\alpha t^c)^\dagger+\mathrm{h.c.} $
\\
\hline\hline
\end{tabular}
}
\end{center}
\caption{\footnotesize Same as in Tab.~\ref{tab-bilinear-mass} but for the  linear spurions in the fundamental representation, namely $\XiF= y_{t_L} P_Q^\alpha Q^\dagger_\alpha$ and $\XiF=y_{t_R} P_t t^{c \dagger}$ and possibly the spurions $\XiF^\alpha= y_{t_L} P_Q^\alpha$ and/or $ \XiF=y_{t_R} P_t$.}
\label{tab-fund-mass}
\end{table}

\newpage

$\bullet$ \textbf{Linear coupling in the adjoint representation}

For a linear top coupling transforming in the adjoint representation, the operators contributing at tree level to the top mass organise as follows:
\begin{itemize}
    \item[(i)] Two top spurions $\XiAd^2$  (class $y_{t_L} y_{t_R}$).
    \item[(ii)] Four top spurions   $\XiAd^4$ (classes $y_{t_L}^3 y_{t_R}$, $y_{t_L} y_{t_R}^3$).
    \item[(iii)] Two top spurions $\XiAd^2$ and one mass spurion $\XiA$ or $\XiA^\dagger$ (class  $y_{t_L} y_{t_R} \chi$).
    \item[(iv)] Two top spurions $\XiAd^2$ and two gauge spurions $\XiAd^2$ (classes  $y_{t_L} y_{t_R} g^2$, $y_{t_L} y_{t_R} g^{\prime 2}$).
\end{itemize}
They are listed in Tab.~\ref{tab-Adj-mass}.

\begin{table}[h!]
\renewcommand{\arraystretch}{1.1}
\begin{center}
\scalebox{0.8}{
\begin{tabular}{ c c  c }
\hline\hline
Class & General form & Operator  \\
\hline 
$y_{t_{L}} y_{t_R}$  ~~
&  $\mathrm{Tr}[\XiAd \Sigma \XiAd^T \Sigma^\dagger]$ 
& $y_{t_{L}} y_{t_R} ~\mathrm{Tr}[P_Q^\alpha \Sigma P_t^T \Sigma^\dagger](Q_\alpha t^c)^\dagger+\mathrm{h.c.}$ 
\\
\hline
$y_{t_{L}}^3 y_{t_R}, y_{t_{L}} y_{t_R}^3$  ~~
&  $\mathrm{Tr}[\XiAd \Sigma \XiAd^T \Sigma^\dagger]^2$ 
& $y_{t_{L}}^3 y_{t_R} ~\mathrm{Tr}[P_Q^\alpha \Sigma P_t^T \Sigma^\dagger]\mathrm{Tr}[P_Q^\beta \Sigma P_{Q\beta}^* \Sigma^\dagger](Q_\alpha t^c)^\dagger+\mathrm{h.c.}$ 
\\

& 
& $y_{t_{L}} y_{t_R}^3 ~\mathrm{Tr}[P_Q^\alpha \Sigma P_t^T \Sigma^\dagger]\mathrm{Tr}[P_t \Sigma P_t^* \Sigma^\dagger](Q_\alpha t^c)^\dagger+\mathrm{h.c.}$ 
\\
&  $\mathrm{Tr}[\XiAd^3 \Sigma \XiAd^T \Sigma^\dagger]$ 
&  $y_{t_{L}}^3 y_{t_R} ~\mathrm{Tr}[P_Q^\alpha P_t  P_Q^\beta \Sigma P_{Q\beta}^* \Sigma^\dagger](Q_\alpha t^c)^\dagger+\mathrm{h.c.}$ 
\\
&  
&  $y_{t_{L}} y_{t_R}^3 ~\mathrm{Tr}[P_Q^\alpha P_t  P_t \Sigma P_t^* \Sigma^\dagger](Q_\alpha t^c)^\dagger+\mathrm{h.c.}$ 
\\
&  $\mathrm{Tr}[\XiAd^2 \Sigma \XiAd^T \XiAd^T \Sigma^\dagger]$ 
&  $y_{t_{L}}^3 y_{t_R} ~\mathrm{Tr}[P_Q^\alpha P_t \Sigma P_Q^{\beta T}  P_{Q\beta}^* \Sigma^\dagger](Q_\alpha t^c)^\dagger+\mathrm{h.c.}$ 
\\
&  
&  $y_{t_{L}} y_{t_R}^3 ~\mathrm{Tr}[P_Q^\alpha P_t \Sigma P_t^{T}  P_{t}^* \Sigma^\dagger](Q_\alpha t^c)^\dagger+\mathrm{h.c.}$ 
\\
&  $\mathrm{Tr}[\XiAd \Sigma \XiAd^T \Sigma^\dagger \XiAd \Sigma \XiAd^T \Sigma^\dagger]$ 
& $y_{t_{L}}^3 y_{t_R} ~\mathrm{Tr}[P_Q^\alpha \Sigma P_t^T \Sigma^\dagger P_Q^{\beta} \Sigma  P_{Q\beta}^* \Sigma^\dagger](Q_\alpha t^c)^\dagger+\mathrm{h.c.}$ 
\\
&  
& $y_{t_{L}} y_{t_R}^3 ~\mathrm{Tr}[P_Q^\alpha \Sigma P_t^T \Sigma^\dagger P_t \Sigma  P_t^* \Sigma^\dagger](Q_\alpha t^c)^\dagger+\mathrm{h.c.}$ 
\\
\hline
$y_{t_{L}} y_{t_R} \chi$ 
& $\mathrm{Tr}[\XiA \Sigma^\dagger  \XiAd^2]+\mathrm{h.c.} $ 
& $y_{t_{L}} y_{t_R} ~\mathrm{Tr}[\chi \Sigma^\dagger P_Q^\alpha P_t](Q_\alpha t^c)^\dagger +\mathrm{h.c.} $ 
\\
& $\mathrm{Tr}[\XiA \XiAd^T \Sigma^\dagger  \XiAd]+\mathrm{h.c.} $ 
& $y_{t_{L}} y_{t_R}  ~\mathrm{Tr}[\chi P_Q^{\alpha T} \Sigma^\dagger  P_t](Q_\alpha t^c)^\dagger+\mathrm{h.c.} $ 
\\
& $\mathrm{Tr}[\XiA \Sigma^\dagger]  \mathrm{Tr}[\XiAd \Sigma  \XiAd^T \Sigma^\dagger]+\mathrm{h.c.} $
 & $y_{t_{L}} y_{t_R} ~\mathrm{Tr}[\chi \Sigma^\dagger]  \mathrm{Tr}[P_Q^\alpha \Sigma  P_t^T \Sigma^\dagger](Q_\alpha t^c)^\dagger+\mathrm{h.c.} $
\\
& $\mathrm{Tr}[\XiA \Sigma^\dagger \XiAd \Sigma  \XiAd^T \Sigma^\dagger]+\mathrm{h.c.} $ & $y_{t_{L}} y_{t_R} ~\mathrm{Tr}[\chi \Sigma^\dagger P_Q^\alpha \Sigma  P_t^T \Sigma^\dagger](Q_\alpha t^c)^\dagger +\mathrm{h.c.} $
\\
\hline
$y_{t_{L}} y_{t_R} g^2$ 
& $\mathrm{Tr}[\XiAd \Sigma \XiAd^T \Sigma^\dagger]^2$ 
& $y_{t_{L}} y_{t_R} g^2 ~\mathrm{Tr}[P_Q^\alpha \Sigma P_t^T \Sigma^\dagger]\mathrm{Tr}[T_L^A \Sigma (T^A_L)^T\Sigma^\dagger](Q_\alpha t^c)^\dagger+\mathrm{h.c.}$ 
\\
&  $\mathrm{Tr}[\XiAd^3 \Sigma \XiAd^T \Sigma^\dagger]$ 
& $y_{t_{L}} y_{t_R} g^2 ~\mathrm{Tr}[P^\alpha_Q P_t T^A_L \Sigma (T^A_L)^T \Sigma^\dagger](Q_\alpha t^c)^\dagger+\mathrm{h.c.}$ 
\\
&  $\mathrm{Tr}[\XiAd^2 \Sigma \XiAd^T \XiAd^T \Sigma^\dagger]$ 
& $y_{t_{L}} y_{t_R} g^2 ~\mathrm{Tr}[P_Q^\alpha P_t \Sigma (T^A_L)^T (T^A_L)^T \Sigma^\dagger](Q_\alpha t^c)^\dagger+\mathrm{h.c.}$ 
\\
&  $\mathrm{Tr}[\XiAd \Sigma \XiAd^T \Sigma^\dagger \XiAd \Sigma \XiAd^T \Sigma^\dagger]$ 
& $y_{t_{L}} y_{t_R} g^2 ~\mathrm{Tr}[P_Q^\alpha \Sigma P_t^T \Sigma^\dagger T^A_L \Sigma (T^A_L)^T \Sigma^\dagger](Q_\alpha t^c)^\dagger+\mathrm{h.c.}$ 
\\
\hline\hline
	\end{tabular}
	}
    \end{center}
    \caption{\footnotesize Same as in Tab.~\ref{tab-bilinear-mass} but for the  linear spurions in the adjoint representation, namely ${\XiAd= y_{t_L} P_Q^\alpha Q^\dagger_\alpha}$ and ${\XiAd=y_{t_R} P_t t^{c \dagger}}$ and possibly the spurions $\XiAd^\alpha= y_{t_L} P_Q^\alpha$ and/or $ \XiAd=y_{t_R} P_t$.}
    \label{tab-Adj-mass}
\end{table}

\newpage

$\bullet$ \textbf{Linear coupling in the symmetric representation}

For a linear top coupling transforming in the symmetric representation, the operators contributing at tree level to the top mass organise as follows:
\begin{itemize}
    \item[(i)] Two top spurions $\XiS^2$  (class $y_{t_L} y_{t_R}$).
    \item[(ii)] Four top spurions   $\XiS^2(\XiS \XiS^\dagger)$ (classes $y_{t_L}^3 y_{t_R}$, $y_{t_L} y_{t_R}^3$).
    \item[(iii)] Two top spurions $\XiS^2$ and one mass spurion $\XiA$ or $\XiA^\dagger$ (class  $y_{t_L} y_{t_R} \chi$).
    \item[(iv)] Two top spurions $\XiS^2$ and two gauge spurions $\XiAd^2$ (classes  $y_{t_L} y_{t_R} g^2$, $y_{t_L} y_{t_R} g^{\prime 2}$).
\end{itemize}
They are listed in Tab.~\ref{tab-Sym-mass}.

\begin{table}[h!]
\renewcommand{\arraystretch}{1.1}
\begin{center}
\scalebox{0.8}{
\begin{tabular}{ c c  c  }
\hline\hline
Class & General form & Operator  
\\
\hline 
$y_{t_{L}} y_{t_R}$  ~~
&  ~~$\mathrm{Tr}[\XiS \Sigma^\dagger \XiS\Sigma^\dagger] +\mathrm{h.c.} $ 
& ~~$y_{t_L} y_{t_R}~ \mathrm{Tr}[P_Q^\alpha \Sigma^\dagger P_t \Sigma^\dagger](Q_\alpha t^c)^\dagger +\mathrm{h.c.} $ 
\\
\hline
$y_{t_{L}}^3 y_{t_R}$  
&  $\mathrm{Tr}[\XiS  \Sigma^\dagger \XiS \Sigma^\dagger \XiS \XiS^\dagger] +\mathrm{h.c.} $ 
& $y_{t_{L}}^3 y_{t_R} ~\mathrm{Tr}[P_Q^\alpha  \Sigma^\dagger P_t \Sigma^\dagger P_Q^\beta P_{Q\beta}^\dagger](Q_\alpha t^c)^\dagger +\mathrm{h.c.} $ 
\\
$y_{t_{L}} y_{t_R}^3$
& 
& $y_{t_{L}} y_{t_R}^3 ~\mathrm{Tr}[P_t  \Sigma^\dagger P_t \Sigma^\dagger P_Q^\alpha P_t^\dagger](Q_\alpha t^c)^\dagger +\mathrm{h.c.} $ 
\\
\hline
$y_{t_{L}} y_{t_R} \chi$ 
& $\mathrm{Tr}[\XiS \Sigma^\dagger  \XiS \XiA^\dagger]+\mathrm{h.c.} $ 
& $y_{t_L} y_{t_R} ~\mathrm{Tr}[P_Q^\alpha \Sigma^\dagger  P_t \chi^\dagger] (Q_\alpha t^c)^\dagger+\mathrm{h.c.} $
\\
& $\mathrm{Tr}[\XiA \Sigma^\dagger] \mathrm{Tr}[ \XiS \Sigma^\dagger \XiS \Sigma^\dagger]+\mathrm{h.c.} $ 
& $y_{t_L} y_{t_R} ~\mathrm{Tr}[\chi \Sigma^\dagger] \mathrm{Tr}[ P^\alpha_Q \Sigma^\dagger P_t\Sigma^\dagger](Q_\alpha t^c)^\dagger+\mathrm{h.c.} $ 
\\
& $\mathrm{Tr}[\Sigma \XiA^\dagger] \mathrm{Tr}[ \XiS \Sigma^\dagger \XiS \Sigma^\dagger]+\mathrm{h.c.} $ 
& $y_{t_L} y_{t_R} ~\mathrm{Tr}[\Sigma \chi^\dagger] \mathrm{Tr}[ P^\alpha_Q \Sigma^\dagger P_t\Sigma^\dagger](Q_\alpha t^c)^\dagger+\mathrm{h.c.}$ 
\\
& $\mathrm{Tr}[\XiS \Sigma^\dagger \XiS \Sigma^\dagger \XiA \Sigma^\dagger]+\mathrm{h.c.} $ 
&$y_{t_L} y_{t_R} ~\mathrm{Tr}[P_Q^\alpha \Sigma^\dagger P_t \Sigma^\dagger \chi \Sigma^\dagger](Q_\alpha t^c)^\dagger+\mathrm{h.c.}$ 
\\
\hline
$y_{t_{L}} y_{t_R} g^2,~ y_{t_{L}} y_{t_R} g^{\prime 2}$ 
 & $\mathrm{Tr}[\XiS \Sigma^\dagger \XiAd ]^2 +\mathrm{h.c.}$ 
& $y_{t_L} y_{t_R} g^2 ~\mathrm{Tr}[P_Q^\alpha \Sigma^\dagger T^A_L ] \mathrm{Tr}[P_t\Sigma^\dagger T^A_L ] (Q_\alpha t^c)^\dagger +\mathrm{h.c.}$ 
\\
& $\mathrm{Tr}[\XiS \Sigma^\dagger \XiS \Sigma^\dagger] \mathrm{Tr}[\XiAd\Sigma \XiAd^T\Sigma^\dagger ] +\mathrm{h.c.}$ 
& ~~~~$y_{t_L} y_{t_R} g^2 ~\mathrm{Tr}[P_Q^\alpha \Sigma^\dagger P_t \Sigma^\dagger] \mathrm{Tr}[T^A_L\Sigma T_L^{A T}\Sigma^\dagger ](Q_\alpha t^c)^\dagger +\mathrm{h.c.}$ 
\\
& $\mathrm{Tr}[\XiS \Sigma^\dagger \XiAd \XiS \Sigma^\dagger \XiAd ] +\mathrm{h.c.}$ 
& $y_{t_L} y_{t_R} g^2 ~ \mathrm{Tr}[P_Q^\alpha \Sigma^\dagger T^A_L P_t \Sigma^\dagger T_L^A ](Q_\alpha t^c)^\dagger +\mathrm{h.c.}$ 
\\
& $\mathrm{Tr}[\XiS \Sigma^\dagger \XiAd \XiS  \XiAd^T \Sigma^\dagger ] +\mathrm{h.c.}$ 
& $y_{t_L} y_{t_R} g^2 ~ \mathrm{Tr}[P_Q^\alpha \Sigma^\dagger T^A_L P_t  T^{AT}_L \Sigma^\dagger ](Q_\alpha t^c)^\dagger +\mathrm{h.c.}$ 
\\
 & $\mathrm{Tr}[\XiS \Sigma^\dagger  \XiS  \XiAd^T \Sigma^\dagger \XiAd] +\mathrm{h.c.}$ 
& $ y_{t_L} y_{t_R} g^2~ \mathrm{Tr}[P_Q^\alpha \Sigma^\dagger  P_t  T_L^{A T} \Sigma^\dagger T_L^A] (Q_\alpha t^c)^\dagger +\mathrm{h.c.}$ 
\\
 & $\mathrm{Tr}[\XiS \Sigma^\dagger  \XiS   \Sigma^\dagger \XiAd^2] +\mathrm{h.c.}$ 
&  $y_{t_L} y_{t_R} g^2~ \mathrm{Tr}[P_Q^\alpha \Sigma^\dagger  P_t   \Sigma^\dagger T^A_L T^A_L] (Q_\alpha t^c)^\dagger +\mathrm{h.c.}$
\\
 & $\mathrm{Tr}[\XiS \Sigma^\dagger  \XiS   \Sigma^\dagger \XiAd \Sigma \XiAd^T \Sigma^\dagger] +\mathrm{h.c.}$ 
& $y_{t_L} y_{t_R} g^2 ~\mathrm{Tr}[P_Q^\alpha \Sigma^\dagger  P_t   \Sigma^\dagger T_L^A \Sigma T_L^{A T} \Sigma^\dagger](Q_\alpha t^c)^\dagger+\mathrm{h.c.}$ 
\\
\hline\hline
\end{tabular}
}
\end{center}
\caption{\footnotesize Same as in Tab.~\ref{tab-bilinear-mass} but for the  linear spurions in the symmetric representation, namely $\XiS= y_{t_L} P_Q^\alpha Q^\dagger_\alpha$ and $\XiS=y_{t_R} P_t t^{c \dagger}$ and possibly the spurions $\XiS^\alpha= y_{t_L} P_Q^\alpha$ and/or $ \XiS=y_{t_R} P_t$.}
\label{tab-Sym-mass}
\end{table}

\newpage

$\bullet$ \textbf{Linear coupling in the antisymmetric representation}

Finally, for a linear top coupling transforming in the antisymmetric representation, the operators contributing at tree level to the top mass organise as follows:
\begin{itemize}
    \item[(i)] Two top spurions $\XiA^2$  (class $y_{t_L} y_{t_R}$).
    \item[(ii)] Four top spurions   $\XiA^2(\XiA \XiA^\dagger)$ (classes $y_{t_L}^3 y_{t_R}$, $y_{t_L} y_{t_R}^3$).
    \item[(iii)] Two top spurions $\XiA^2$ and one mass spurion $\XiA$ or $\XiA^\dagger$ (class  $y_{t_L} y_{t_R} \chi$).
    \item[(iv)] Two top spurions $\XiA^2$ and two gauge spurions $\XiAd^2$ (classes  $y_{t_L} y_{t_R} g^2$, $y_{t_L} y_{t_R} g^{\prime 2}$).
\end{itemize}
They are listed in Tab.~\ref{tab-Anti-mass}.

\begin{table}[h!]
\renewcommand{\arraystretch}{1.1}
\begin{center}
\scalebox{0.8}{
\begin{tabular}{ c c  c }
\hline\hline
Class & General form & Operator 
\\
\hline 
$y_{t_{L}} y_{t_R}$  
&  $\mathrm{Tr}[\XiA \Sigma^\dagger]^2 +\mathrm{h.c.} $ 
&  $y_{t_{L}} y_{t_R} ~\mathrm{Tr}[P_Q^\alpha \Sigma^\dagger] \mathrm{Tr}[P_t\Sigma^\dagger] (Q_\alpha t^c)^\dagger+\mathrm{h.c.} $
\\
&  $\mathrm{Tr}[\XiA \Sigma^\dagger \XiA \Sigma^\dagger] +\mathrm{h.c.} $ 
& $y_{t_{L}} y_{t_R}~ \mathrm{Tr}[P_Q^\alpha \Sigma^\dagger P_t \Sigma^\dagger](Q_\alpha t^c)^\dagger +\mathrm{h.c.} $ 
\\
\hline
$y_{t_{L}}^3 y_{t_R},~ y_{t_{L}} y_{t_R}^3$  ~~
&  $\mathrm{Tr}[\XiA \Sigma^\dagger]^3 \mathrm{Tr}[\Sigma \XiA ^\dagger] +\mathrm{h.c.} $ 
& $y_{t_{L}}^3 y_{t_R} ~ \mathrm{Tr}[P_Q^\alpha \Sigma^\dagger]\mathrm{Tr}[P_t \Sigma^\dagger] \mathrm{Tr}[P_Q^\beta \Sigma^\dagger] \mathrm{Tr}[\Sigma P_{Q\beta}^\dagger](Q_\alpha t^c)^\dagger +\mathrm{h.c.} $
\\
&  
& $y_{t_{L}} y_{t_R}^3 \mathrm{Tr}[P_Q^\alpha \Sigma^\dagger]\mathrm{Tr}[P_t \Sigma^\dagger] \mathrm{Tr}[P_t \Sigma^\dagger] \mathrm{Tr}[\Sigma P_t^\dagger](Q_\alpha t^c)^\dagger +\mathrm{h.c.} $
\\
&  $\mathrm{Tr}[\XiA \Sigma^\dagger] \mathrm{Tr}[\Sigma \XiA^\dagger]  \mathrm{Tr}[\XiA \Sigma^\dagger\XiA \Sigma^\dagger] +\mathrm{h.c.} $ 
& $y_{t_{L}}^3 y_{t_R} ~  \mathrm{Tr}[P_Q^\alpha \Sigma^\dagger] \mathrm{Tr}[\Sigma P_{Q\beta}^\dagger]  \mathrm{Tr}[P_Q^\beta \Sigma^\dagger P_t \Sigma^\dagger](Q_\alpha t^c)^\dagger  +\mathrm{h.c.} $ 
\\
&  
& $y_{t_{L}} y_{t_R}^3 ~  \mathrm{Tr}[P_Q^\alpha \Sigma^\dagger] \mathrm{Tr}[\Sigma P_t^\dagger]  \mathrm{Tr}[P_t \Sigma^\dagger P_t \Sigma^\dagger](Q_\alpha t^c)^\dagger  +\mathrm{h.c.} $ 
\\
&  $\mathrm{Tr}[\XiA \Sigma^\dagger]   \mathrm{Tr}[\XiA \Sigma^\dagger\XiA \XiA^\dagger] +\mathrm{h.c.} $ 
& $y_{t_L}^3 y_{t_R} \mathrm{Tr}[P_Q^\alpha \Sigma^\dagger]   \mathrm{Tr}[P_Q^\beta \Sigma^\dagger P_t P_{Q\beta}^\dagger](Q_\alpha t^c)^\dagger +\mathrm{h.c.} $ 
\\
&  
& $y_{t_L} y_{t_R}^3 \mathrm{Tr}[P_Q^\alpha \Sigma^\dagger]   \mathrm{Tr}[P_t \Sigma^\dagger P_t P_t^\dagger](Q_\alpha t^c)^\dagger +\mathrm{h.c.} $ 
\\
\hline
$y_{t_{L}} y_{t_R} \chi$ 
& $\mathrm{Tr}[\XiA \Sigma^\dagger  \XiA \XiA^\dagger]+\mathrm{h.c.} $ 
& $y_{t_L} y_{t_R}~ \mathrm{Tr}[P_Q^\alpha \Sigma^\dagger  P_t \chi^\dagger](Q_\alpha t^c)^\dagger+\mathrm{h.c.} $ 
\\
& $\mathrm{Tr}[\XiA \Sigma^\dagger]^3+\mathrm{h.c.} $ 
&$y_{t_L} y_{t_R}~ \mathrm{Tr}[P_Q^\alpha \Sigma^\dagger]\mathrm{Tr}[P_t \Sigma^\dagger] \mathrm{Tr}[\chi \Sigma^\dagger](Q_\alpha t^c)^\dagger+\mathrm{h.c.} $ 
\\
& $\mathrm{Tr}[\XiA \Sigma^\dagger]^2 \mathrm{Tr}[ \Sigma \XiA^\dagger]+\mathrm{h.c.} $ 
& $y_{t_L} y_{t_R}~ \mathrm{Tr}[P_Q^\alpha \Sigma^\dagger]  \mathrm{Tr}[P_t \Sigma^\dagger]\mathrm{Tr}[ \Sigma \chi^\dagger](Q_\alpha t^c)^\dagger+\mathrm{h.c.}$ 
\\
& $\mathrm{Tr}[\XiA \Sigma^\dagger] \mathrm{Tr}[\XiA \Sigma^\dagger\XiA \Sigma^\dagger]+\mathrm{h.c.} $ 
& $y_{t_L} y_{t_R}~ \mathrm{Tr}[\chi \Sigma^\dagger] \mathrm{Tr}[P_Q^\alpha \Sigma^\dagger P_t \Sigma^\dagger](Q_\alpha t^c)^\dagger+\mathrm{h.c.} $ 
\\
& $\mathrm{Tr}[\Sigma \XiA ^\dagger] \mathrm{Tr}[\XiA \Sigma^\dagger \XiA \Sigma ^\dagger]+\mathrm{h.c.} $ 
& $y_{t_L} y_{t_R}~ \mathrm{Tr}[\Sigma \chi^\dagger] \mathrm{Tr}[P_Q^\alpha \Sigma^\dagger P_t \Sigma ^\dagger](Q_\alpha t^c)^\dagger +\mathrm{h.c.}$
\\
& $\mathrm{Tr}[\XiA \Sigma^\dagger \XiA \Sigma^\dagger\XiA \Sigma^\dagger]+\mathrm{h.c.} $ 
& $y_{t_L} y_{t_R}~ \mathrm{Tr}[P_Q^\alpha \Sigma^\dagger P_t \Sigma^\dagger\chi \Sigma^\dagger](Q_\alpha t^c)^\dagger+\mathrm{h.c.} $ 
\\
\hline
$y_{t_{L}} y_{t_R} g^2,~ y_{t_{L}} y_{t_R} g^{\prime 2} $ 
& $\mathrm{Tr}[\XiA \Sigma^\dagger]^2 \mathrm{Tr}[\XiAd \Sigma \Xi_{Ad}^T \Sigma^\dagger] +\mathrm{h.c.}$ 
& $y_{t_L} y_{t_R} g^2 \mathrm{Tr}[P_Q^\alpha \Sigma^\dagger] \mathrm{Tr}[P_t\Sigma^\dagger] \mathrm{Tr}[T^A_L \Sigma T^{AT}_L \Sigma^\dagger](Q_\alpha t^c)^\dagger +\mathrm{h.c.}$ 
\\
& $\mathrm{Tr}[\XiA \Sigma^\dagger \XiAd]^2+\mathrm{h.c.}$ 
& $y_{t_{L}} y_{t_R} g^2 ~ \mathrm{Tr}[P_Q^\alpha \Sigma^\dagger T^A_L] \mathrm{Tr}[P_t \Sigma^\dagger T^A_L](Q_\alpha t^c)^\dagger+\mathrm{h.c.}$ 
\\
& $\mathrm{Tr}[\XiA \Sigma^\dagger]  \mathrm{Tr}[\XiA \Sigma^\dagger \XiAd^2]+\mathrm{h.c.}$ 
& $y_{t_{L}} y_{t_R} g^2 ~ \mathrm{Tr}[P_Q^\alpha \Sigma^\dagger]  \mathrm{Tr}[P_t \Sigma^\dagger T^A_L T^A_L](Q_\alpha t^c)^\dagger +\mathrm{h.c.}$ 
\\
& $\mathrm{Tr}[\XiA \Sigma^\dagger]  \mathrm{Tr}[\XiA \XiAd ^T \Sigma^\dagger \XiAd]+\mathrm{h.c.}$ 
& $y_{t_{L}} y_{t_R} g^2 ~ \mathrm{Tr}[P_Q^\alpha \Sigma^\dagger]  \mathrm{Tr}[P_t T_L^{AT} \Sigma^\dagger T^A_L](Q_\alpha t^c)^\dagger +\mathrm{h.c.}$ 
\\
& $\mathrm{Tr}[\XiA \Sigma^\dagger\XiA \Sigma^\dagger]  \mathrm{Tr}[\XiAd \Sigma \XiAd^T \Sigma^\dagger]+\mathrm{h.c.}$ 
& $y_{t_{L}} y_{t_R} g^2 ~  \mathrm{Tr}[P_Q^\alpha \Sigma^\dagger P_t \Sigma^\dagger]  \mathrm{Tr}[T^A_L \Sigma T_L^{AT} \Sigma^\dagger](Q_\alpha t^c)^\dagger +\mathrm{h.c.}$ 
\\
& $\mathrm{Tr}[\XiA \Sigma^\dagger]  \mathrm{Tr}[\XiA \Sigma^\dagger \XiAd \Sigma \XiAd^T \Sigma^\dagger]+\mathrm{h.c.}$ 
& $y_{t_{L}} y_{t_R} g^2 ~ \mathrm{Tr}[P_Q^\alpha \Sigma^\dagger]  \mathrm{Tr}[P_t\Sigma^\dagger T^A_L \Sigma T_L^{AT} \Sigma^\dagger](Q_\alpha t^c)^\dagger+\mathrm{h.c.}$ 
\\
& $\mathrm{Tr}[\XiA \Sigma^\dagger  \XiAd \XiA \Sigma^\dagger \XiAd]+\mathrm{h.c.}$ 
& $y_{t_{L}} y_{t_R} g^2 ~ \mathrm{Tr}[P_Q^\alpha \Sigma^\dagger  T^A_L P_t \Sigma^\dagger T^A_L](Q_\alpha t^c)^\dagger+\mathrm{h.c.}$ 
\\
& $\mathrm{Tr}[\XiA \Sigma^\dagger  \XiAd \XiA  \XiAd^T \Sigma^\dagger]+\mathrm{h.c.}$ 
& $y_{t_{L}} y_{t_R} g^2 ~  \mathrm{Tr}[P_Q^\alpha \Sigma^\dagger  T^A_L P_t  T_L^{AT} \Sigma^\dagger](Q_\alpha t^c)^\dagger+\mathrm{h.c.}$ 
\\
& $\mathrm{Tr}[\XiA \Sigma^\dagger \XiA \XiAd^T \Sigma^\dagger \XiAd]+\mathrm{h.c.}$ 
& $y_{t_{L}} y_{t_R} g^2 ~  \mathrm{Tr}[P_Q^\alpha \Sigma^\dagger P_t T_L^{AT} \Sigma^\dagger T^A_L](Q_\alpha t^c)^\dagger+\mathrm{h.c.}$ 
\\
& $\mathrm{Tr}[\XiA \Sigma^\dagger \XiA \Sigma^\dagger \XiAd \Sigma  \XiAd^T \Sigma^\dagger]+\mathrm{h.c.}$ 
& $y_{t_{L}} y_{t_R} g^2 ~ \mathrm{Tr}[P_Q^\alpha \Sigma^\dagger P_t \Sigma^\dagger T^A_L \Sigma  T_L^{AT} \Sigma^\dagger](Q_\alpha t^c)^\dagger+\mathrm{h.c.}$ 
\\
\hline\hline
\end{tabular}
}
\end{center}
\caption{\footnotesize Same as in Tab.~\ref{tab-bilinear-mass} but for the  linear spurions in the antisymmetric representation, namely $\XiA= y_{t_L} P_Q^\alpha Q^\dagger_\alpha$ and $\XiA=y_{t_R} P_t t^{c \dagger}$ and possibly the spurions $\XiA^\alpha= y_{t_L} P_Q^\alpha$ and/or $ \XiA=y_{t_R} P_t$.}
\label{tab-Anti-mass}
\end{table}

\clearpage


\section{Generic classification of spurionic operators}
\label{General classification of the spurionic operators}

The purpose of this appendix is to provide  details about the general classification discussed in 
Sec.~\ref{Introduction of spurions}. 
We derive a complete set of non-derivative operators involving up to four spurions in a two-index representation ($\XiSA$ and $\XiAd$) of the flavour symmetry. An the end of the appendix, we outline how the discussion can be extended to derivative operators.

This general set of operators can then be used, once the explicit breaking sources are specified, 
to construct all the operators up to NLO that explicitly break $\GF$.
Since only the transformation properties under the global symmetry, $\GF$, are fixed, while the chiral counting as well as the properties dictated by the UV theory are not yet imposed, 
the classification below is completely general and can be applied to a wide range of theories where spurions transform in two-index representations.

A concrete application to composite-Higgs models is presented in 
Sec.~\ref{Explicit breaking sources in composite Higgs models} and the details are reported in 
App.~\ref{Operators involving mass, gauge and top bilinear spurions}.
The restriction to four spurions in the same operator is justified by the chiral counting associated with the composite-Higgs spurions.
Indeed, in this specific example, all spurions appear at least  at order ${\cal O}(p)$
\footnote{Except for the partial-compositeness spurions with no elementary fields where $y_{t_{\mathrm{L}}} P_Q^\alpha$ and $y_{t_{\mathrm{R}}} P_t$ appear at ${\cal O}(\sqrt{p})$.
However, as discussed in Sec.~\ref{Top couplings} , one can still restrict to four spurions.}.

\begin{table}[h!]
    \scriptsize
    \renewcommand{\arraystretch}{1.}
    \begin{center}
	\begin{tabular}{ c  c c c }
	    \hline\hline
	     No $\Sigma$ & Linear in $\Sigma$ & Quadratic in $\Sigma$ & Three $\Sigma$ \\
	    \hline 
	    $\mathrm{Tr}[\XiSA \XiSA^\dagger \XiAd] $   & $\mathrm{Tr}[\XiA \Sigma^\dagger] \mathrm{Tr}[ \XiSA  \XiSA^\dagger]+ \mathrm{h.c.}$ 
		& $\mathrm{Tr}[\XiA \Sigma^\dagger] \mathrm{Tr}[\XiSA \Sigma^\dagger \XiAd]+ \mathrm{h.c.}$ 
		& $\mathrm{Tr}[\XiA \Sigma^\dagger]^3 +\mathrm{h.c.}$
	    \\
	    $\mathrm{Tr}[\XiS \XiA^\dagger \XiAd]+\mathrm{h.c.} $  & $\mathrm{Tr}[\XiA \Sigma^\dagger] \mathrm{Tr}[ \XiAd^2]+ \mathrm{h.c.}$ 
		& $\mathrm{Tr}[\XiA \Sigma^\dagger] \mathrm{Tr}[\Sigma \XiSA^\dagger \XiAd ]+ \mathrm{h.c.}$ 
		& $\mathrm{Tr}[\XiA \Sigma^\dagger]^2 \mathrm{Tr}[\Sigma \XiA^\dagger]+\mathrm{h.c.}$
	     \\
	    $\mathrm{Tr}[\XiAd^3]$  & $\mathrm{Tr}[\XiS \Sigma^\dagger \XiSA \XiAS^\dagger ]+\mathrm{h.c.} $ 
		& $\mathrm{Tr}[\XiSA \Sigma^\dagger \XiSA \Sigma^\dagger \XiAd ]+\mathrm{h.c.} $
		& $\mathrm{Tr}[\XiA \Sigma^\dagger] \mathrm{Tr}[\XiSA\Sigma^\dagger \XiSA\Sigma^\dagger]+ \mathrm{h.c.}$
	    \\
	    & $\mathrm{Tr}[\XiA \Sigma^\dagger \XiSA \XiSA^\dagger ]+\mathrm{h.c.} $ 
		& $\mathrm{Tr}[\XiSA \Sigma^\dagger \XiAS \Sigma^\dagger \XiAd ]+\mathrm{h.c.} $ 
		& $\mathrm{Tr}[\XiA \Sigma^\dagger] \mathrm{Tr}[ \Sigma \XiSA^\dagger \Sigma \XiSA^\dagger]+ \mathrm{h.c.}$
	    \\
	    & $\mathrm{Tr}[\XiSA \Sigma^\dagger  \XiAd^2 ]+\mathrm{h.c.} $ 
		& $\mathrm{Tr}[\XiSA \Sigma^\dagger \XiAd  \Sigma  \XiSA^\dagger ]+\mathrm{h.c.} $ 
		& $\mathrm{Tr}[\XiA \Sigma^\dagger] \mathrm{Tr}[  \XiAd \Sigma \XiAd^T \Sigma^\dagger]+ \mathrm{h.c.}$
	    \\
	    & $\mathrm{Tr}[\XiSA \XiAd^T\Sigma^\dagger \XiAd]+\mathrm{h.c.} $ 
		& $\mathrm{Tr}[\XiS \Sigma^\dagger \XiAd  \Sigma  \XiA^\dagger ]+\mathrm{h.c.} $
		& $\mathrm{Tr}[\XiSA \Sigma^\dagger \XiSA \Sigma^\dagger \XiA \Sigma^\dagger ]+\mathrm{h.c.} $
	    \\
	    & & $\mathrm{Tr}[\XiAd^2 \Sigma  \XiAd^T \Sigma^\dagger ]$ 
		& $\mathrm{Tr}[\XiSA \Sigma^\dagger \XiAd \Sigma \XiAd^T \Sigma^\dagger ]+\mathrm{h.c.} $
	    \\
	    \hline\hline
	\end{tabular}
    \end{center}
    \caption{Non-derivative operators involving three two-index spurions. As explained in the text, the operators divide into the three following classes: $\mathrm{Tr}[X_1]\mathrm{Tr}[X_2]\mathrm{Tr}[X_3]$, $\mathrm{Tr}[X_1]\mathrm{Tr}[X_2 X_3]$ and $\mathrm{Tr}[X_1 X_2 X_3]$.}
    \label{tab2}
\end{table}

To simplify the classification, instead of considering the two-index spurions that transform differently under $\GF$, we construct objects (see Tab.~\ref{tabspurions}) transforming in the same way as 
$X_i \rightarrow g X_i g^\dagger$ where $X_i=\{ \XiSA \Sigma^\dagger, \Sigma \XiSA^\dagger, \XiAd, \Sigma \XiAd^T \Sigma^\dagger \}$.
As explained in Sec.~\ref{Introduction of spurions}, we restrict to the pseudo-real case (coset $\SU(\NF)/\Sp(\NF)$) since the real case (coset $\SU(\NF)/\SO(\NF)$)  is easily recovered via $\XiS\leftrightarrow \XiA$.
The general procedure is as follows:
we first divide the operators according to the number of spurions and flavour traces.
Then, we construct all the possible combinations involving the objects $X_i$.
Using the cyclic properties of the traces as well as the symmetry properties  of the spurions (transpositions, traceless), we remove some redundant operators.
As an example, in the case of two spurions we have $\mathrm{Tr}[\XiSA \Sigma^\dagger  \Sigma \XiAd^T \Sigma^\dagger] = \mp  \mathrm{Tr} [\XiSA \Sigma^\dagger  \XiAd]$.
The operators involving one and two spurions are listed in Tab.~\ref{tab1}, the ones with three spurions in Tab.~\ref{tab2}, and the ones with four spurions in Tabs~\ref{tab3}, \ref{tab4}, \ref{tab5}, and~\ref{tab6}.

Finally, let us discuss how we can extend the above basis of non-derivative operators to derivative ones.
As by definition the covariant derivatives transform like the fields themselves, it is trivial to construct objects such as $D_\mu X_i$ or $D^2 X_i$ with the desired properties of transformations.
From these objects, one can follow the procedure described previously.
In general, we get a large number of operators and some of them are redundant.
They can be eliminated~\cite{Fearing:1994ga} using 
\begin{equation}
\mathrm{Tr}[(D_\mu A_1)A_2 \cdots A_n+ \cdots+ A_1 A_2 \cdots(D_\mu A_n)]=\partial_\mu \mathrm{Tr}[A_1 A_2 \cdots A_n]=0~,
\label{trace-relation1}
\end{equation}
where $A_1 \cdots A_n \rightarrow  g A_1 \cdots A_n g^\dagger$.
Note that it is enough to restrict to two derivatives in order to get all the NLO operators in composite-Higgs models.

\begin{table}[tb]
    \scriptsize
    \renewcommand{\arraystretch}{1.}
    \begin{center}
	\begin{tabular}{  c c c }
	    \hline\hline
	     Quadratic in $\Sigma$ & Three $\Sigma$ & Four $\Sigma$ \\
	    \hline 
	     $\mathrm{Tr}[\XiA \Sigma^\dagger]^2 \mathrm{Tr}[\XiSA \XiSA^\dagger]+\mathrm{h.c.}$ 
		& $\mathrm{Tr}[\XiA \Sigma^\dagger]^2 \mathrm{Tr}[\XiSA \Sigma^\dagger \XiAd]+\mathrm{h.c.}$ 
		& $\mathrm{Tr}[\XiA \Sigma^\dagger]^4 +\mathrm{h.c.}$
	    \\
	     $\mathrm{Tr}[\XiA \Sigma^\dagger] \mathrm{Tr}[\Sigma \XiA^\dagger] \mathrm{Tr}[\XiSA \XiSA^\dagger]$ 
		& $\mathrm{Tr}[\XiA \Sigma^\dagger]^2 \mathrm{Tr}[ \Sigma \XiSA^\dagger \XiAd]+\mathrm{h.c.}$ 
		& $\mathrm{Tr}[\XiA \Sigma^\dagger]^3 \mathrm{Tr}[ \Sigma \XiA^\dagger] +\mathrm{h.c.}$
	    \\
	     $\mathrm{Tr}[\XiA \Sigma^\dagger]^2 \mathrm{Tr}[\XiAd^2]+\mathrm{h.c.}$ 
		& $\mathrm{Tr}[\XiA \Sigma^\dagger] \mathrm{Tr}[\Sigma \XiA^\dagger] \mathrm{Tr}[\XiSA \Sigma^\dagger \XiAd]+\mathrm{h.c.}$ 
		& $\mathrm{Tr}[\XiA \Sigma^\dagger]^2 \mathrm{Tr}[ \Sigma \XiA^\dagger]^2$
	    \\
	     $\mathrm{Tr}[\XiA \Sigma^\dagger] \mathrm{Tr}[\Sigma \XiA^\dagger]  \mathrm{Tr}[\XiAd^2]$ & 
		& $\mathrm{Tr}[\XiA \Sigma^\dagger]^2 \mathrm{Tr}[\XiSA \Sigma^\dagger \XiSA \Sigma^\dagger]+\mathrm{h.c.}$
	    \\
	     & & $\mathrm{Tr}[\XiA \Sigma^\dagger]^2 \mathrm{Tr}[ \Sigma \XiSA^\dagger \Sigma \XiSA^\dagger]+\mathrm{h.c.}$
	    \\
	     & & $\mathrm{Tr}[\XiA \Sigma^\dagger] \mathrm{Tr}[\Sigma \XiA^\dagger] \mathrm{Tr}[\XiSA \Sigma^\dagger \XiSA \Sigma^\dagger]+\mathrm{h.c.}$
	    \\
	     & & $\mathrm{Tr}[\XiA \Sigma^\dagger]^2 \mathrm{Tr}[\XiAd \Sigma \XiAd^T \Sigma^\dagger]+\mathrm{h.c.}$
	    \\
	     & & $\mathrm{Tr}[\XiA \Sigma^\dagger] \mathrm{Tr}[\Sigma \XiA^\dagger]  \mathrm{Tr}[\XiAd \Sigma \XiAd^T \Sigma^\dagger]$
	    \\
	     \hline\hline
	    \end{tabular}
	\end{center}
    \caption{Same as in Tab.~\ref{tab2} but for the operators involving four spurions and belonging to the classes $\mathrm{Tr}[X_1]\mathrm{Tr}[X_2]\mathrm{Tr}[X_3]\mathrm{Tr}[X_4]$ or  $\mathrm{Tr}[X_1] \mathrm{Tr}[X_2] \mathrm{Tr}[X_3 X_4]$.}
    \label{tab3}
\end{table}

\clearpage

\begin{table}[tb]
    \scriptsize
    \renewcommand{\arraystretch}{1.}
    \begin{center}
	\begin{tabular}{ c  c c  }
	    \hline\hline
	     No $\Sigma$ & Linear in $\Sigma$ & Quadratic in $\Sigma$  \\
	     \hline
	    $\mathrm{Tr}[\XiSA \XiSA^\dagger]^2$ & $\mathrm{Tr}[\XiSA \XiSA^\dagger] \mathrm{Tr}[\XiSA \Sigma^\dagger \XiAd]+\mathrm{h.c.}$ 
		& $\mathrm{Tr}[\XiAd \Sigma^\dagger]\mathrm{Tr}[ \XiSA  \Sigma^\dagger \XiSA \XiAd^\dagger ]+\mathrm{h.c.}$
	    \\
	    $\mathrm{Tr}[\XiAd \XiAd^\dagger] \mathrm{Tr}[\XiAd \XiAd^\dagger]$
		& $\mathrm{Tr}[\XiSA \XiSA^\dagger] \mathrm{Tr}[\XiAS \Sigma^\dagger \XiAd]+\mathrm{h.c.}$ 
		& $\mathrm{Tr}[\XiSA \Sigma^\dagger \XiSA \Sigma^\dagger]\mathrm{Tr}[\XiSA \XiSA^\dagger]+ \mathrm{h.c.}$
	    \\
	    $\mathrm{Tr}[\XiSA \XiSA^\dagger]\mathrm{Tr}[\XiAd^2] $& $\mathrm{Tr}[\XiSA\Sigma^\dagger \XiAd] \mathrm{Tr}[\XiAd^2]+\mathrm{h.c.}$ 
		& $\mathrm{Tr}[\XiSA \Sigma^\dagger \XiSA \Sigma^\dagger]\mathrm{Tr}[\XiAS \XiAS^\dagger]+ \mathrm{h.c.}$
	    \\
	    $\mathrm{Tr}[\XiAd^2]^2 $ & $\mathrm{Tr}[\XiAd \Sigma^\dagger]\mathrm{Tr}[\XiSA  \XiSA^\dagger \XiAd ]+\mathrm{h.c.}$ 
		& $\mathrm{Tr}[\XiSA \Sigma^\dagger \XiSA \Sigma^\dagger]\mathrm{Tr}[\XiAd^2]+ \mathrm{h.c.}$
	    \\
	     & $\mathrm{Tr}[\XiAd \Sigma^\dagger]\mathrm{Tr}[\XiSA  \XiAS^\dagger \XiAd ]+\mathrm{h.c.}$ 
		& $\mathrm{Tr}[\XiSA \XiSA^\dagger] \mathrm{Tr}[\XiAd \Sigma \XiAd^T \Sigma^\dagger]$
	    \\
	    & $\mathrm{Tr}[\XiAd \Sigma^\dagger]\mathrm{Tr}[ \XiAd^3 ]+\mathrm{h.c.}$ 
		& $\mathrm{Tr}[\XiSA \Sigma^\dagger \XiAd] \mathrm{Tr}[ \Sigma \XiSA^\dagger \XiAd]$
	    \\
	    & & $\mathrm{Tr}[\XiAd \Sigma^\dagger \XiAd]  \mathrm{Tr}[\XiAd \Sigma^\dagger \XiAd]+\mathrm{h.c.}$
	    \\
	    & & $\mathrm{Tr}[\XiAd \Sigma^\dagger \XiAd] \mathrm{Tr}[ \Sigma \XiAd^\dagger \XiAd]+\mathrm{h.c.}$
	    \\
	    & & $\mathrm{Tr}[\XiSA \Sigma^\dagger \XiAd]^2 +\mathrm{h.c.}$
	    \\
	    & & $\mathrm{Tr}[\XiAd^2] \mathrm{Tr}[\XiAd \Sigma \XiAd^T \Sigma^\dagger ]$
	    \\
	    & & $\mathrm{Tr}[\XiAd \Sigma^\dagger]\mathrm{Tr}[ \XiSA  \XiAd^\dagger  \XiAS \Sigma^\dagger]+\mathrm{h.c.}$
	    \\
	    & & $\mathrm{Tr}[\XiAd \Sigma^\dagger]\mathrm{Tr}[ \XiSA  \XiAd^\dagger \Sigma \XiAS^\dagger]+\mathrm{h.c.}$
	    \\
	    & & $\mathrm{Tr}[\XiAd \Sigma^\dagger]\mathrm{Tr}[ \XiSA  \XiAd^\dagger \Sigma \XiSA^\dagger]+\mathrm{h.c.}$
	    \\
	    & & $\mathrm{Tr}[\XiAd \Sigma^\dagger]\mathrm{Tr}[ \XiSA  \Sigma^\dagger \XiAd^2]+\mathrm{h.c.}$
	    \\
	    & & $\mathrm{Tr}[\XiAd \Sigma^\dagger]\mathrm{Tr}[ \XiSA \XiAd^T \Sigma^\dagger \XiAd]+\mathrm{h.c.}$
	    \\
	    & & $\mathrm{Tr}[\XiAd \Sigma^\dagger]\mathrm{Tr}[\Sigma \XiSA^\dagger \Sigma \XiSA^\dagger \XiAd]+\mathrm{h.c.}$
	    \\
	    & & $\mathrm{Tr}[\XiAd \Sigma^\dagger]\mathrm{Tr}[\Sigma \XiSA^\dagger  \XiAd^2]+\mathrm{h.c.}$
	    \\
	    & & $\mathrm{Tr}[\XiAd \Sigma^\dagger]\mathrm{Tr}[\XiSA^\dagger  \XiAd \Sigma \XiAd^T ]+\mathrm{h.c.}$
	    \\
	    \hline\hline
	    Three $\Sigma$ & Four $\Sigma$ &
	    \\
	    \hline
	    $\mathrm{Tr}[\XiSA \Sigma^\dagger \XiSA \Sigma^\dagger] \mathrm{Tr}[\XiSA \Sigma^\dagger \XiAd]+\mathrm{h.c.}$ 
		& $\mathrm{Tr}[\XiSA \Sigma^\dagger \XiSA \Sigma^\dagger] \mathrm{Tr}[ \Sigma \XiSA^\dagger  \Sigma \XiSA^\dagger]$ &
	    \\
	    $\mathrm{Tr}[\XiSA \Sigma^\dagger \XiSA \Sigma^\dagger] \mathrm{Tr}[\Sigma \XiSA^\dagger \XiAd]+\mathrm{h.c.}$ 
		& $\mathrm{Tr}[\XiAd \Sigma^\dagger \XiAd \Sigma^\dagger] \mathrm{Tr}[ \XiAd \Sigma^\dagger \XiAd \Sigma^\dagger]+\mathrm{h.c.}$ &
	    \\
	    $\mathrm{Tr}[\XiSA \Sigma^\dagger \XiSA \Sigma^\dagger] \mathrm{Tr}[\XiAS \Sigma^\dagger \XiAd]+\mathrm{h.c.}$ 
		& $\mathrm{Tr}[\XiAd \Sigma^\dagger \XiAd \Sigma^\dagger] \mathrm{Tr}[ \Sigma \XiAd^\dagger  \Sigma \XiAd^\dagger]+\mathrm{h.c.}$  &
	    \\
	    $\mathrm{Tr}[\XiSA \Sigma^\dagger \XiSA \Sigma^\dagger] \mathrm{Tr}[\Sigma \XiAS^\dagger \XiAd]+\mathrm{h.c.}$ 
		& $\mathrm{Tr}[\XiSA \Sigma^\dagger \XiSA \Sigma^\dagger] \mathrm{Tr}[\XiAd \Sigma \XiAd^T  \Sigma^\dagger]+\mathrm{h.c.}$ &
	    \\
	    $\mathrm{Tr}[\XiSA \Sigma^\dagger \XiAd] \mathrm{Tr}[\XiAd \Sigma \XiAd^T \Sigma^\dagger]+\mathrm{h.c.}$ 
		& $\mathrm{Tr}[\XiAd \Sigma \XiAd^T  \Sigma^\dagger]^2$ &
	    \\
	    $\mathrm{Tr}[\XiAd \Sigma^\dagger]\mathrm{Tr}[ \XiSA \Sigma^\dagger \XiSA \Sigma^\dagger \XiAd]+\mathrm{h.c.}$ 
		&  $\mathrm{Tr}[\XiAd \Sigma^\dagger]\mathrm{Tr}[\XiSA \Sigma^\dagger \XiSA \Sigma^\dagger \XiAd \Sigma^\dagger]+\mathrm{h.c.}$ &
	    \\
	    $\mathrm{Tr}[\XiAd \Sigma^\dagger]\mathrm{Tr}[ \XiSA \Sigma^\dagger \XiAS \Sigma^\dagger \XiAd]+\mathrm{h.c.}$ 
		&  $\mathrm{Tr}[\XiAd \Sigma^\dagger]\mathrm{Tr}[\Sigma\XiSA^\dagger \Sigma \XiSA^\dagger \Sigma \XiAd^\dagger]+\mathrm{h.c.}$ &
	    \\
	    $\mathrm{Tr}[\XiAd \Sigma^\dagger]\mathrm{Tr}[ \XiSA \Sigma^\dagger \XiAd \Sigma \XiSA^\dagger]+\mathrm{h.c.}$ 
		&  $\mathrm{Tr}[\XiAd \Sigma^\dagger]\mathrm{Tr}[\XiSA \Sigma^\dagger \XiAd \Sigma \XiAd^T \Sigma^\dagger]+\mathrm{h.c.}$ &
	    \\
	    $\mathrm{Tr}[\XiAd \Sigma^\dagger]\mathrm{Tr}[ \XiSA \Sigma^\dagger \XiAd \Sigma \XiAS^\dagger]+\mathrm{h.c.}$ 
		&  $\mathrm{Tr}[\XiAd \Sigma^\dagger]\mathrm{Tr}[\Sigma \XiSA^\dagger \Sigma \XiAd^T \Sigma^\dagger \XiAd]+\mathrm{h.c.}$ &
	    \\
	    $\mathrm{Tr}[\XiAd \Sigma^\dagger]\mathrm{Tr}[\Sigma \XiSA^\dagger \Sigma  \XiAS^\dagger \XiAd ]+\mathrm{h.c.}$
		& $\mathrm{Tr}[\XiSA \Sigma^\dagger \XiSA \Sigma^\dagger]^2+ \mathrm{h.c.}$ & 
	    \\
	    $\mathrm{Tr}[\XiAd \Sigma^\dagger]\mathrm{Tr}[ \XiAd^2 \Sigma  \XiAd^T \Sigma^\dagger ]+\mathrm{h.c.}$ & &
	    \\
	    \hline\hline
	\end{tabular}
    \end{center}
    \caption{Same as in Tab.~\ref{tab2} but for the operators involving four spurions and belonging to the classes $\mathrm{Tr}[X_1]\mathrm{Tr}[X_2 X_3 X_4]$ or  $\mathrm{Tr}[X_1 X_2] \mathrm{Tr}[X_3 X_4 ]$.}
    \label{tab4}
\end{table}

\begin{table}[tb]
    \scriptsize
    \renewcommand{\arraystretch}{1.}
    \begin{center}
	\begin{tabular}{ c  c c  }
	    \hline\hline
	     No $\Sigma$ & Linear in $\Sigma$ & Quadratic in $\Sigma$ 
	     \\
	     \hline
	    $\mathrm{Tr}[\XiSA \XiSA^\dagger \XiSA \XiSA^\dagger]$  & $\mathrm{Tr}[\XiSA \Sigma^\dagger \XiSA \XiSA^\dagger \XiAd]+\mathrm{h.c.}$  
		& $\mathrm{Tr}[\XiSA  \Sigma^\dagger \XiSA \Sigma^\dagger \XiSA  \XiSA^\dagger]+\mathrm{h.c.}$ 
	    \\
	    $\mathrm{Tr}[\XiAd^4]$ & $\mathrm{Tr}[\XiSA  \XiSA^\dagger \XiSA \Sigma^\dagger \XiAd]+\mathrm{h.c.}$  
		& $\mathrm{Tr}[\XiSA  \Sigma^\dagger \XiSA  \XiSA^\dagger \Sigma \XiSA^\dagger]+\mathrm{h.c.}$ 
	    \\
	    $\mathrm{Tr}[\XiSA \XiSA^\dagger \XiAd^2]$  & $\mathrm{Tr}[\XiSA \Sigma^\dagger \XiAd^3]+\mathrm{h.c.}$  
		& $\mathrm{Tr}[\XiAd^3 \Sigma \XiAd^T \Sigma^\dagger ]$ 
	    \\
	    $\mathrm{Tr}[\XiSA \XiAd^T \XiSA^\dagger \XiAd]$  & $\mathrm{Tr}[\XiSA \XiAd^T \Sigma^\dagger \XiAd^2]+\mathrm{h.c.}$ 
		& $\mathrm{Tr}[\XiAd^2 \Sigma \XiAd^T \XiAd^T \Sigma^\dagger ]$ 
	    \\
	    $\mathrm{Tr}[\XiAd \XiSA^\dagger \XiAd \XiAS^\dagger]$  & $\mathrm{Tr}[\XiSA \Sigma^\dagger \XiSA \XiAS^\dagger \XiAd]+\mathrm{h.c.}$ 
		& $\mathrm{Tr}[\XiSA \XiSA^\dagger \XiAd \Sigma \XiAd^T \Sigma^\dagger ]+\mathrm{h.c.}$ 
	    \\
	    $\mathrm{Tr}[\XiAd \XiAd^\dagger \XiAd \XiAd^\dagger]+\mathrm{h.c.}$  
		& $\mathrm{Tr}[\XiSA \Sigma^\dagger \XiAS \XiSA^\dagger \XiAd]+\mathrm{h.c.}$  
		& $\mathrm{Tr}[\XiSA \Sigma^\dagger  \XiAd \XiSA \Sigma^\dagger \XiAd]+\mathrm{h.c.}$ 
	    \\
	    $\mathrm{Tr}[\XiAd \XiAd^\dagger \XiAd^2]+\mathrm{h.c.}$  
		& $\mathrm{Tr}[\XiSA \Sigma^\dagger \XiAS \XiAS^\dagger \XiAd]+\mathrm{h.c.}$  
		& $\mathrm{Tr}[\XiSA \Sigma^\dagger \XiAd \XiSA \XiAd^T \Sigma^\dagger ]+\mathrm{h.c.}$ 
	    \\
	    $\mathrm{Tr}[\XiAd \XiAd^T \XiAd^\dagger \XiAd]+\mathrm{h.c.}$  
		& $\mathrm{Tr}[\XiSA  \XiAS^\dagger \XiSA \Sigma^\dagger \XiAd]+\mathrm{h.c.}$  
		& $\mathrm{Tr}[\XiSA \Sigma^\dagger \XiAd \Sigma \XiSA^\dagger \XiAd]$ 
	    \\
	    & $\mathrm{Tr}[\XiSA  \XiAS^\dagger \XiAS \Sigma^\dagger \XiAd]+\mathrm{h.c.}$ 
		& $\mathrm{Tr}[\XiSA \Sigma^\dagger \XiAd^2\Sigma \XiSA^\dagger]$ 
	    \\
	    & $\mathrm{Tr}[\XiSA \Sigma^\dagger  \XiAd \XiAS  \XiAS^\dagger]+\mathrm{h.c.}$ & 
	    $\mathrm{Tr}[\XiSA \Sigma^\dagger \XiSA \XiAd^T \Sigma^\dagger \XiAd ]+\mathrm{h.c.}$ 
	    \\
	    & & $\mathrm{Tr}[\XiSA \Sigma^\dagger \XiSA \Sigma^\dagger \XiAd^2 ]+\mathrm{h.c.}$ 
	    \\
	    & & $\mathrm{Tr}[\XiSA \Sigma^\dagger \XiSA \Sigma^\dagger \XiAS \XiAS^\dagger]+\mathrm{h.c.}$ 
	    \\
	    & & $\mathrm{Tr}[\XiSA \Sigma^\dagger \XiSA  \XiAS^\dagger \XiAS \Sigma^\dagger  ]+\mathrm{h.c.}$ 
	    \\
	    & & $\mathrm{Tr}[\XiAd \Sigma^\dagger \XiSA  \XiAS^\dagger \Sigma \XiAd^\dagger  ]+\mathrm{h.c.}$ 
	    \\
	    & & $\mathrm{Tr}[\XiSA  \XiSA^\dagger \Sigma  \XiAS^\dagger \XiAS  \Sigma^\dagger  ]$ 
	    \\
	    & & $\mathrm{Tr}[\XiSA \Sigma^\dagger  \XiAS \Sigma^\dagger   \XiSA \XiAS ^\dagger]+\mathrm{h.c.}$ 
	    \\
	    & & $\mathrm{Tr}[\XiSA \Sigma^\dagger  \XiAS \Sigma^\dagger  \XiAd^2]+\mathrm{h.c.}$ 
	    \\
	    & & $\mathrm{Tr}[\XiAd \Sigma^\dagger  \XiAd   \XiAd^T \Sigma^\dagger \XiAd]+\mathrm{h.c.}$ 
	    \\
	    & & $\mathrm{Tr}[\XiSA   \XiAS^\dagger   \XiAd \Sigma \XiAd^T \Sigma^\dagger]+\mathrm{h.c.}$ 
	    \\
	    & & $\mathrm{Tr}[\XiAd \Sigma^\dagger   \XiAd   \XiAd \Sigma^\dagger \XiAd]+\mathrm{h.c.}$ 
	    \\
	    & & $\mathrm{Tr}[\XiSA \Sigma^\dagger \XiAd  \XiAS \XiAd^T \Sigma^\dagger]+\mathrm{h.c.}$ 
	    \\
	    & & $\mathrm{Tr}[\XiAd \Sigma^\dagger \XiAd \Sigma \XiAd^\dagger \XiAd]+\mathrm{h.c.}$ 
	    \\
	    & & $\mathrm{Tr}[\XiAd \Sigma^\dagger \XiAd^2 \Sigma \XiAd^\dagger]+\mathrm{h.c.}$ 
	    \\
	     \hline\hline
	     Three $\Sigma$ & Four $\Sigma$  &
	     \\
	     \hline
	    $\mathrm{Tr}[\XiSA  \Sigma^\dagger \XiSA  \Sigma^\dagger \XiSA  \Sigma^\dagger \XiAd  ]+\mathrm{h.c.}$ 
		& $\mathrm{Tr}[\XiSA  \Sigma^\dagger \XiSA  \Sigma^\dagger \XiSA  \Sigma^\dagger \XiSA  \Sigma^\dagger]+\mathrm{h.c.}$ &
	    \\
	    $\mathrm{Tr}[\XiSA  \Sigma^\dagger \XiSA \Sigma^\dagger \XiSA  \Sigma^\dagger \XiAd]+\mathrm{h.c.}$ 
		& $\mathrm{Tr}[\XiAd  \Sigma \XiAd^T  \Sigma^\dagger \XiAd  \Sigma \XiAd^T  \Sigma^\dagger]$ &
	    \\
	    $\mathrm{Tr}[\XiSA  \Sigma^\dagger \XiSA \Sigma^\dagger \XiAd \Sigma  \XiSA^\dagger]+\mathrm{h.c.}$  
		& $\mathrm{Tr}[\XiSA  \Sigma^\dagger \XiAd \Sigma \XiSA^\dagger  \Sigma \XiAd^T  \Sigma^\dagger]$ &
	    \\
	    $\mathrm{Tr}[\XiSA  \Sigma^\dagger \XiAd^2 \Sigma \XiAd^T  \Sigma^\dagger]+\mathrm{h.c.}$ 
		& $\mathrm{Tr}[\XiSA  \Sigma^\dagger \XiSA  \Sigma^\dagger \XiAd \Sigma \XiAd^T  \Sigma^\dagger]+\mathrm{h.c.}$ & 
	    \\
	    $\mathrm{Tr}[\XiSA  \Sigma^\dagger \XiAd \Sigma \XiAd^T \Sigma^\dagger  \XiAd]+\mathrm{h.c.}$ 
		& $\mathrm{Tr}[\XiAd  \Sigma^\dagger \XiSA  \Sigma^\dagger \XiAS  \Sigma^\dagger \XiAd  \Sigma^\dagger]+\mathrm{h.c.}$& 
	    \\
	    $\mathrm{Tr}[\XiSA  \Sigma^\dagger \XiSA  \Sigma^\dagger \XiAS  \Sigma^\dagger \XiAd]+\mathrm{h.c.}$ 
		& $\mathrm{Tr}[\XiAd  \Sigma^\dagger \XiAd  \Sigma^\dagger \XiAd  \Sigma \XiAd^T  \Sigma^\dagger]+\mathrm{h.c.}$ & 
	    \\
	    $\mathrm{Tr}[\XiSA  \Sigma^\dagger \XiSA  \Sigma^\dagger \XiAd \XiAS  \Sigma^\dagger ]+\mathrm{h.c.}$ 
		& $\mathrm{Tr}[\XiAd  \Sigma^\dagger \XiAd  \Sigma \XiAd^\dagger  \Sigma \XiAd^T  \Sigma^\dagger]+\mathrm{h.c.}$ & 
	    \\
	    $\mathrm{Tr}[\XiSA  \Sigma^\dagger \XiSA  \Sigma^\dagger \XiAd  \Sigma \XiAS^\dagger ]+\mathrm{h.c.}$ & & 
	    \\
	    $\mathrm{Tr}[\XiSA  \Sigma^\dagger \XiAS  \Sigma^\dagger   \XiSA\Sigma^\dagger \XiAd ]+\mathrm{h.c.}$ & & 
	    \\
	    $\mathrm{Tr}[\XiSA  \Sigma^\dagger \XiAS  \Sigma^\dagger \XiAd  \Sigma \XiSA^\dagger ]+\mathrm{h.c.}$ & & 
	    \\
	    $\mathrm{Tr}[\XiSA  \Sigma^\dagger \XiAS  \Sigma^\dagger \XiAd  \Sigma \XiAS^\dagger ]+\mathrm{h.c.}$ & & 
	    \\
	    \hline\hline
	\end{tabular}
    \end{center}
    \caption{Same as in Tab.~\ref{tab2} but for the operators involving four spurions and belonging to the class
	$\mathrm{Tr}[X_1 X_2 X_3 X_4]$.}
    \label{tab5}
\end{table}

\clearpage

\begin{table}[tb]
    \scriptsize
    \renewcommand{\arraystretch}{1.}
    \begin{center}
	\begin{tabular}{ c  c c c }
\hline\hline
No $\Sigma$ & Linear in $\Sigma$ & Quadratic in $\Sigma$ &  Three $\Sigma$
\\
\hline
$\mathrm{Tr}[\XiF\cdot \XiF^\dagger]$  
& $\mathrm{Tr}[\XiF\cdot \XiF^T \Sigma^\dagger]+\mathrm{h.c.}$ 
& $\mathrm{Tr}[\Sigma \XiAd^T \Sigma^\dagger \XiF\cdot \XiF^\dagger]$  
& $\mathrm{Tr}[\Xi_A \Sigma^\dagger]^2 \mathrm{Tr}[\XiF\cdot \XiF^T \Sigma^\dagger]+\mathrm{h.c.}$ 
\\
$\mathrm{Tr}[\XiAd \XiF\cdot \XiF^\dagger]$    
& $\mathrm{Tr}[\XiSA \Sigma^\dagger\XiF\cdot \XiF^\dagger]+\mathrm{h.c.}$  
& $\mathrm{Tr}[\XiSA \Sigma^\dagger \XiF\cdot \XiF^T \Sigma^\dagger]+\mathrm{h.c.}$ 
& $\mathrm{Tr}[\Sigma \Xi_A^\dagger]^2 \mathrm{Tr}[\XiF\cdot \XiF^T \Sigma^\dagger]+\mathrm{h.c.}$ 
\\
$\mathrm{Tr}[\XiSA^\dagger \XiF\cdot \XiF^T]+\mathrm{h.c.}$	
& $\mathrm{Tr}[\XiAd \XiF\cdot \XiF^T \Sigma^\dagger]+\mathrm{h.c.}$  
& $\mathrm{Tr}[\Xi_A \Sigma^\dagger] \mathrm{Tr}[\XiF\cdot \XiF^T \Sigma^\dagger]+\mathrm{h.c.}$ 
& $\mathrm{Tr}[\Xi_A \Sigma^\dagger] \mathrm{Tr}[\Sigma \Xi_A^\dagger] \mathrm{Tr}[\XiF\cdot \XiF^T \Sigma^\dagger]+\mathrm{h.c.}$ 
\\
$\mathrm{Tr}[\XiF\cdot \XiF^\dagger]^2$ 
&  $\mathrm{Tr}[\Xi_A \Sigma^\dagger] \mathrm{Tr}[\XiF\cdot \XiF^\dagger]+\mathrm{h.c.}$ 
& $\mathrm{Tr}[ \Sigma \Xi_A^\dagger] \mathrm{Tr}[\XiF\cdot \XiF^T \Sigma^\dagger]+\mathrm{h.c.}$ 
& $\mathrm{Tr}[\Xi_A \Sigma^\dagger] \mathrm{Tr}[\Sigma \XiAd^T \Sigma^\dagger \XiF\cdot \XiF^\dagger]+\mathrm{h.c.}$ 
\\
$\mathrm{Tr}[\XiF\cdot \XiF^\dagger] \mathrm{Tr}[\XiSA \XiSA^\dagger] $ 	
& $\mathrm{Tr}[\Xi_A \Sigma^\dagger] \mathrm{Tr}[\XiAd \XiF\cdot \XiF^\dagger]+\mathrm{h.c.}$ 
& $\mathrm{Tr}[\Xi_A \Sigma^\dagger]^2 \mathrm{Tr}[\XiF\cdot \XiF^\dagger]+\mathrm{h.c.}$ 
& $\mathrm{Tr}[\Xi_A \Sigma^\dagger] \mathrm{Tr}[\XiSA \Sigma^\dagger \XiF\cdot \XiF^T\Sigma^\dagger]+\mathrm{h.c.}$ 
\\
$\mathrm{Tr}[\XiF\cdot \XiF^\dagger] \mathrm{Tr}[\XiAd^2] $ 	
& $\mathrm{Tr}[\XiF\cdot \XiF^\dagger] \mathrm{Tr}[\XiF \cdot \XiF^T \Sigma^\dagger]+\mathrm{h.c.} $  
& $\mathrm{Tr}[\Xi_A \Sigma^\dagger] \mathrm{Tr}[\Sigma \Xi_A^\dagger] \mathrm{Tr}[\XiF\cdot \XiF^\dagger]$ 
& $\mathrm{Tr}[\Sigma \Xi_A ^\dagger] \mathrm{Tr}[ \XiSA \Sigma^\dagger \XiF\cdot \XiF^T \Sigma^\dagger]+\mathrm{h.c.}$ 
\\
$\mathrm{Tr}[\XiF\cdot \XiF^\dagger~ \XiF\cdot \XiF^\dagger] $ 
& $\mathrm{Tr}[\XiF\cdot \XiF^\dagger] \mathrm{Tr}[\XiSA \Sigma^\dagger \XiAd]+\mathrm{h.c.} $ 
& $\mathrm{Tr}[\XiF\cdot \XiF^T \Sigma^\dagger]^2+\mathrm{h.c.}$ 
& $\mathrm{Tr}[\Sigma \XiSA^\dagger \Sigma \XiAd^T \Sigma^\dagger \XiF\cdot \XiF^\dagger]+\mathrm{h.c.}$ 
\\
$\mathrm{Tr}[\XiF\cdot \XiF^T~ \XiF^*\cdot \XiF^\dagger] $ 
& $\mathrm{Tr}[\XiF\cdot \XiF^T\Sigma^\dagger] \mathrm{Tr}[\XiSA \XiSA^\dagger ]+\mathrm{h.c.} $  & $\mathrm{Tr}[\XiF\cdot \XiF^T \Sigma^\dagger] \mathrm{Tr}[\Sigma \XiF^* \cdot \XiF^\dagger]$ 
& $\mathrm{Tr}[\XiF \cdot \XiF^T \Sigma^\dagger] \mathrm{Tr}[\XiSA \Sigma^\dagger  \XiSA \Sigma^\dagger] +\mathrm{h.c.}$
\\
$\mathrm{Tr}[\XiSA \XiSA^\dagger ~\XiF\cdot \XiF^\dagger] $ 
& $\mathrm{Tr}[\XiF\cdot \XiF^T\Sigma^\dagger] \mathrm{Tr}[\XiAd^2]+\mathrm{h.c.} $  
& $\mathrm{Tr}[\Xi_A \Sigma^\dagger] \mathrm{Tr}[\XiSA \Sigma^\dagger \XiF\cdot \XiF^\dagger]+\mathrm{h.c.}$ 
&  $\mathrm{Tr}[\XiF \cdot \XiF^T \Sigma^\dagger] \mathrm{Tr}[\Sigma \XiSA^\dagger \Sigma \XiSA^\dagger] +\mathrm{h.c.}$
\\
$\mathrm{Tr}[\XiAd \XiAd^\dagger ~\XiF\cdot \XiF^\dagger]+\mathrm{h.c.} $ 
& $\mathrm{Tr}[\XiF\cdot \XiF^\dagger~ \XiF \cdot \XiF^T \Sigma^\dagger] +\mathrm{h.c.} $ 
&$\mathrm{Tr}[\Xi_A \Sigma^\dagger] \mathrm{Tr}[\XiAd \XiF\cdot \XiF^T \Sigma^\dagger]+\mathrm{h.c.}$ 
& $\mathrm{Tr}[\XiF \cdot \XiF^T \Sigma^\dagger] \mathrm{Tr}[\XiAd\Sigma  \XiAd^T \Sigma^\dagger] +\mathrm{h.c.}$
\\
$\mathrm{Tr}[\XiAd^2 ~\XiF\cdot \XiF^\dagger] $ 
& $\mathrm{Tr}[\XiSA \Sigma^\dagger \XiAd~ \XiF \cdot \XiF^\dagger] +\mathrm{h.c.} $ 
& $\mathrm{Tr}[\Sigma \Xi_A^\dagger] \mathrm{Tr}[\XiAd  \XiF\cdot \XiF^T \Sigma^\dagger]+\mathrm{h.c.}$ 
& $\mathrm{Tr}[\XiAd \Sigma^\dagger \XiAd \Sigma^\dagger \XiF\cdot \XiF^T \Sigma^\dagger]+\mathrm{h.c.}$ 
\\
$\mathrm{Tr}[\XiSA^\dagger \XiAd ~ \XiF \cdot \XiF^T] +\mathrm{h.c.} $	
& $\mathrm{Tr}[\XiSA \XiAd^T \Sigma^\dagger ~ \XiF \cdot \XiF^\dagger] +\mathrm{h.c.} $ 
& $\mathrm{Tr}[\XiF \cdot \XiF^\dagger] \mathrm{Tr}[\XiSA \Sigma^\dagger \XiSA \Sigma^\dagger] +\mathrm{h.c.} $ 
& $\mathrm{Tr}[\XiAd \Sigma \XiAd^T \Sigma^\dagger \XiF\cdot \XiF^T \Sigma^\dagger]+\mathrm{h.c.}$ 
\\
$\mathrm{Tr}[\Xi_A \Sigma^\dagger] \mathrm{Tr}[ \XiSA^\dagger \XiF\cdot \XiF^T]+\mathrm{h.c.}$ 
& $\mathrm{Tr}[\Sigma \XiSA^\dagger \XiAd  ~ \XiF \cdot \XiF^\dagger] +\mathrm{h.c.} $ 
& $\mathrm{Tr}[\XiF \cdot \XiF^\dagger] \mathrm{Tr}[\XiAd \Sigma^\dagger \XiAd \Sigma^\dagger] +\mathrm{h.c.} $ 
& 
\\
	& $\mathrm{Tr}[ \XiSA \XiSA^\dagger   ~ \XiF \cdot \XiF^T \Sigma^\dagger] +\mathrm{h.c.} $ 
& $\mathrm{Tr}[\XiF \cdot \XiF^\dagger] \mathrm{Tr}[\XiAd \Sigma  \XiAd^T \Sigma^\dagger] $
&  
\\
	& $\mathrm{Tr}[ \XiSA \XiAS^\dagger   ~ \XiF \cdot \XiF^T \Sigma^\dagger] +\mathrm{h.c.} $ 
& $\mathrm{Tr}[\XiF \cdot \XiF^T \Sigma^\dagger] \mathrm{Tr}[\XiSA \Sigma^\dagger  \XiAd] +\mathrm{h.c.}$
& 
\\
	& $\mathrm{Tr}[\XiSA^\dagger \Sigma  \XiSA ^\dagger   ~ \XiF \cdot \XiF^T ] +\mathrm{h.c.} $ & $\mathrm{Tr}[\XiF \cdot \XiF^T \Sigma^\dagger] \mathrm{Tr}[\Sigma \XiSA ^\dagger  \XiAd] +\mathrm{h.c.}$
&
\\
	& $\mathrm{Tr}[\XiAd^\dagger \Sigma  \XiAd ^\dagger   ~ \XiF \cdot \XiF^T ] +\mathrm{h.c.} $ 
& $\mathrm{Tr}[\XiAd \Sigma^\dagger] \mathrm{Tr}[\Sigma  \XiSA^\dagger   ~ \XiF \cdot \XiF^\dagger ] +\mathrm{h.c.} $ 
& 
\\
	& $\mathrm{Tr}[\XiAd^2    ~ \XiF \cdot \XiF^T ~\Sigma^\dagger] +\mathrm{h.c.} $ 
& $\mathrm{Tr}[\XiF \cdot \XiF^T \Sigma^\dagger~ \XiF \cdot \XiF^T \Sigma^\dagger] +\mathrm{h.c.} $ 
& 
\\
& $\mathrm{Tr}[\XiAd^T \Sigma^\dagger \XiAd   ~ \XiF \cdot \XiF^T ] +\mathrm{h.c.} $ 
& $\mathrm{Tr}[\XiSA \Sigma^\dagger  \XiSA \Sigma^\dagger~ \XiF \cdot \XiF^\dagger] +\mathrm{h.c.} $ 
& 
\\
& $\mathrm{Tr}[\Sigma \Xi_A ^\dagger] \mathrm{Tr}[ \XiSA^\dagger \XiF\cdot \XiF^T]+\mathrm{h.c.}$ 
& $\mathrm{Tr}[\XiSA \Sigma^\dagger  \XiAS \Sigma^\dagger~ \XiF \cdot \XiF^\dagger] +\mathrm{h.c.} $ 
&
\\
&
& $\mathrm{Tr}[\Sigma \XiSA^\dagger  \XiSA \Sigma^\dagger~ \XiF \cdot \XiF^\dagger]$ 
&
\\
&
& $\mathrm{Tr}[\Sigma \XiAd^\dagger  \XiAd \Sigma^\dagger~ \XiF \cdot \XiF^\dagger]+\mathrm{h.c.}$ 
&
\\
&
& $\mathrm{Tr}[\Sigma \XiAd^T \XiAd^T \Sigma^\dagger  ~ \XiF \cdot \XiF^\dagger]$ 
&
\\
&
& $\mathrm{Tr}[ \XiAd \Sigma \XiAd^T \Sigma^\dagger  ~ \XiF \cdot \XiF^\dagger]+\mathrm{h.c.}$ 
&
\\
&
& $\mathrm{Tr}[ \XiSA \Sigma^\dagger \XiSA \Sigma^\dagger   ~ \XiF \cdot \XiF^T \Sigma^\dagger]+\mathrm{h.c.}$ 
&
\\
&
& $\mathrm{Tr}[ \XiSA \Sigma^\dagger \XiAd   ~ \XiF \cdot \XiF^T \Sigma^\dagger]+\mathrm{h.c.}$ 
&
\\
&
&$\mathrm{Tr}[ \XiSA  \XiAd^T \Sigma^\dagger   ~ \XiF \cdot \XiF^T \Sigma^\dagger]+\mathrm{h.c.}$ 
&
\\
&
&$\mathrm{Tr}[ \XiSA^\dagger \Sigma  \XiAd^T \Sigma^\dagger   ~ \XiF \cdot \XiF^T]+\mathrm{h.c.}$ 
&
\\
\hline\hline
	\end{tabular}
    \end{center}
    \caption{Same as in Tab.~\ref{tab2} but for the operators involving spurions in the fundamental representation}
    \label{tab6}
\end{table}

\clearpage

\bibliography{Breaking}


\end{document}